\documentclass[iop,apj]{emulateapj}

\usepackage{natbib}
\usepackage{afterpage}
\usepackage{amsmath,amssymb,amstext}

\usepackage[colorlinks=blue,linkcolor=blue,urlcolor=blue,citecolor=blue]{hyperref} 

\usepackage[all]{hypcap}
\usepackage{tabularx}
\usepackage{float}
\usepackage{natbib}
\usepackage[usenames, dvipsnames]{color}

\usepackage[T1]{fontenc}
\usepackage{times} 
\usepackage[utf8]{inputenc}
\usepackage{mathptmx}

\newcommand{\kms}{km\,s$^{-1}$}

\begin{document}
\title{Quasi-Periodic Radio Bursts Associated with Fast-mode Waves near a Magnetic Null Point}
\author{Pankaj Kumar\altaffilmark{1,2\footnote{NPP Fellow}}, Valery M. Nakariakov\altaffilmark{3,4,5}, Kyung-Suk Cho\altaffilmark{2,6}}
\affil{$^1$Heliophysics Science Division, NASA Goddard Space Flight Center, Greenbelt, MD, 20771, USA}
\affil{$^2$Korea Astronomy and Space Science Institute (KASI), Daejeon, 305-348, Republic of Korea}
\affil{$^3$Centre for Fusion, Space and Astrophysics, Department of Physics, University of Warwick, CV4 7AL, UK}
\affil{$^4$School of Space Research, Kyung Hee University, Yongin, 446-701, Gyeonggi, Republic of Korea}
\affil{$^5$St Petersburg Branch, Special Astrophysical Observatory, Russian Academy of Sciences, 196140, St Petersburg, Russia}
\affil{$^6$University of Science and Technology, Daejeon 305-348, Republic of Korea}
\email{pankaj.kumar@nasa.gov}

\begin{abstract}
This paper presents an observation of quasi-periodic rapidly-propagating waves observed in the AIA 171/193 \AA~ channels during the impulsive phase of an M1.9 flare occurred on 7 May 2012. The instant period was found to decrease from 240~s to 120~s, the speed of the wave fronts was in the range of $\sim$664--1416~\kms. Almost simultaneously, quasi-periodic bursts with similar instant periods, $\sim$70~s and $\sim$140~s, occur in the microwave emission and in decimetric type IV, and type III radio bursts, and in the soft X-ray emission. The magnetic field configuration of the flare site was consistent with a breakout topology, i.e., a quadrupolar field along with a magnetic null point. 
 The quasi-periodic rapidly-propagating wavefronts of the EUV emission are interpreted as a fast magnetoacoustic wave train. The observations suggest that the fast-mode waves are generated during the quasi-periodic magnetic reconnection in the cusp-region above the flare arcade loops. For the first time, we provide the evidence of a tadpole wavelet signature at about 70--140~s in decimetric (245/610~MHz) radio bursts, along with the direct observation of a coronal fast-mode wave train in EUV. In addition, at AIA 131/193 \AA\ we observed quasi-periodic EUV disturbances with the periods of 95~s and 240~s propagating downward at the apparent speed of 172--273~\kms. The nature of these downward propagating disturbances is not revealed, but they could be connected with magnetoacoustic waves or periodically shrinking loops.       
\end{abstract}

\section{INTRODUCTION}
Rapidly propagating quasi-periodic EUV disturbances moving approximately at the Alfv\'en speed
have been theoretically predicted as impulsively generated fast magnetoacoustic waves (the fast mode), which are guided by high-density plasma structures of the solar corona (e.g., loops) serving as a waveguide \citep{roberts1983,roberts1984}. The quasi-periodic pattern develops because of the geometrical dispersion caused by a transverse non-uniformity of the fast speed. Fast-mode wave trains may be excited by impulsive energy releases in solar flares. Fast-mode wave trains are important for understanding magnetic connectivity channels, quasi-periodic energy release, particle acceleration, and plasma heating in solar flares.

Possibly the first evidence of fast-mode wave trains was found in the solar corona during eclipses. \citet{williams2001,williams2002} discovered a 6-s oscillation propagating at the speed of $\sim$2100~\kms~ in an active-region loop observed with the Solar Eclipse Corona Imaging System (SECIS), and interpreted it as an impulsively generated fast magnetoacoustic wave train. \citet{nakariakov2004} numerically modelled the characteristic time-evolution of these short-period fast wave trains along a coronal loop. They found that these wave trains have a characteristic tadpole wavelet signature where a narrowband tail precedes a broadband head. An important feature of the wave train signature is the decrease in the oscillation period in time, which is connected with the decrease in the phase and group speeds of fast magnetoacoustic modes of a field-aligned plasma waveguide with the wave number \citep{nakariakov2004}. In other words, fast modes with longer periods propagate faster than the shorter period modes. A similar tadpole feature was detected in the wavelet power spectrum of the signal observed with SECIS, which supported the interpretation in terms of guided fast-mode wave trains. Further modelling of the dispersive evolution of broadband fast wave trains includes the study of this effect in cylindrical waveguides \citep{2015ApJ...806...56O, 2015ApJ...814..135S},  consideration of the finite-beta effects \citep{2016ApJ...833..114C}, and effects of the transverse profile of the fast magnetoacoustic speed \citep{2017ApJ...836....1Y}.

More recently, fast-mode wave trains were discovered in the high-resolution SDO/AIA images.  Using SDO/AIA 171 and 193~\AA~ observations, \citet{Liu2011,liu2012} reported the direct imaging of quasi-periodic fast propagating waves at the speed $\sim$2000~\kms and period of $\sim$2--3~min. \citet{Kumar2013blob} detected fast-wave fronts with a three minute periodicity, propagating at the speed $\sim$1000~\kms~along open structures in an active region behind a primary fast shock wave. These wave trains were shown to be consistent with periodically excited fast magnetoacoustic waves \citep{2011ApJ...740L..33O}. \citet{nistico2014} observed  fast wave trains (speed of $\sim$1000~\kms, period $\sim$1~min) along two different structures in an active region, and demonstrated the similarity between the detected wave patterns and the results of numerical modelling of the evolution of a broadband fast magnetoacoustic pulse \citep[see also][]{2013A&A...560A..97P, 2016ApJ...833...51Y}. Similar propagating fast wave trains have also been detected in other flare events \citep{shen2013,yuan2013,kumar2014,kumar2016}. These fast-mode wave trains are closely associated with flare quasi-periodic energy releases, and generally detected behind the coronal mass ejection (CME) front. \citet{2015A&A...581A..78Z} presented the first simultaneous observation of fast and slow waves propagating along the same coronal structure, interpreted as co-existing fast and slow magnetoacoustic waves generated by different mechanisms.

Evidence of similar fast-mode wave trains was also found in the radio band. The non-thermal electrons trapped in the flare loops generally produce microwave bursts \citep{2017LRSP...14....2B}. The hard X-ray/radio observations have shown that the particle acceleration sites are generally located in the reconnection region (i.e., the cusp) above the soft X-ray flare loop. The non-thermal electrons moving in the upward direction from the reconnection region produce type III (metric) radio bursts, whereas the downward electrons produce reverse-slope (RS) and decimetric (DCIM) bursts \citep{asc2004}. \citet{mes2009b,mes2009a} reported drifting tadpoles in the wavelet power spectra of DCIM type IV radio bursts produced by a flare, and suggested that the drifting tadpoles are produced by a propagating fast wave train guided by the flare loop. In the event analysed by \citet{mes2009b} the wavelet tadpoles with the mean period of about 80~s were detected at all DCIM frequencies (1.1--4.5~GHz). Another event \citep{mes2009a} was found to have a similar drifting wavelet tadpole with a period of 70.9~s at 1.60--1.78~GHz. If the heads of the tadpole wavelet feature at different frequencies of the 1.1--4.5~GHz band are fixed, the emission mechanism is expected to be gyrosynchrotron \citep{mes2009a}, which is modulated by a compressive wave train. But, if there is a frequency drift in the fiber burst (-6.8~MHz~s$^{-1}$ at 1.60--1.78~GHz), the emission mechanism is supposed to be the plasma emission \citep{mes2009b}, which is modulated by a compressive wave train propagating through a low density plasma. Furthermore, using the Giant Metrewave Radio Telescope (GMRT) observations, \citet{mes2013} reported similar tadpoles, with the mean period of 10--83~s, at metric/decimetric frequencies (244/611~MHz), and interpreted it as an evidence of a fast-mode wave train propagating in a fan structure above a coronal null point. Note that \citet{mes2009b,mes2009a} could not observe any wave train in the EUV images, because of the insufficient time resolution of available EUV imagers. 

\citet{Kumar2015w} reported an interesting observation of fast-mode waves generated by an impulsive flare at one of the footpoints of the flare's arcade loops, where one front was trapped within the arcade loop and reflected back from the opposite footpoint of the loop, while another front (i.e., shock) moved radially outward at the speed $>$1000~\kms, and produced a metric type II radio burst. Recently, \citet{goddard2016} detected a fast wave train (speed of $\sim$1200~\kms, period of $\sim$~1.7~min) in the EUV emission, and linked this phenomenon with quasi-periodic radio sparks observed at the 40--100~MHz frequency (1.7~min period) ahead of a metric type II radio burst. The origin of this fast wave train, and its association with the radio bursts is not well understood yet, and needs further investigations. 

Despite several confident detections, observations of rapidly-propagating quasi-periodic waves remain rare, i.e., only several events have been reported in the literature so far. Therefore, the addition of a new observation is valuable for understanding their origin/properties and theoretical modelling. In this paper, we report a new observation of a coronal rapidly-propagating quasi-periodic wave train of the EUV emission variation, associated with quasi-periodic radio bursts (metric, decimetric, and microwave), which occurred during a long duration M1.9 flare in active region NOAA 11471 on 7 May 2012. In Section 2, we present the observations, and in the last section, we discuss and summarise the results.
\begin{figure}
\centering{
\includegraphics[width=9cm]{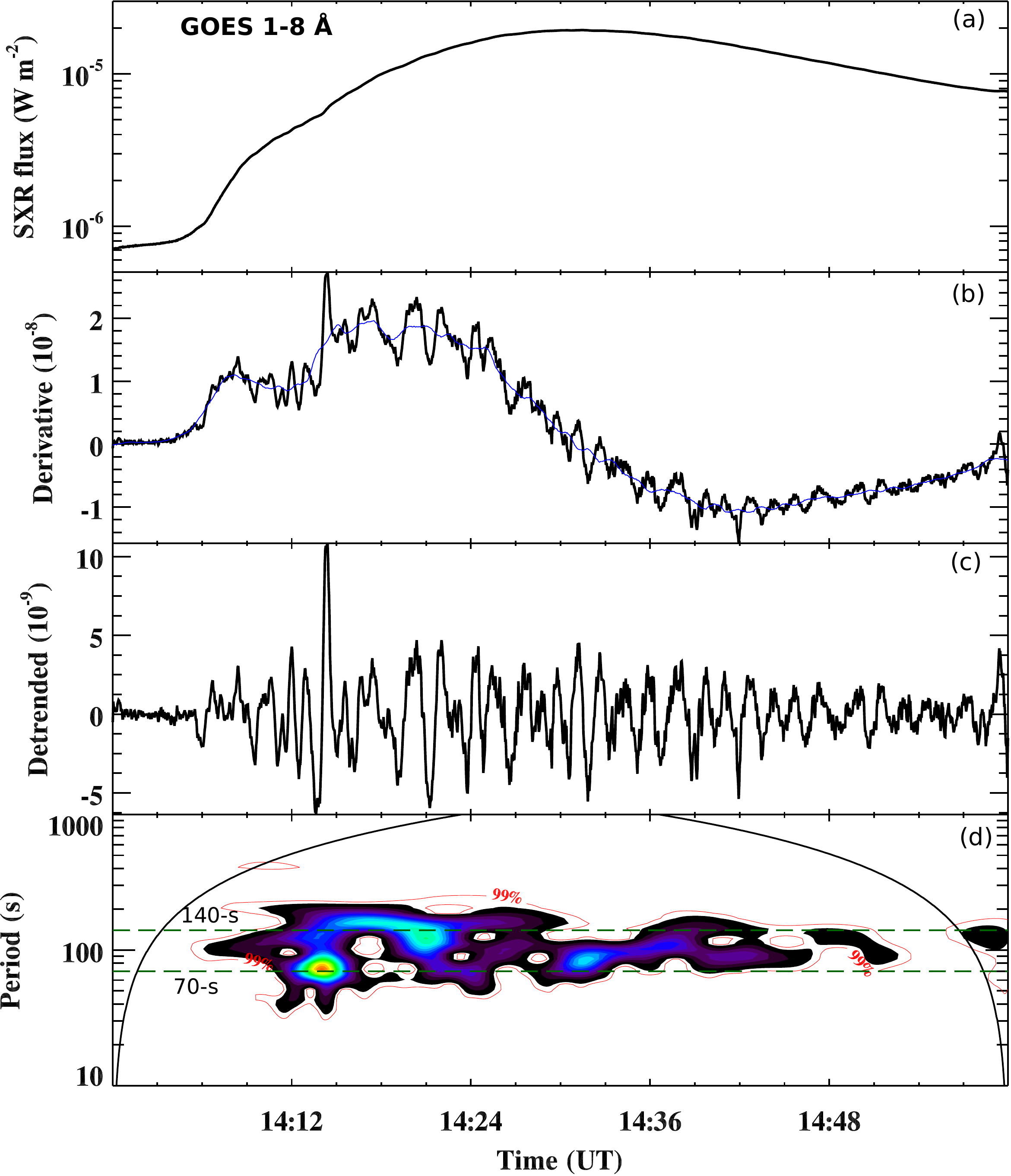}
}
\caption{(a-b) GOES soft X-ray flux (2-s cadence) in the 1--8~\AA~ channel, and its time derivative. (c) Smoothed/detrended time derivative signal after subtracting a smoothed curve (blue). (d) Wavelet power spectrum of the detrended signal.
} 
\label{goes}
\end{figure}

\begin{figure*}
\centering{
\includegraphics[width=5.0cm]{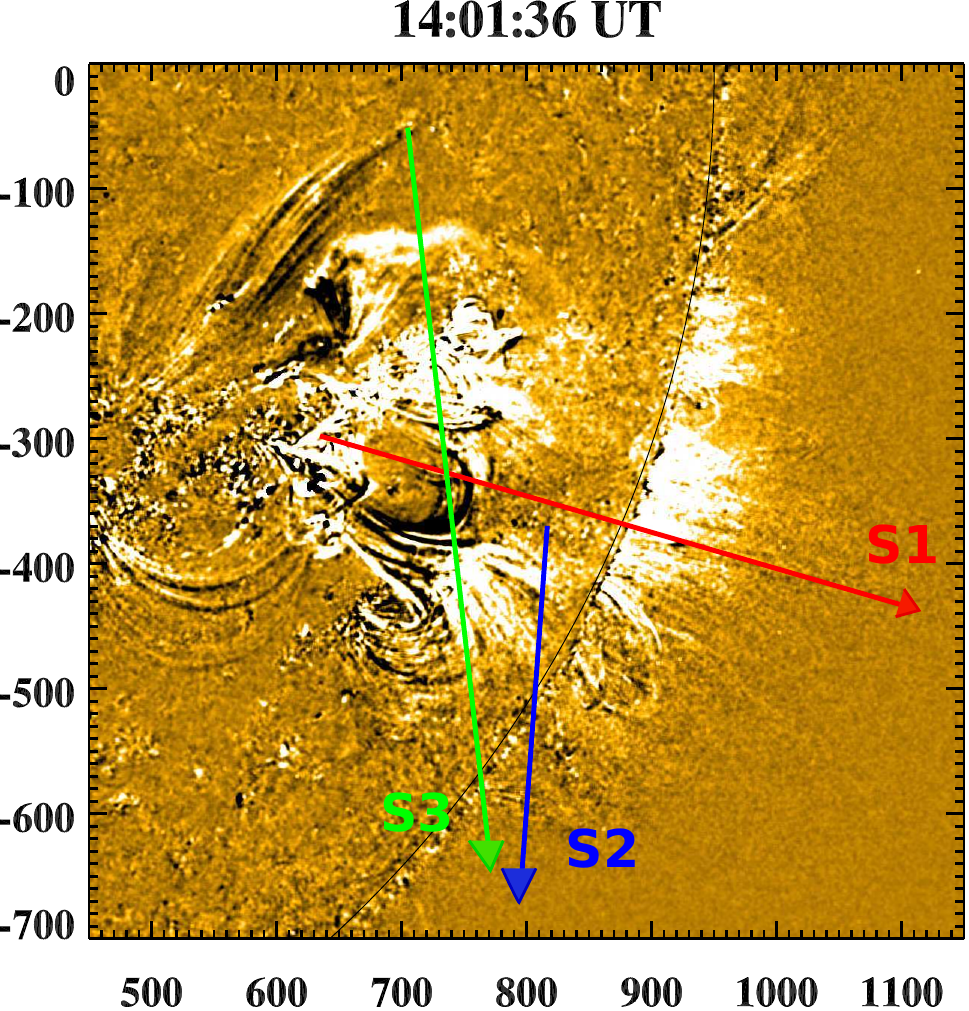}
\includegraphics[width=5.0cm]{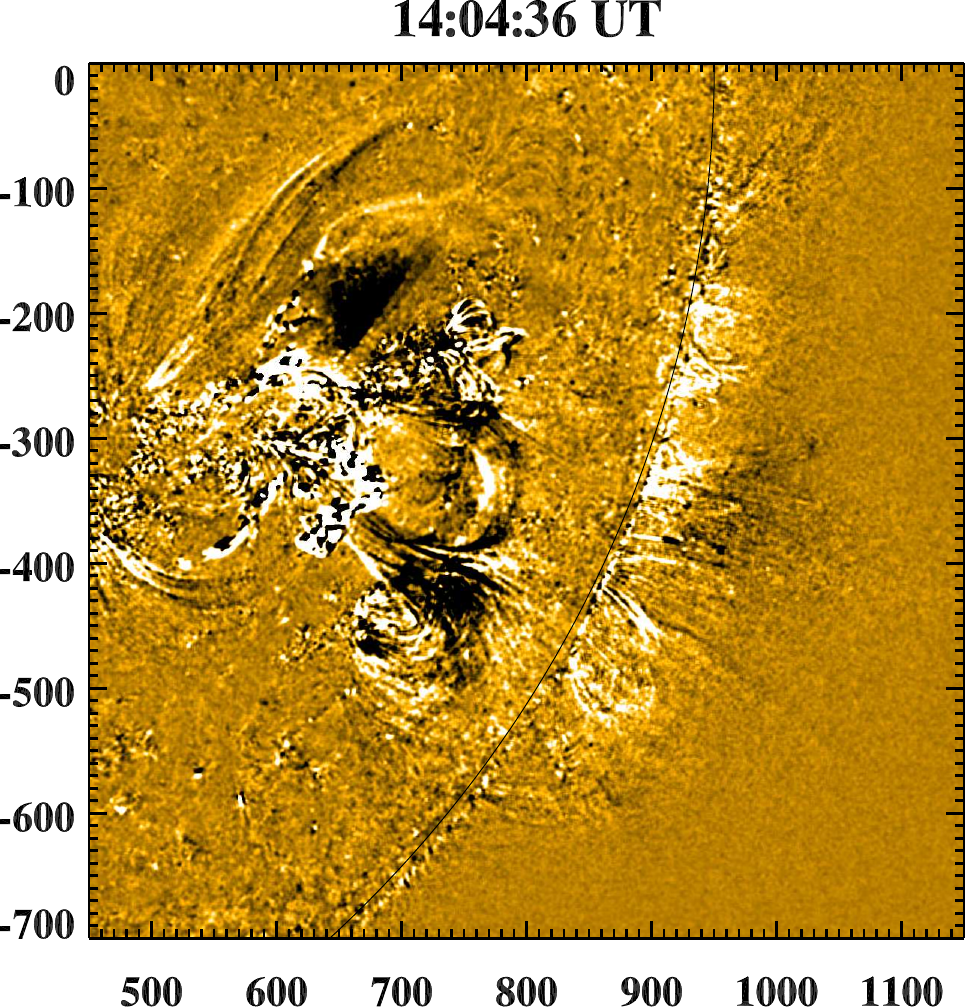}
\includegraphics[width=5.0cm]{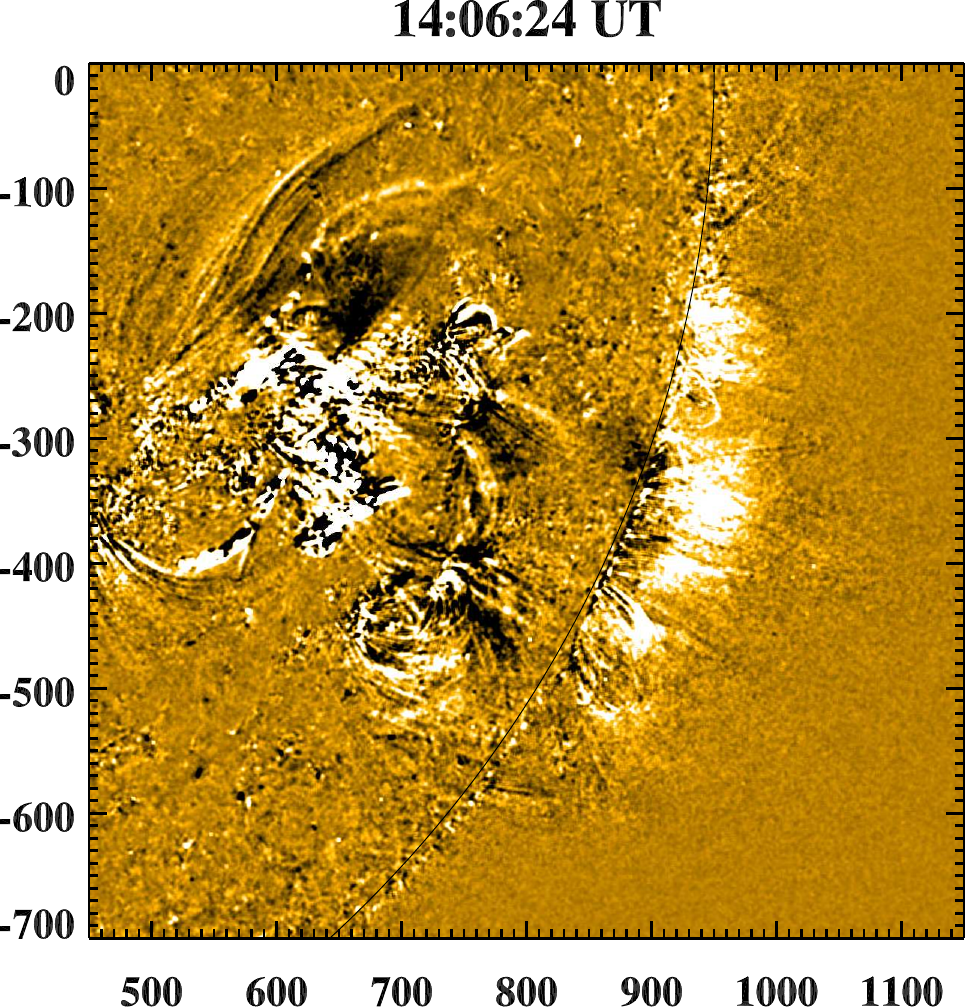}

\includegraphics[width=5.0cm]{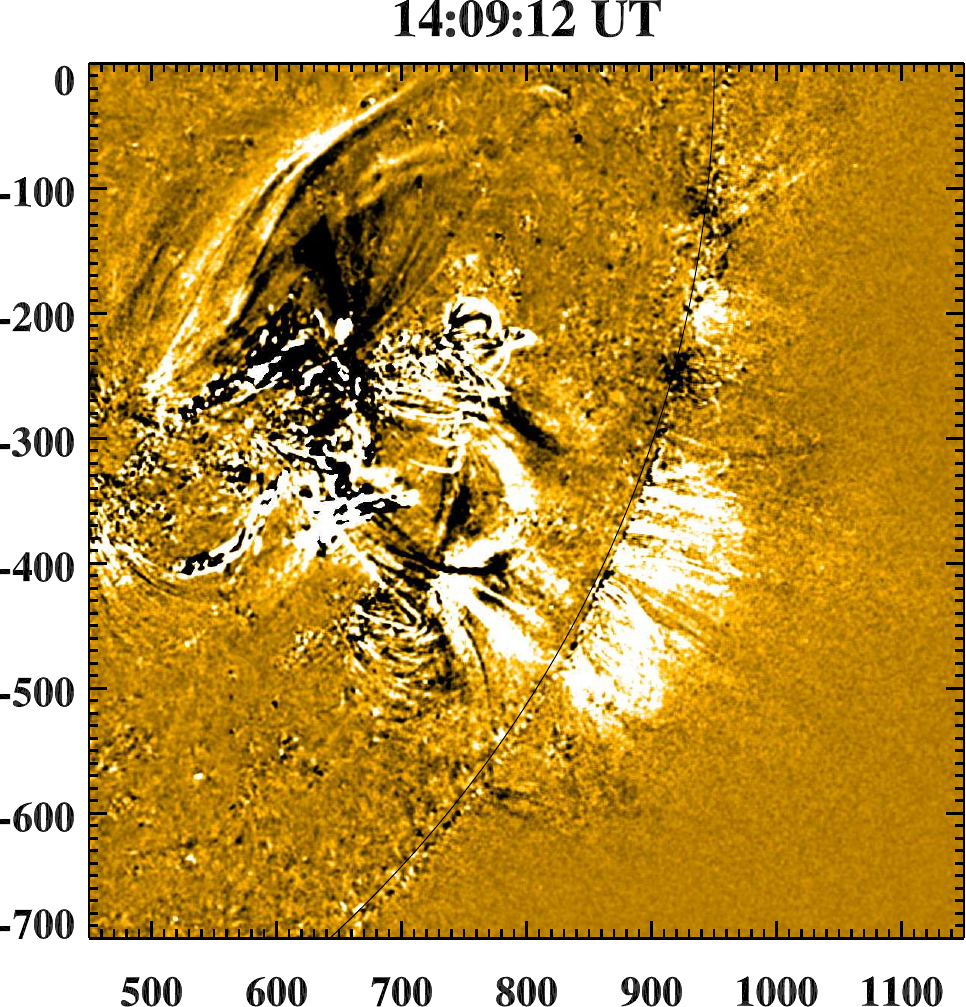}
\includegraphics[width=5.0cm]{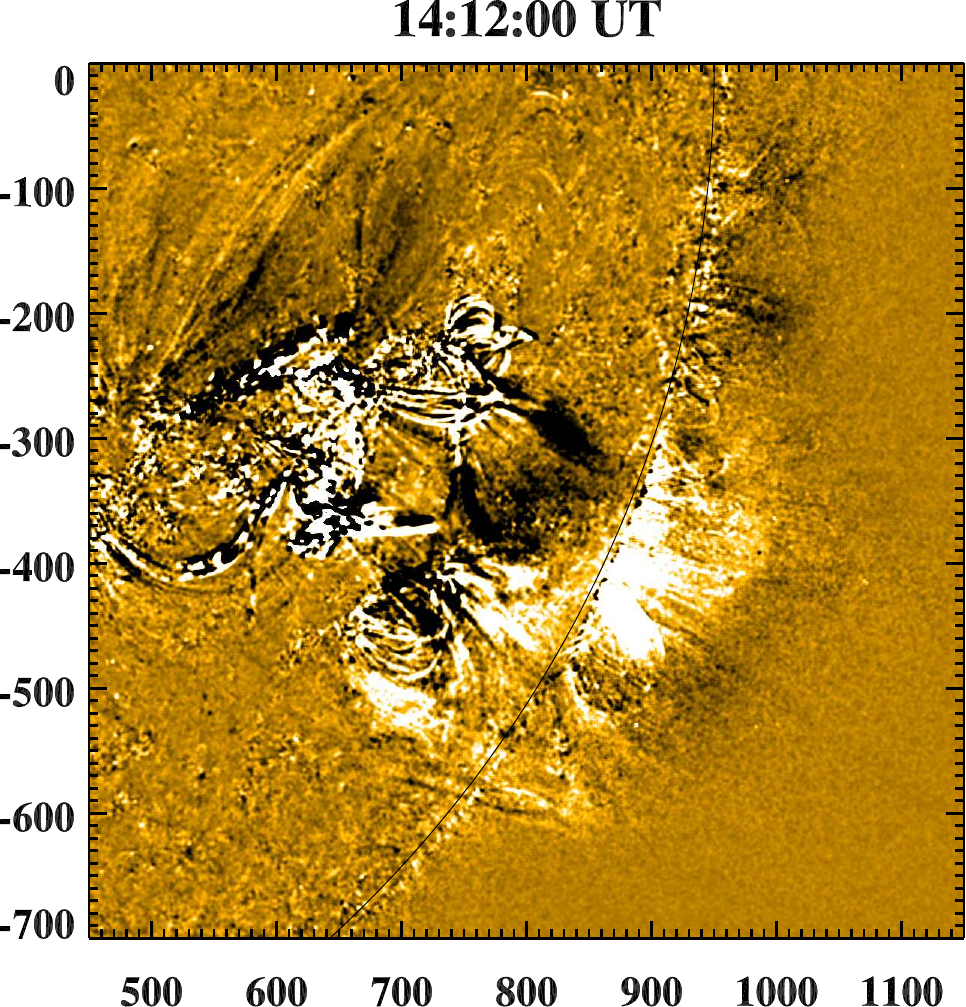}
\includegraphics[width=5.0cm]{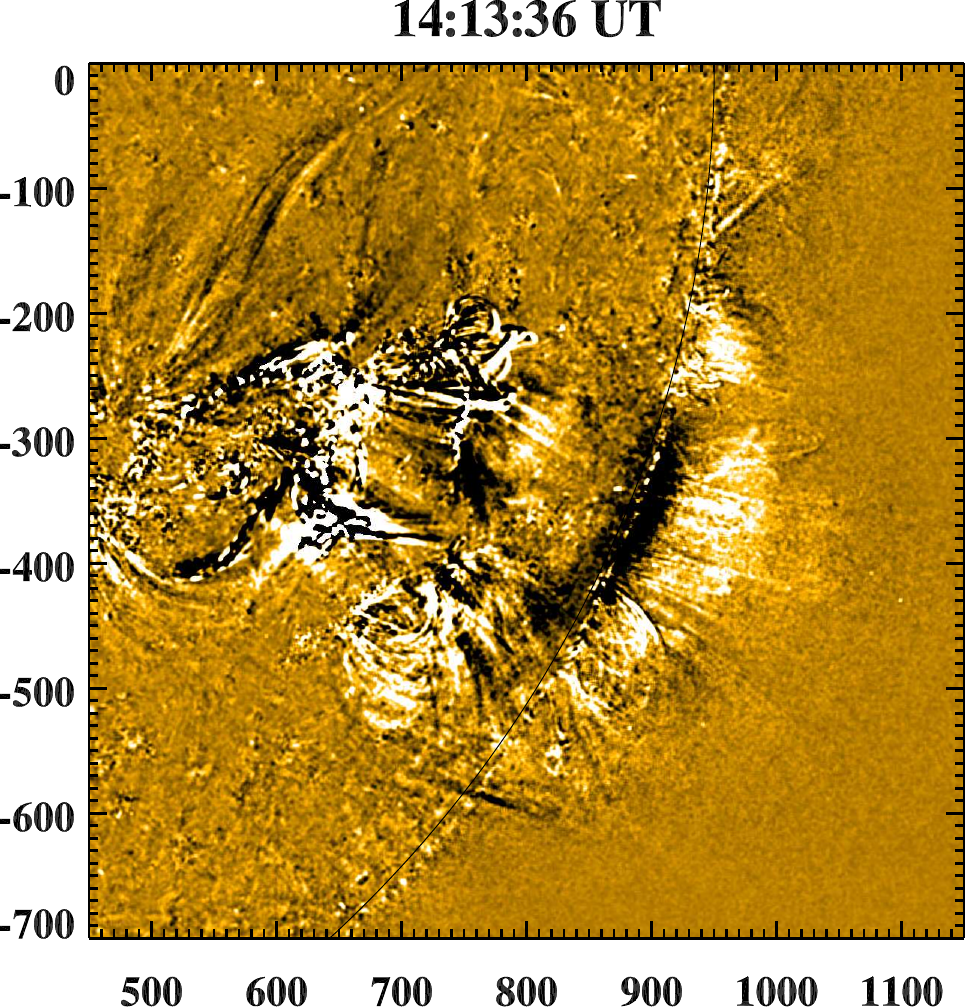}

\includegraphics[width=5.0cm]{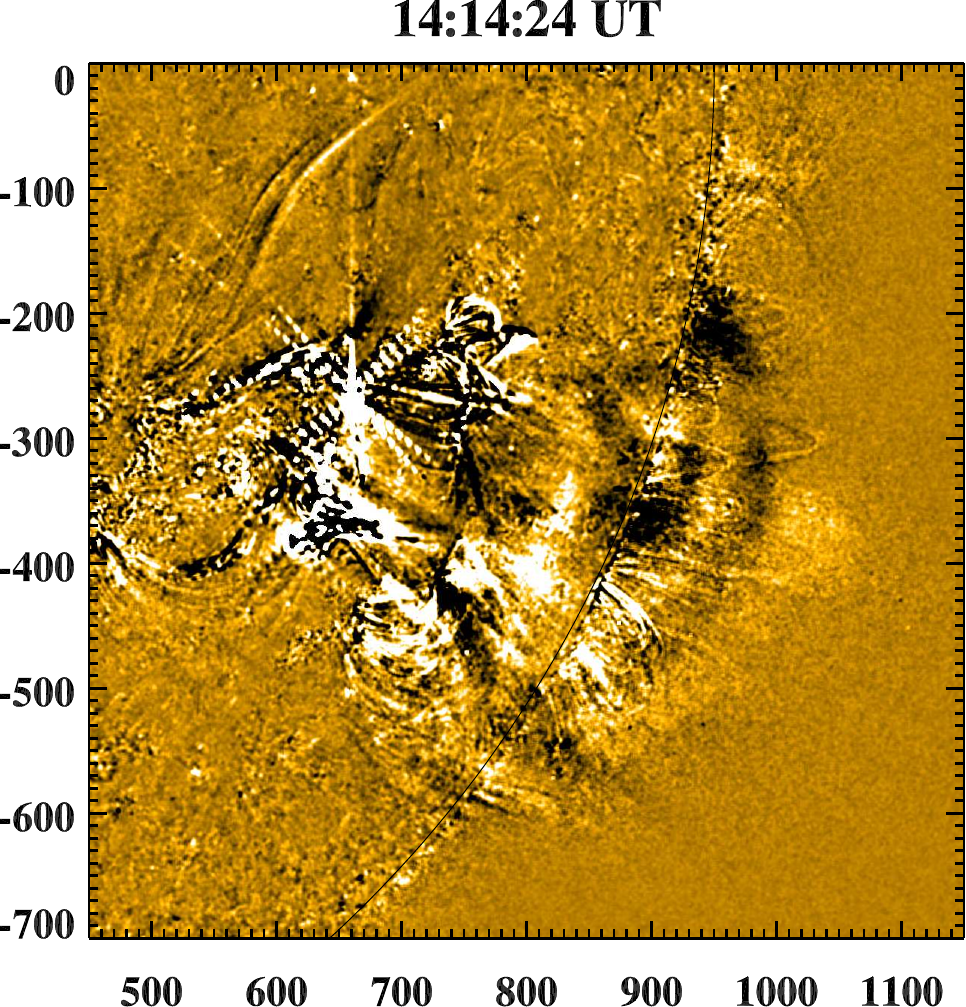}
\includegraphics[width=5.0cm]{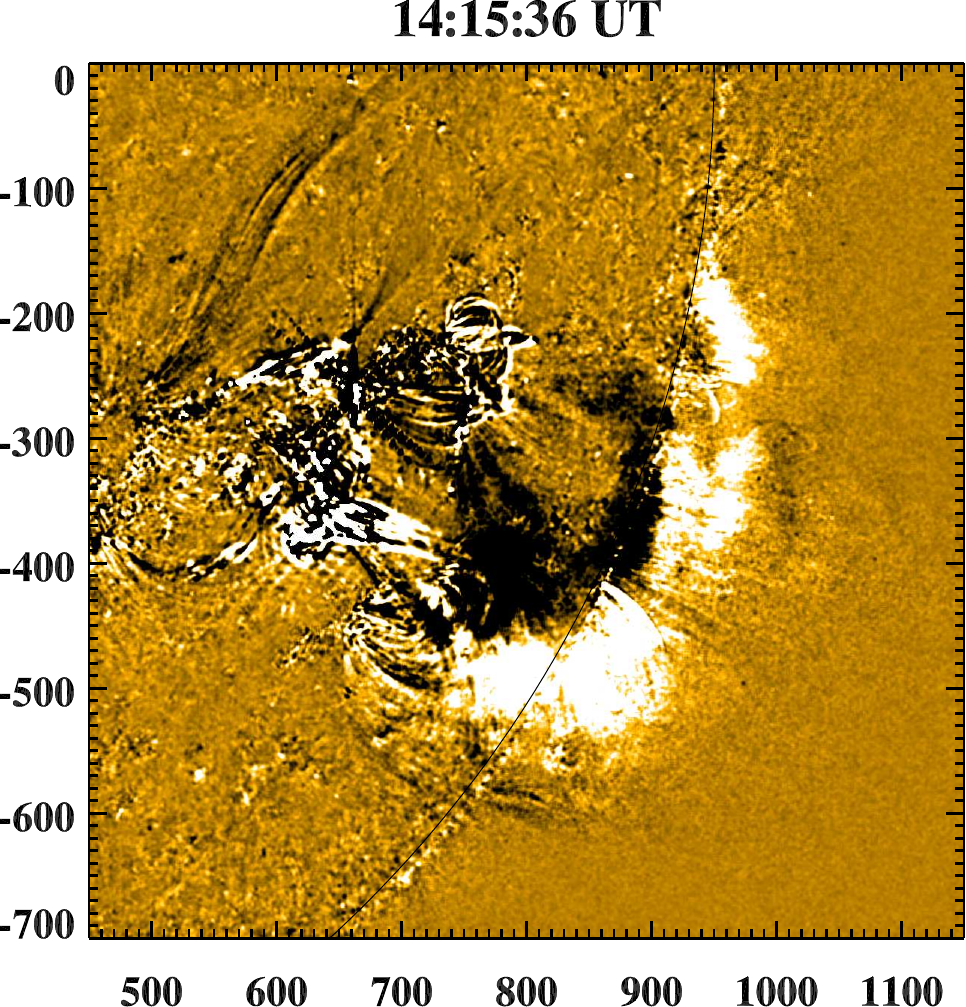}
\includegraphics[width=5.0cm]{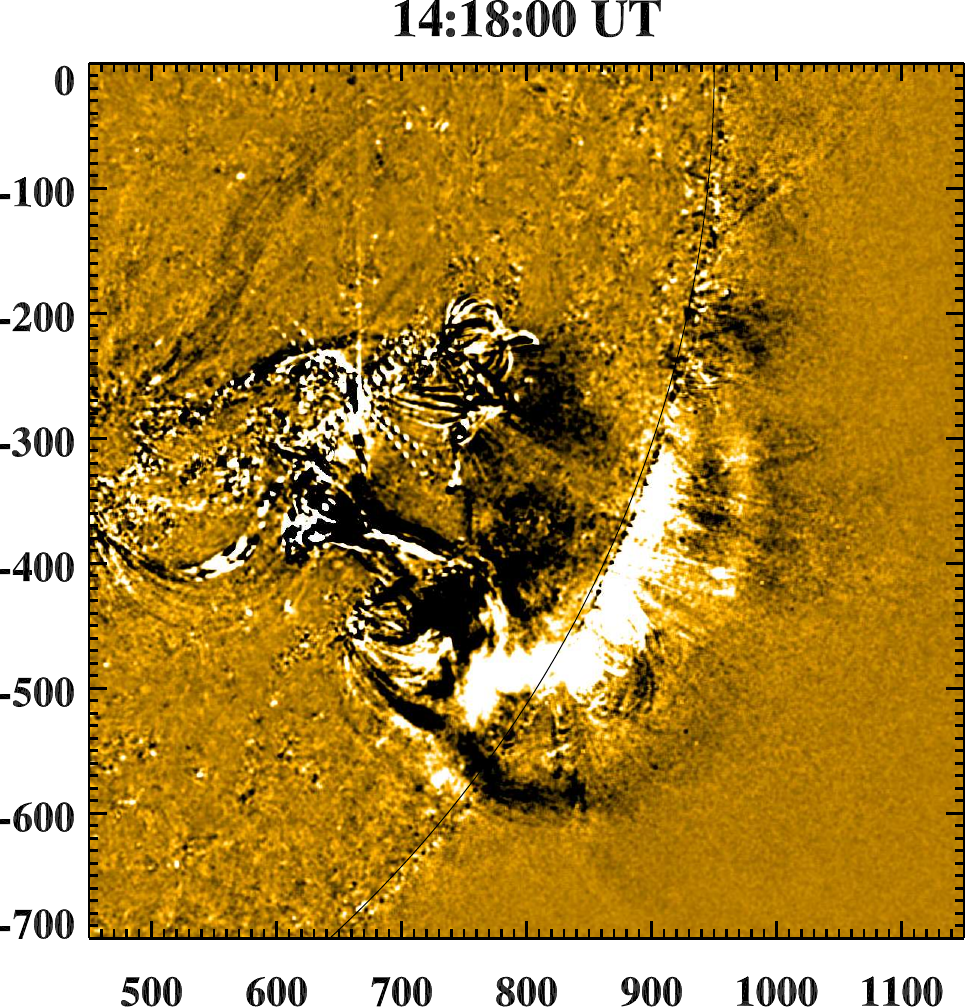}
}
\caption{AIA 171 \AA~ running difference ($\Delta$t=1 min) images showing the propagation of multiple EUV wavefronts during the M1.9 flare. Each panel shows the appearance of a new upward propagating wavefront, which is clearly observed near the limb. S1, S2, S3 are the slices used to create the time-distance (TD) intensity plots. The X and Y axes are labeled in arcsecs. (An animation of this figure is available online).} 
\label{aia171}
\end{figure*}
\begin{figure*}
\centering{
\includegraphics[width=5.0cm]{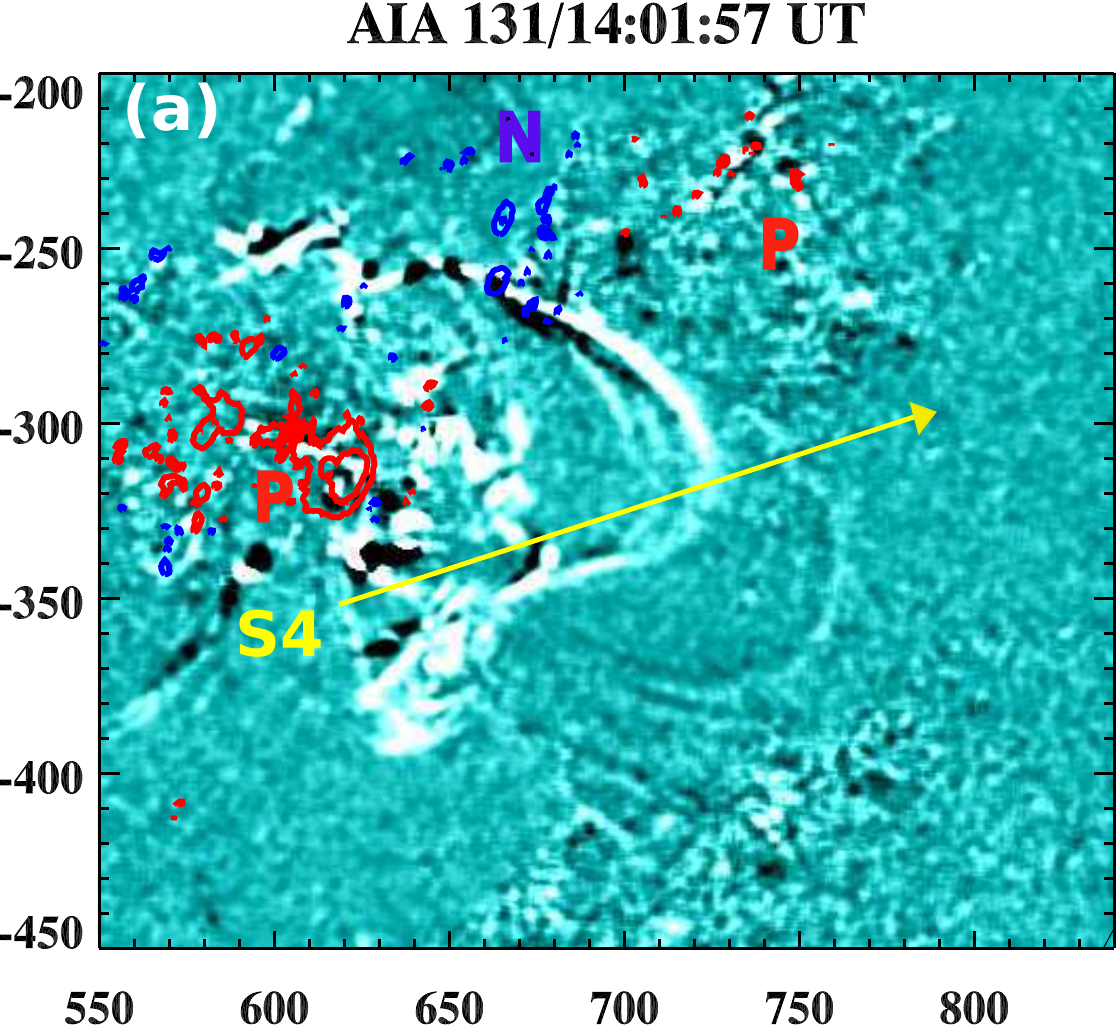}
\includegraphics[width=5.0cm]{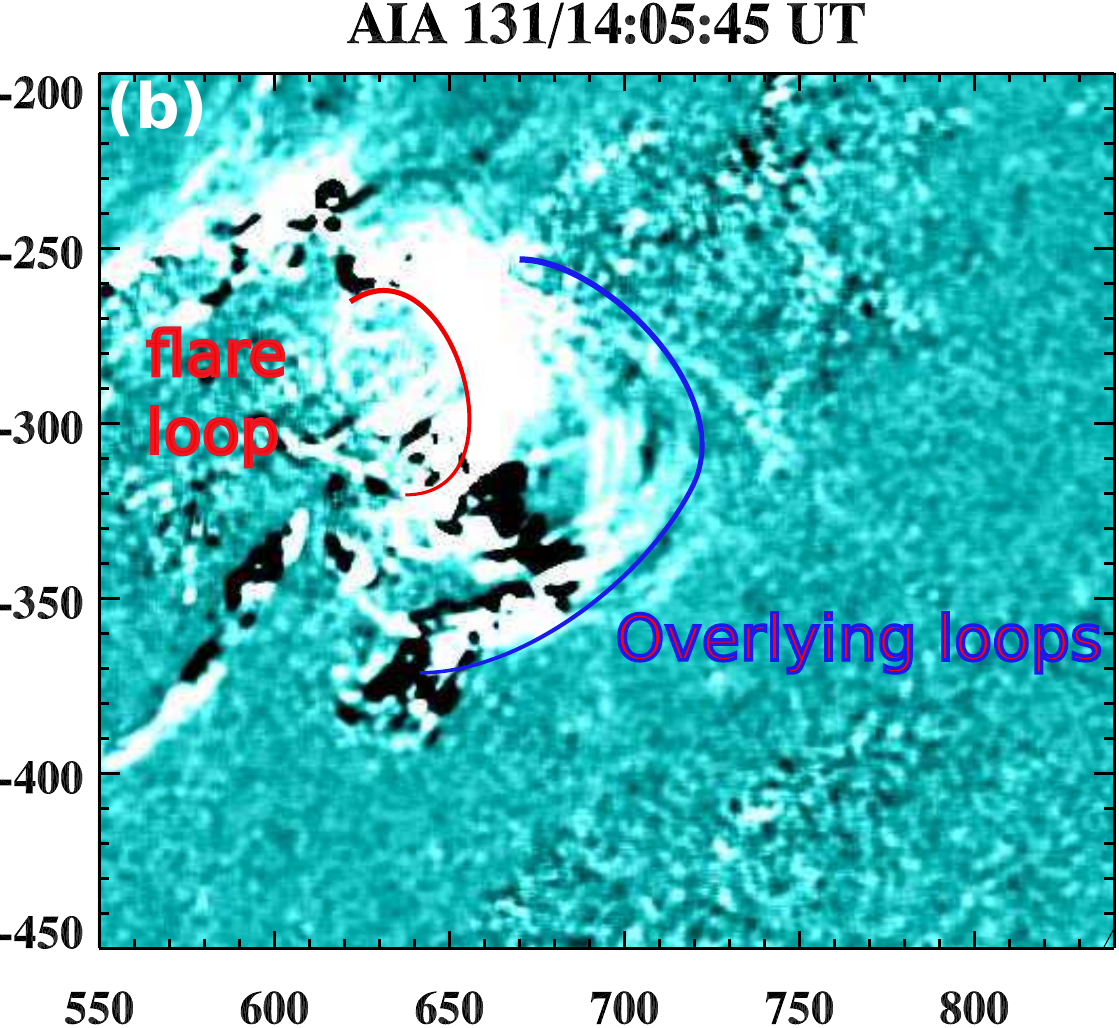}
\includegraphics[width=5.0cm]{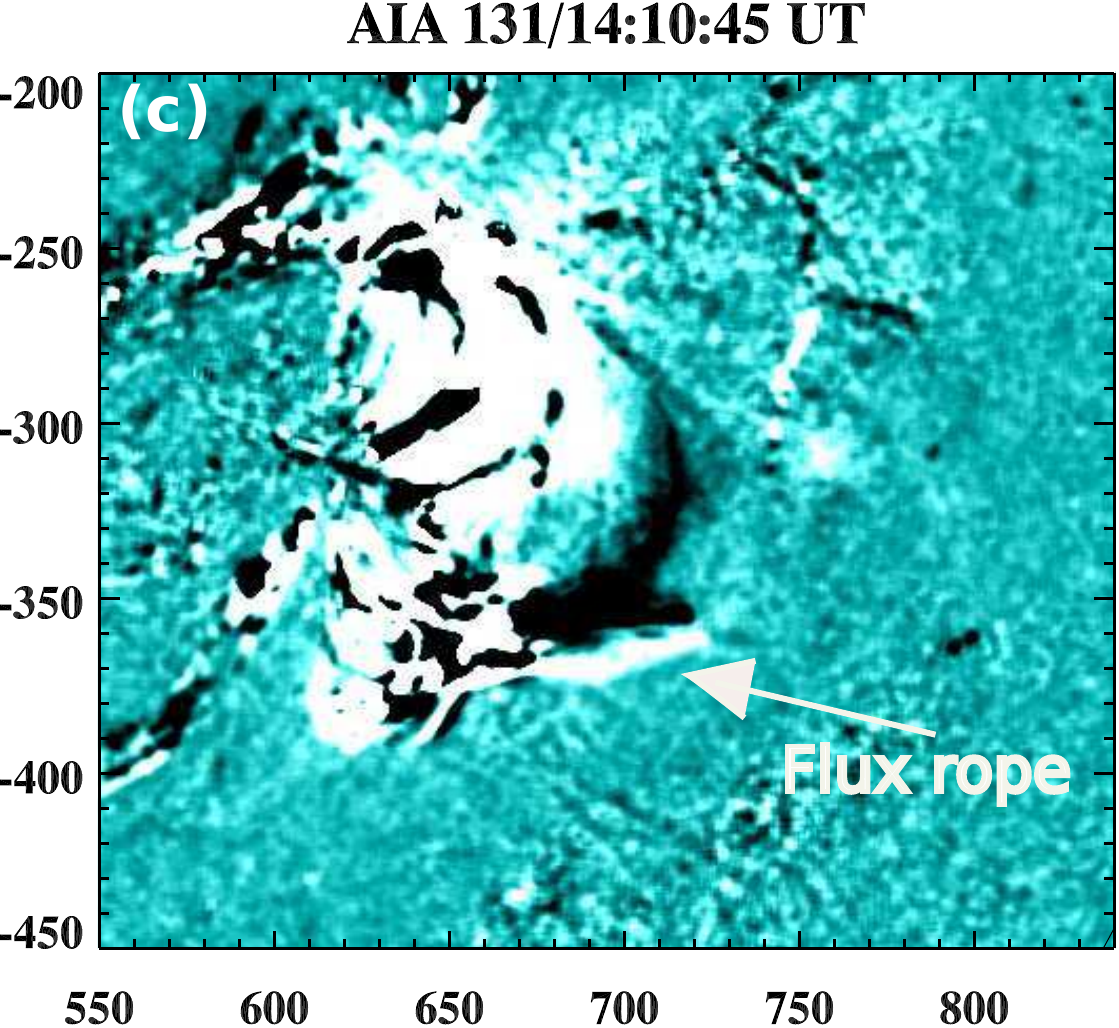}

\includegraphics[width=5.0cm]{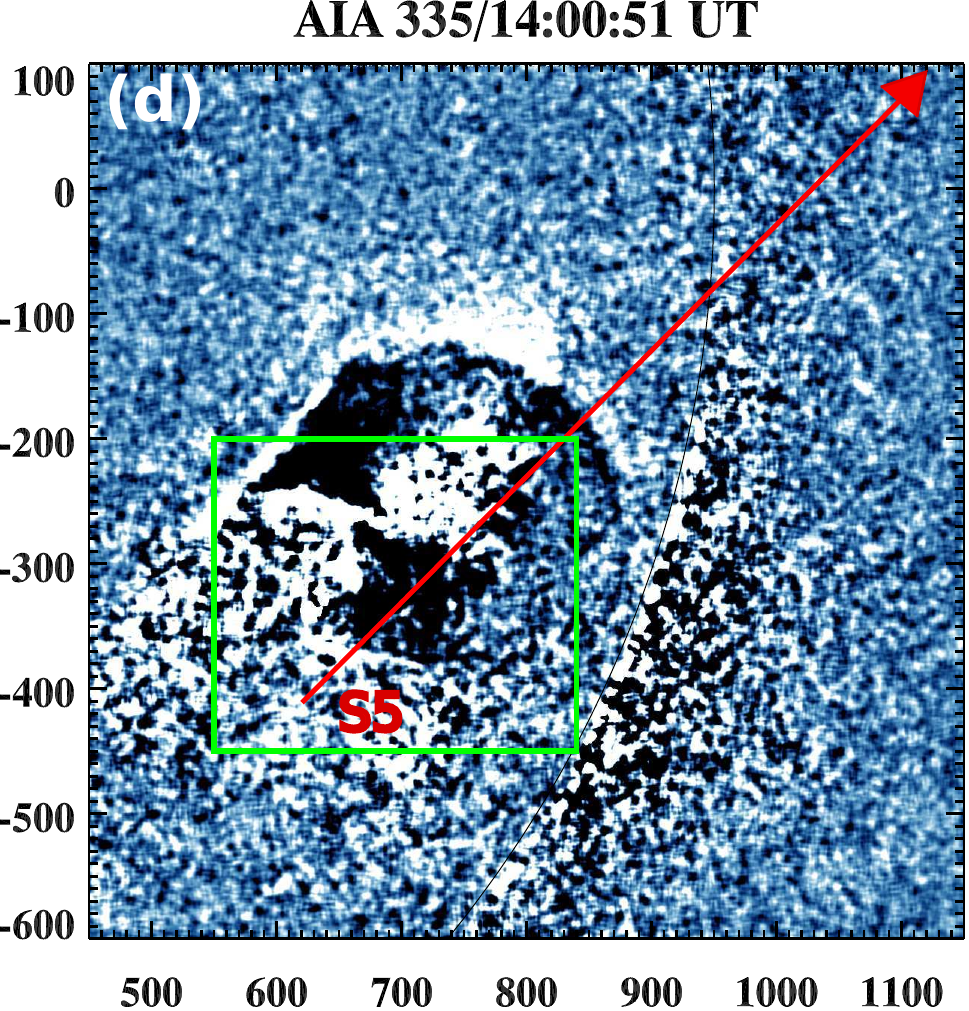}
\includegraphics[width=5.0cm]{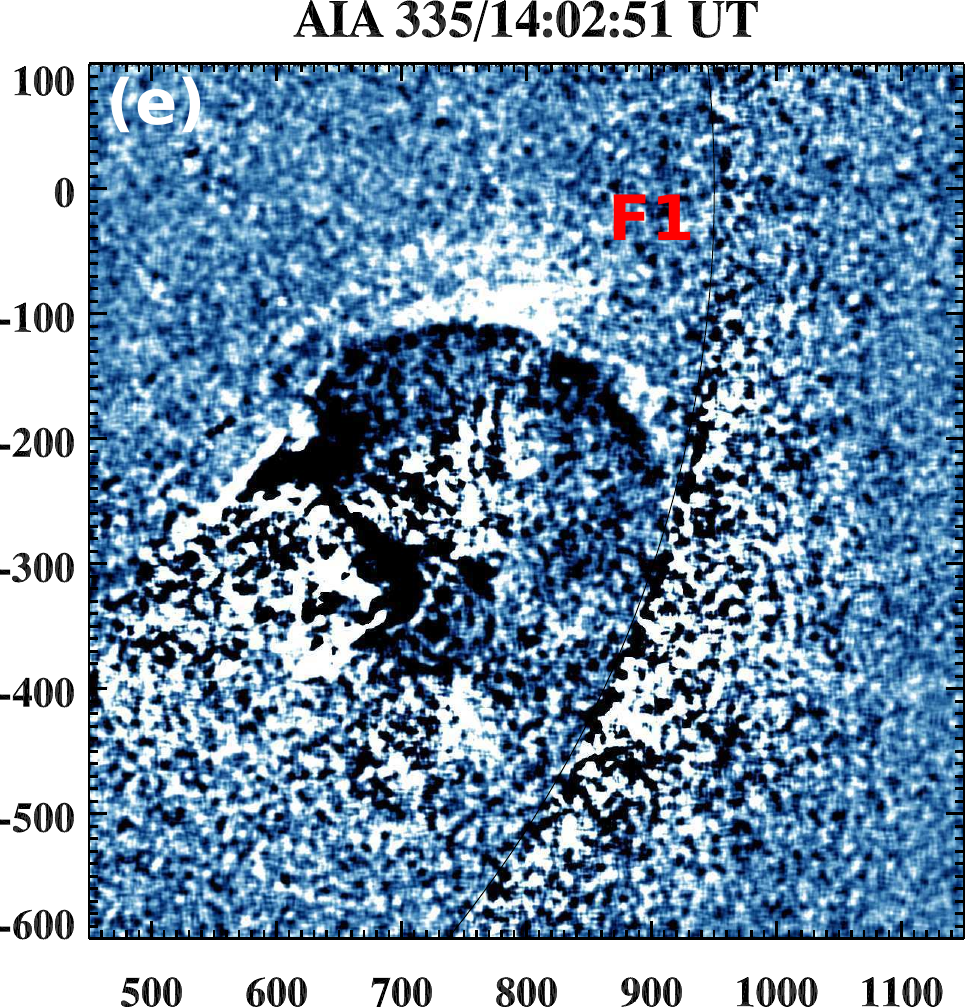}
\includegraphics[width=5.0cm]{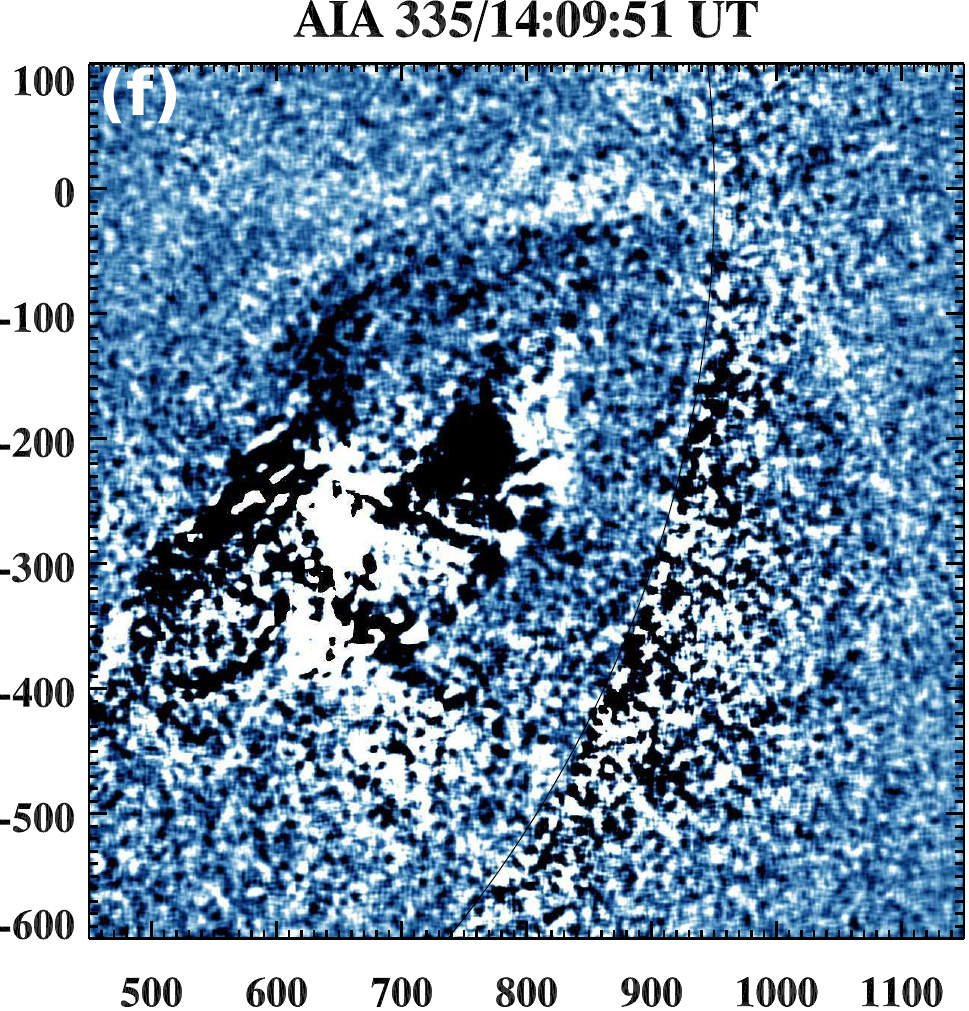}

\includegraphics[width=5.0cm]{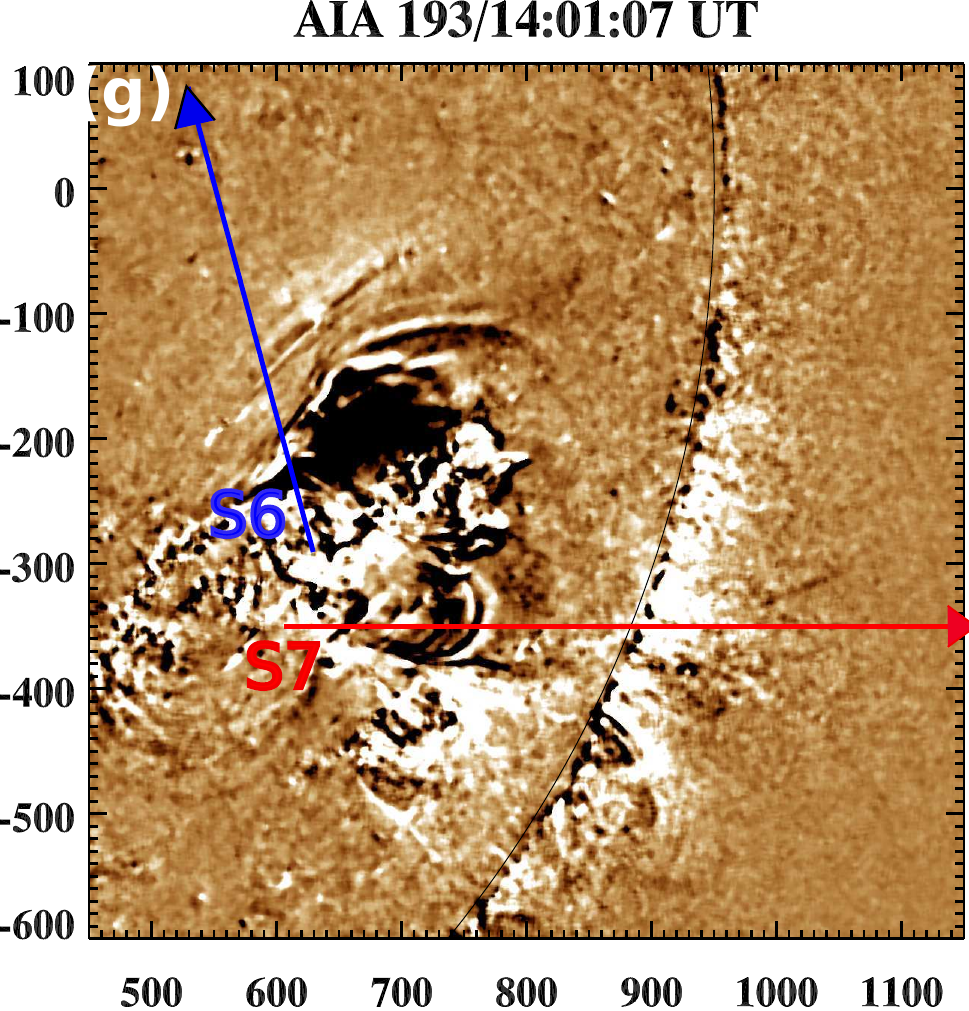}
\includegraphics[width=5.0cm]{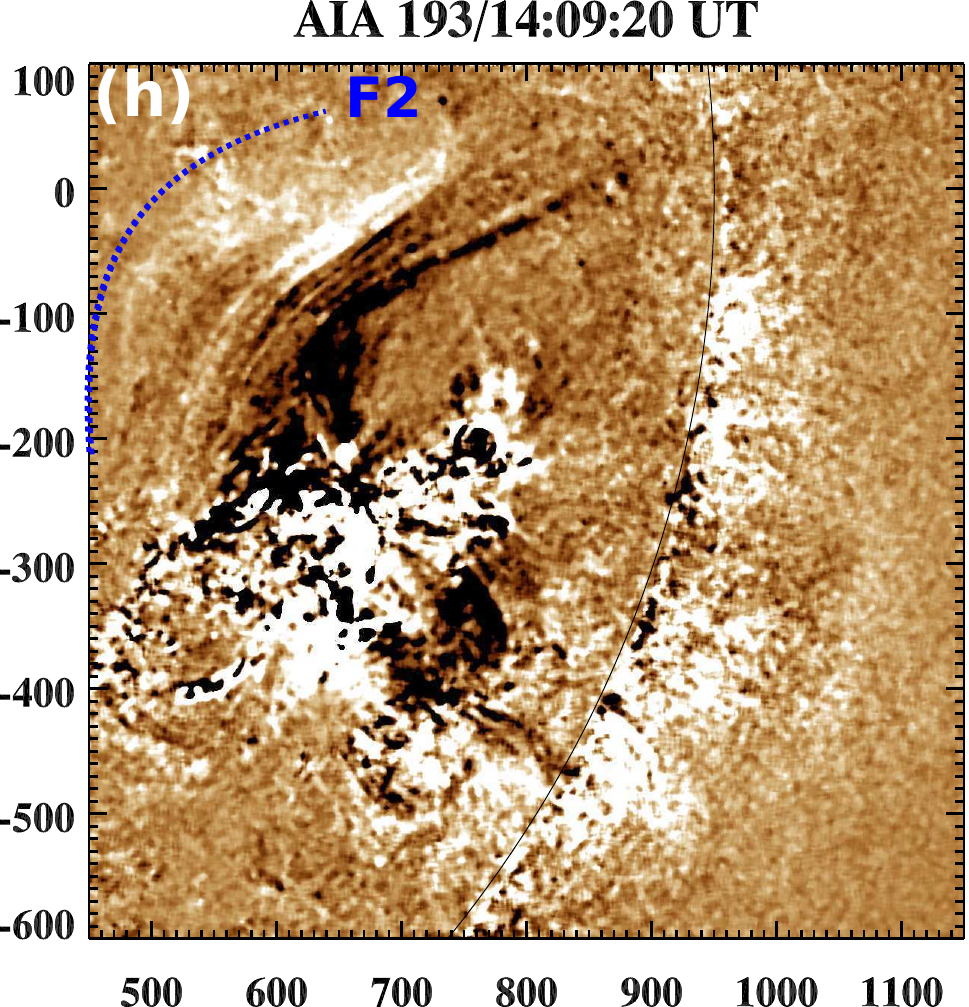}
\includegraphics[width=5.0cm]{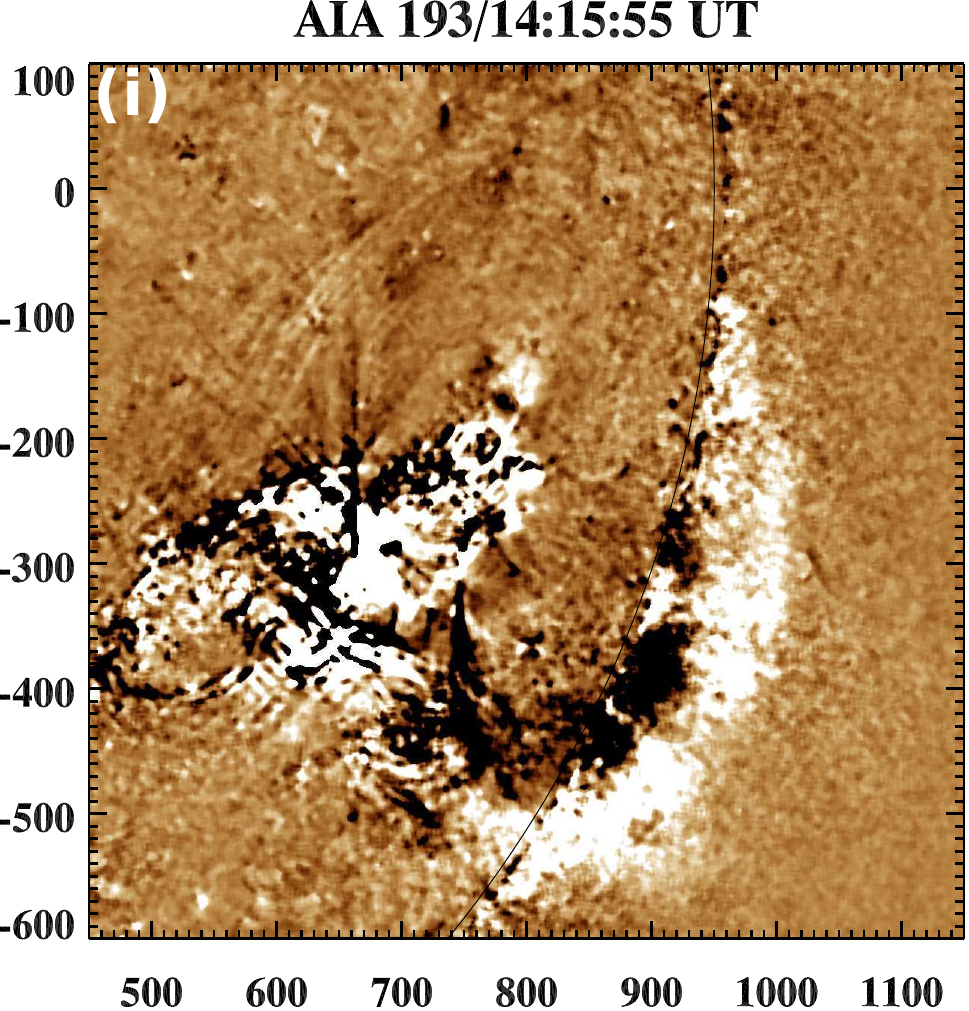}
}
\caption{(a-c) AIA 131~\AA~ running difference ($\Delta$t=1 min) images showing the appearance of a flux rope during the M1.9 flare. Panel (a) is overlaid by HMI magnetogram contours of positive (P, red) and negative (N, blue) polarities. The contour levels are $\pm$400, $\pm$1000, $\pm$2000~G. (d-f) AIA 335~\AA~ running difference images showing the expanding loop in the northwest direction. The green rectangular box in panel (d) represents the size of AIA~131 \AA~ panels. (g-i) AIA 193~\AA~ running difference images showing multiple wavefronts originating from the flare site.  Labels S4, S5, S6, S7 indicate the slices chosen to create the time-distance intensity plots. The X and Y axes are labeled in arcsecs. (An animation of this figure is available online).} 
\label{aia3}
\end{figure*}
\begin{figure*}
\centering{
\includegraphics[width=8.1cm]{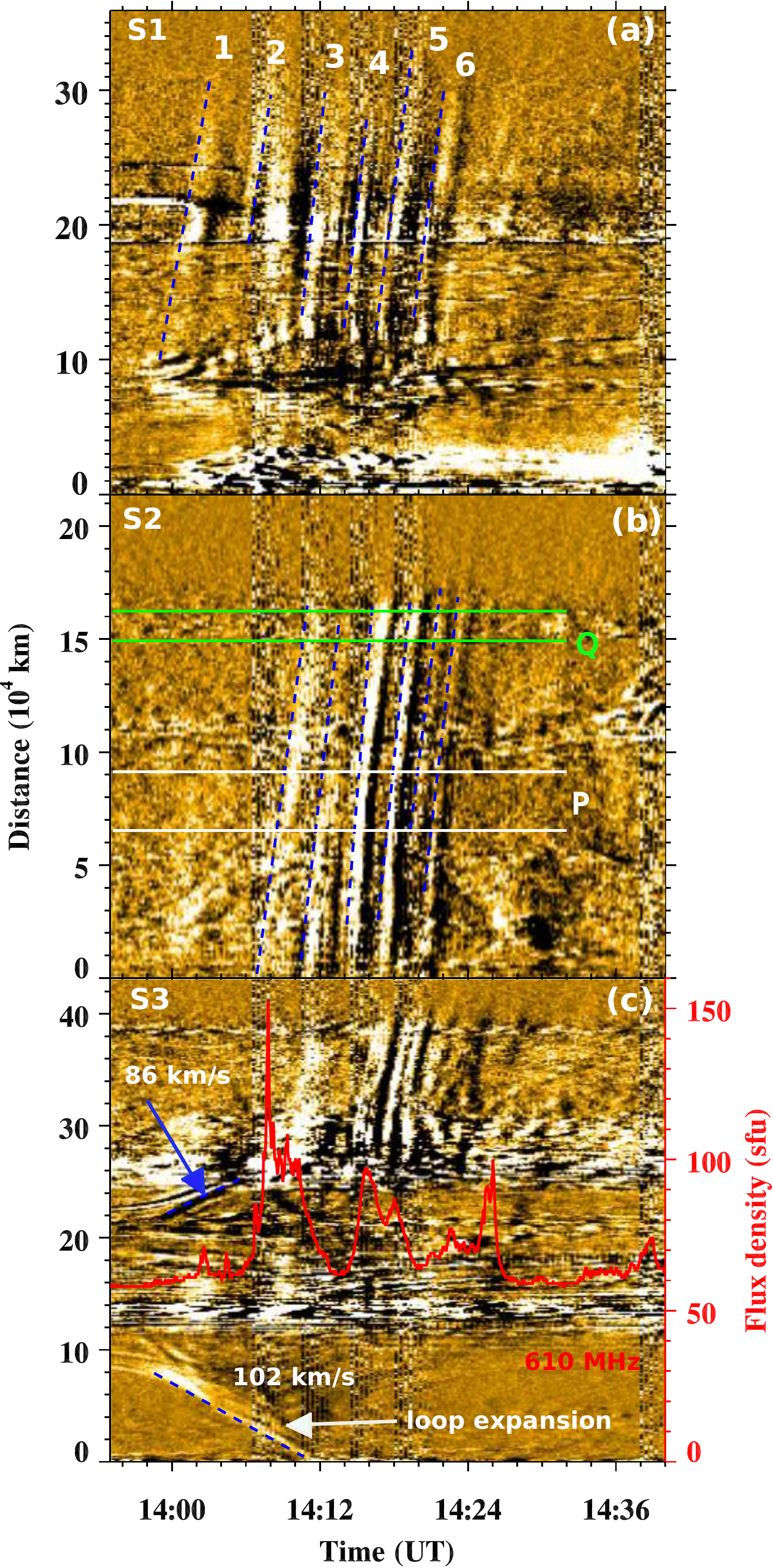}
\includegraphics[width=8cm]{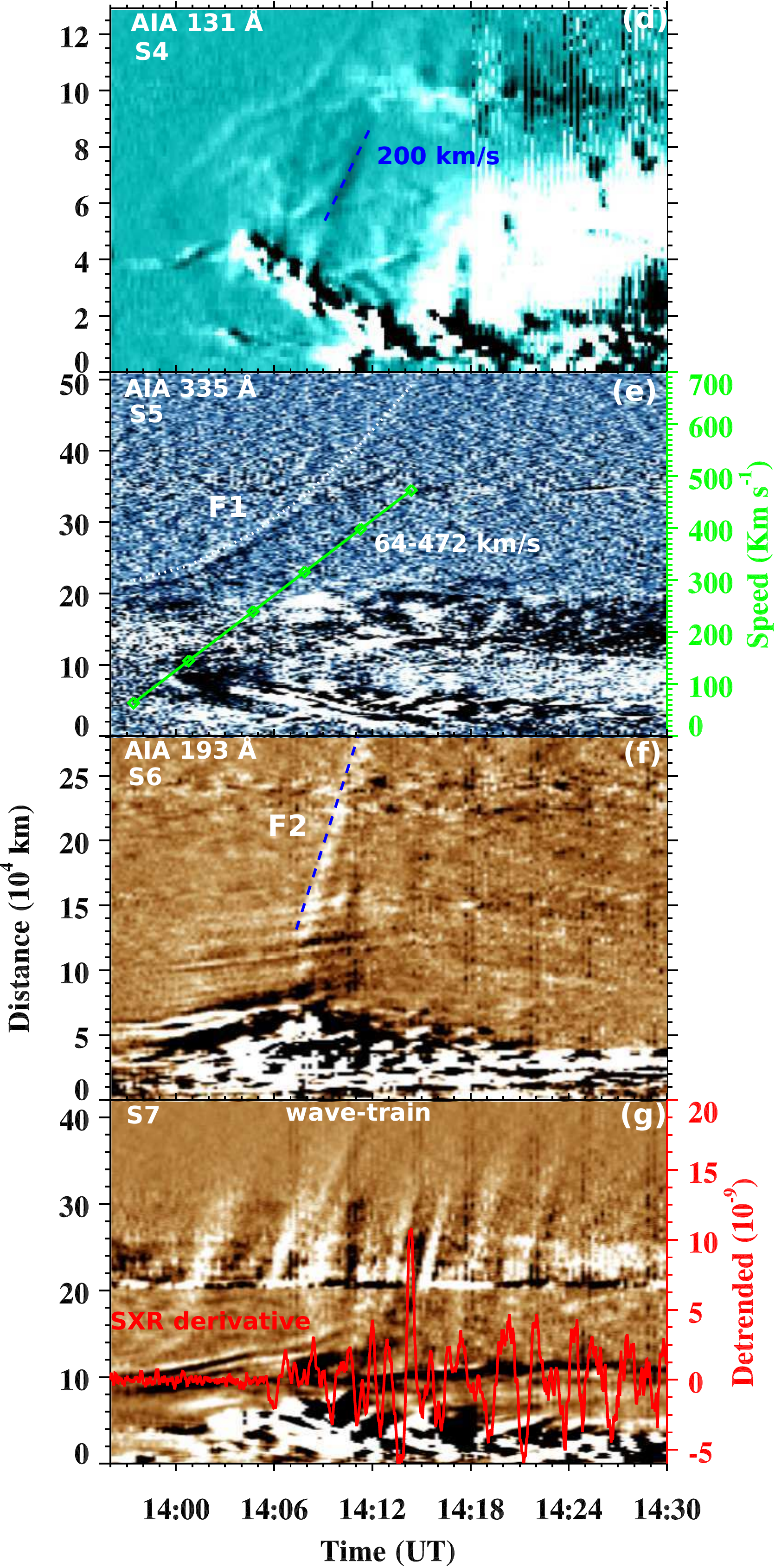}
}
\caption{(a-c) Time--distance intensity maps along slices S1, S2, and S3 created using AIA 171~\AA~ images (running difference/smoothed). The vertical dashed lines (blue) in panel (a) and (b) denote multiple EUV wavefronts (i.e., the wave train). The clearly observed wavefronts are  marked by 1 to 6. Labels P and Q indicate the regions used to extract the average/total intensity profile. Panel (c) displays the lateral expansion of the loops at the speed of $\sim$86--102~\kms. The red curve is the radio flux profile at 610~MHz, showing the repetitive bursts during the rapidly propagating wave fronts. (d-g) AIA 131, 335, 193~\AA~ time--distance intensity (running difference) plots along slices S4, S5, S6, and S7. F1 and F2 are the propagating fronts (north) observed in the AIA 335 and 193 \AA~ channels. The speed profile of F1 is shown by the green curve.} 
\label{st}
\end{figure*}
\begin{figure*}
\centering{
\includegraphics[width=8cm]{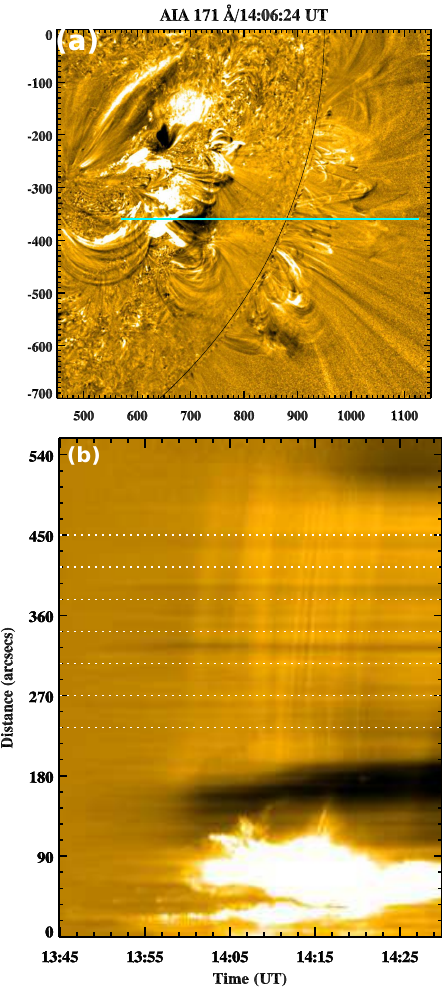}
\includegraphics[width=7.4cm]{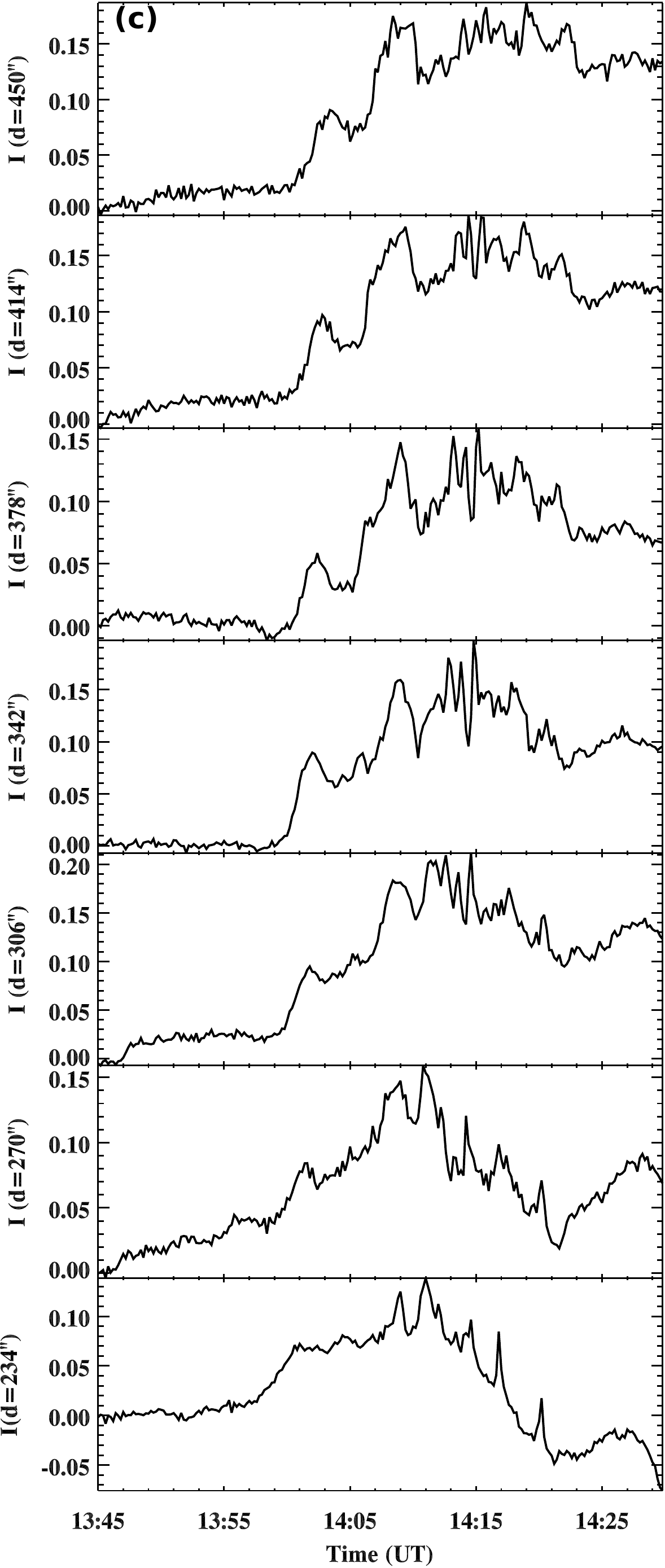}
}
\caption{(a,b) Time--distance intensity map along a slice cut (cyan) created using AIA 171~\AA~ base ratio images. (c) Relative intensity of the radially outward propagating multiple EUV fronts at different distances (arcsecs) along the horizontal dotted lines marked in panel (b).} 
\label{base_rat}
\end{figure*}
\begin{figure*}
\centering{
\includegraphics[width=4.8cm]{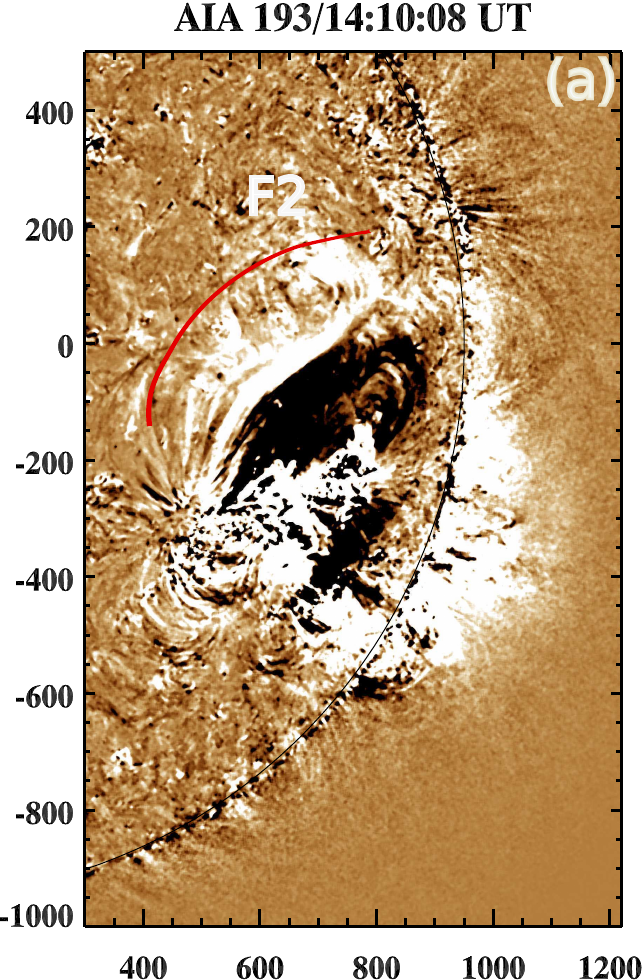}
\includegraphics[width=5cm]{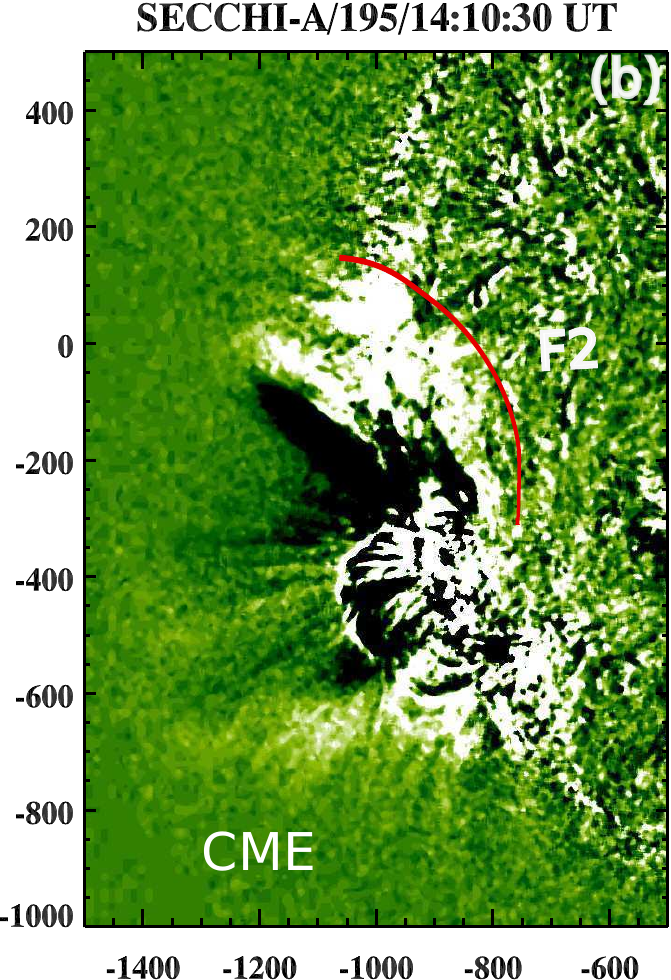}
\includegraphics[width=5cm]{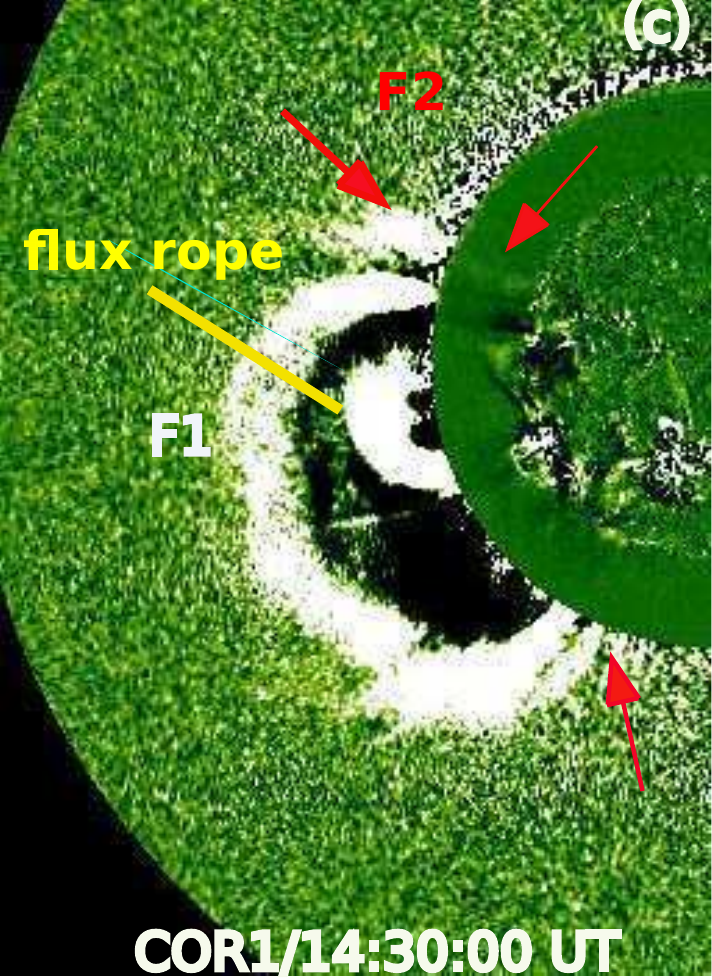}
}
\caption{AIA 193~\AA, STEREO-A SECCHI 195~\AA, and COR1 running difference ($\Delta$t=5 min) images showing the CME association with the low coronal eruption. The red curve is the EUV front F2 moving in the north direction. F1 is the CME frontal loop. The red arrows show the separation of fronts (north and south) from the CME flanks. (An animation of this figure is available online).} 
\label{stereo}
\end{figure*}
\begin{figure*}
\centering{
\includegraphics[width=7.1cm]{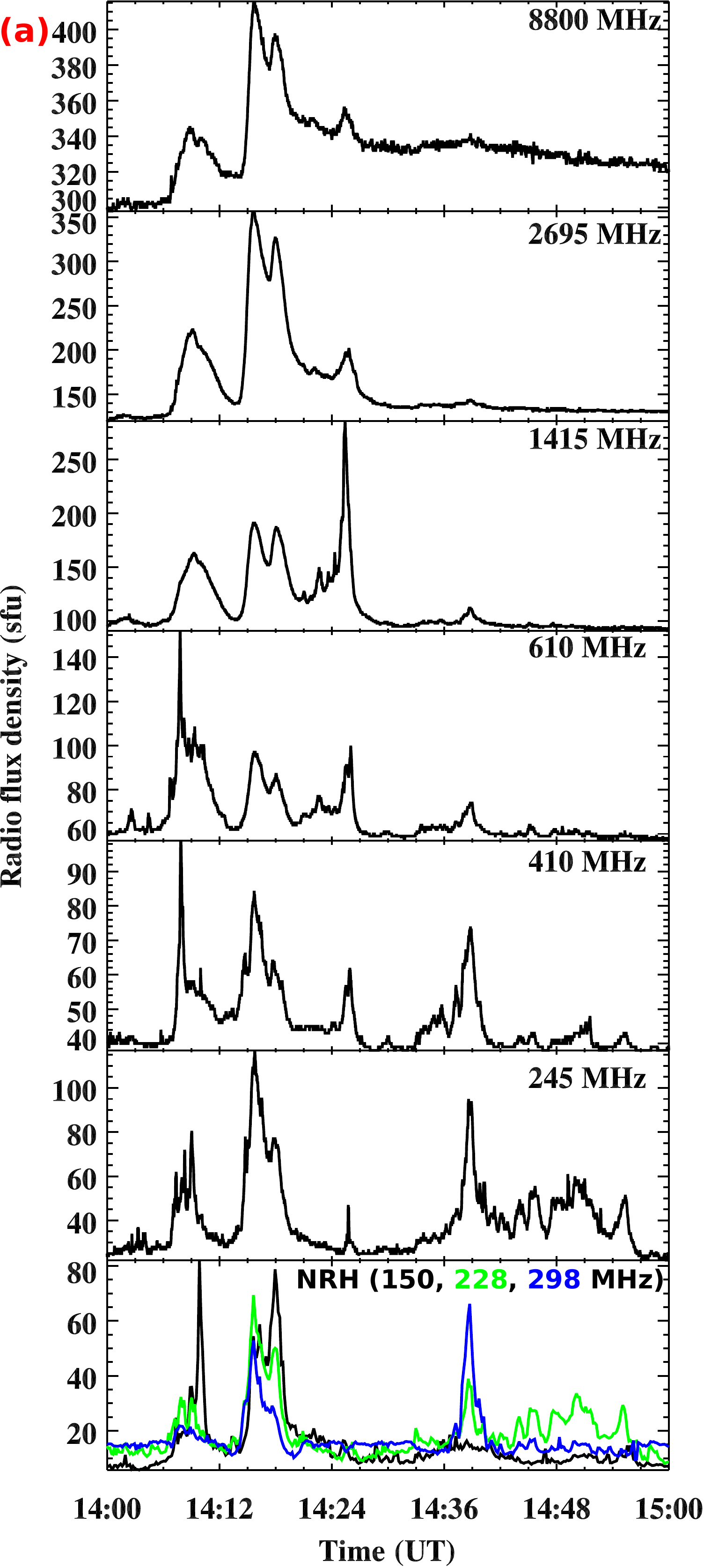}
\includegraphics[width=8.6cm]{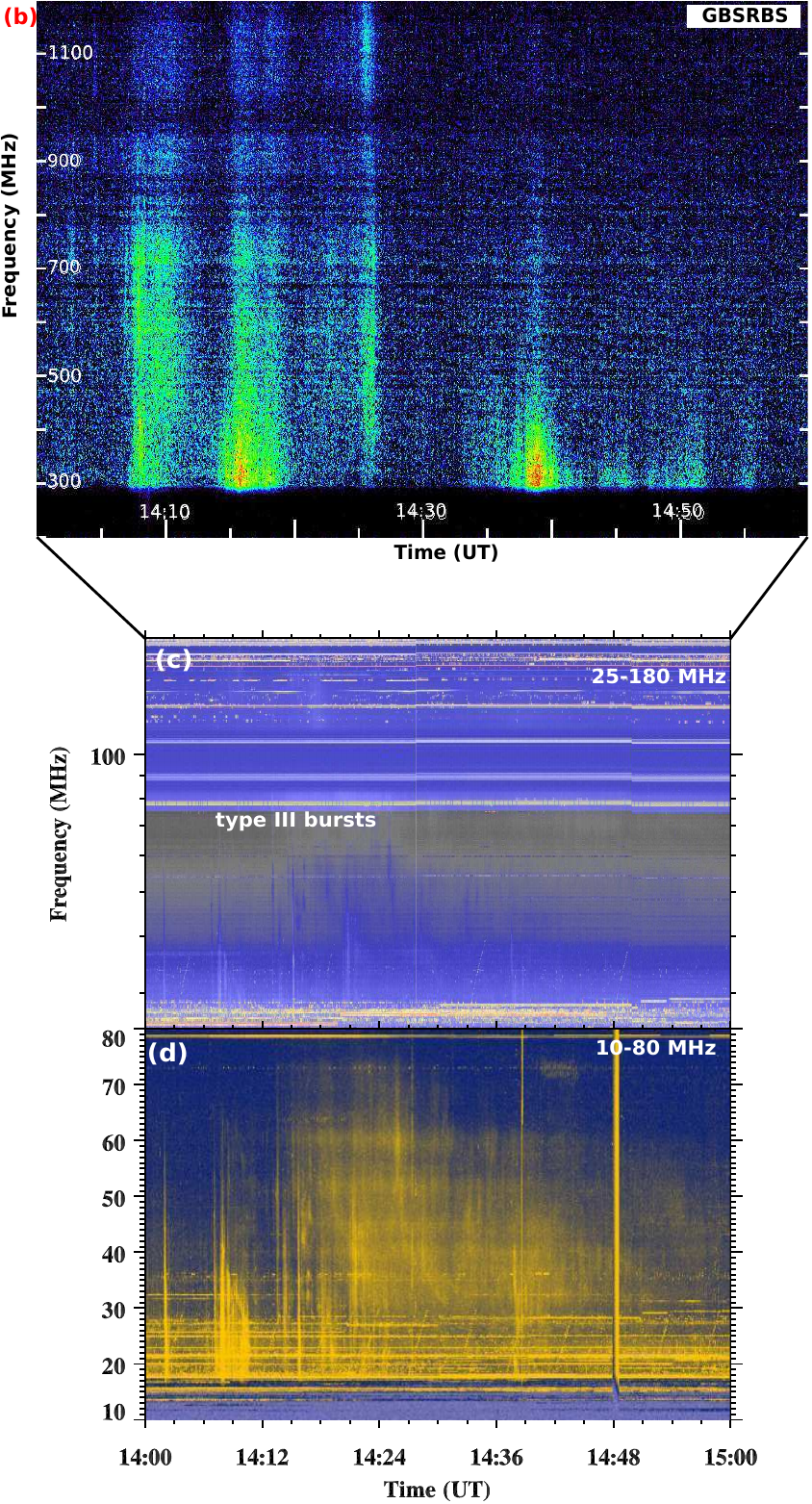}
}
\caption{(a) Radio flux density profiles (1-s cadence) in different frequency bands observed at the Sagamore Hill station of RSTN. The bottom panel is the NRH 10-s integrated flux profiles in 150, 228 (green), 298 (blue) MHz. (b,c,d) Dynamic radio spectrum from the Green Bank Solar Radio Burst Spectrometer (300--1200~MHz), San-Vito radio station (25-180 MHz), and Nan\c{c}ay decametric array (10-80 MHz).  
(1~sfu=10$^{-22}$~W~m$^{-2}$ Hz$^{-1}$).} 
\label{rstn}
\end{figure*}
\begin{figure*}
\centering{
\includegraphics[width=14cm]{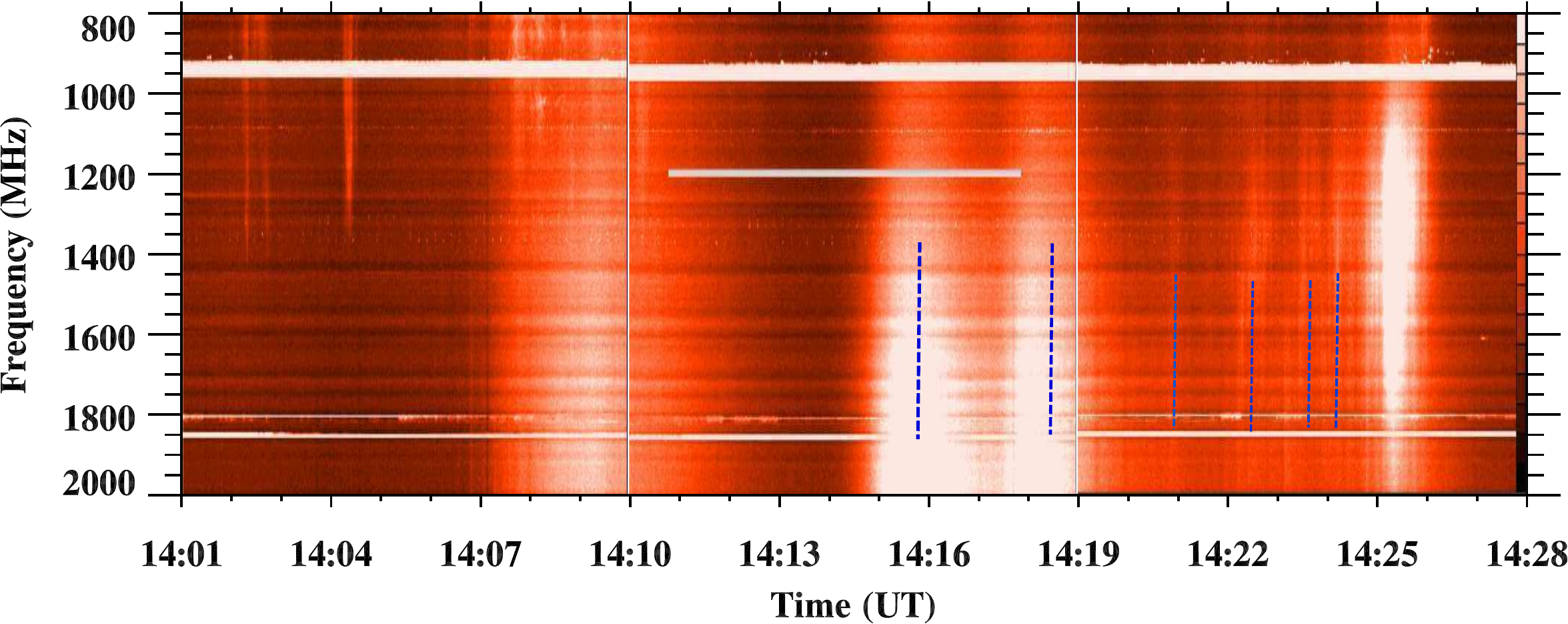}
}
\caption{Dynamic radio spectrum obtained at 800--2000~MHz with the Ondrejov radio-spectrograph. 
The vertical blue dashed lines indicate the multiple decimetric bursts during the propagation of the fast-mode wave train.} 
\label{ondr}
\end{figure*}
\begin{figure*}
\centering{
\includegraphics[width=7.5cm]{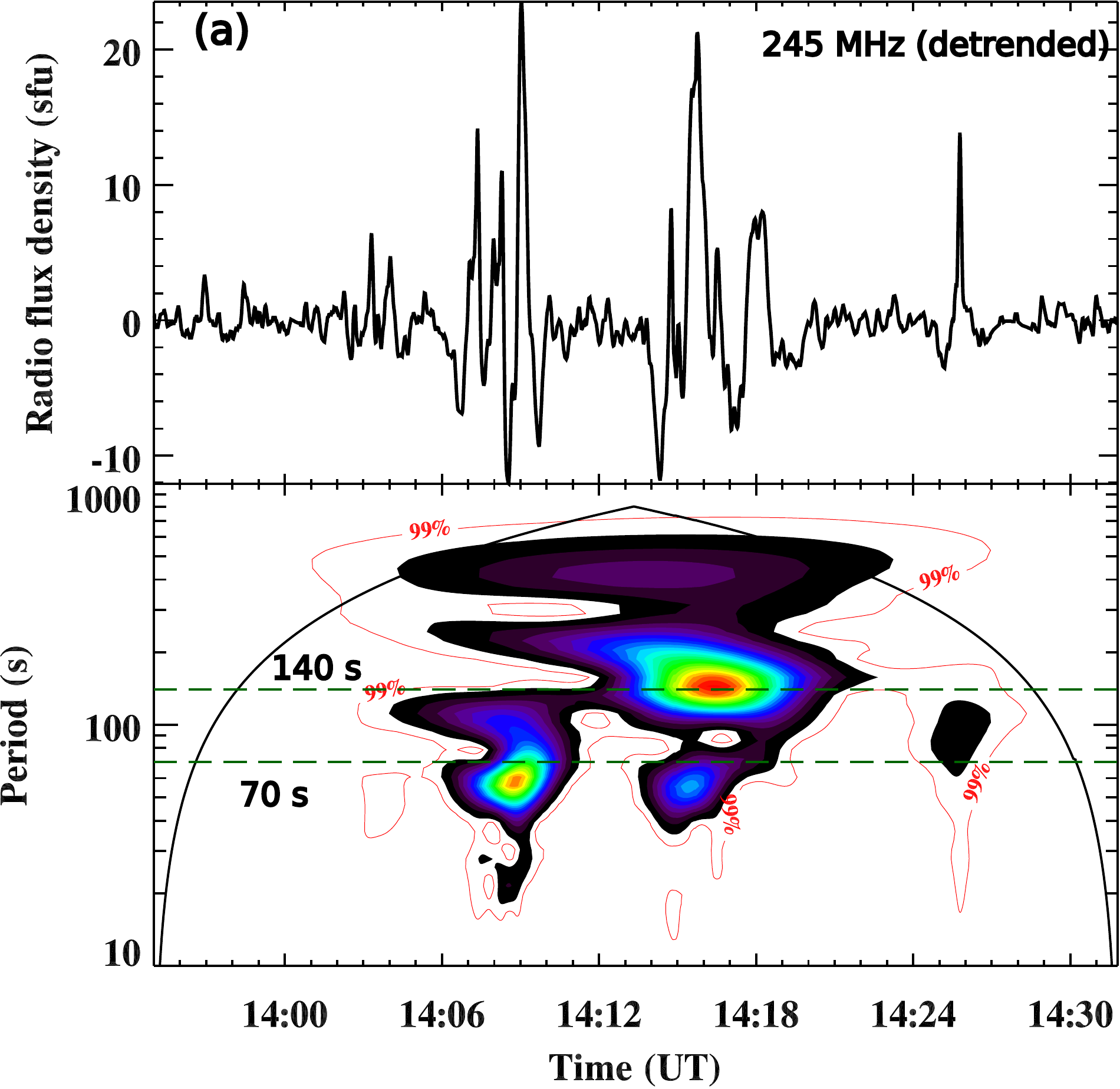}
\includegraphics[width=7.5cm]{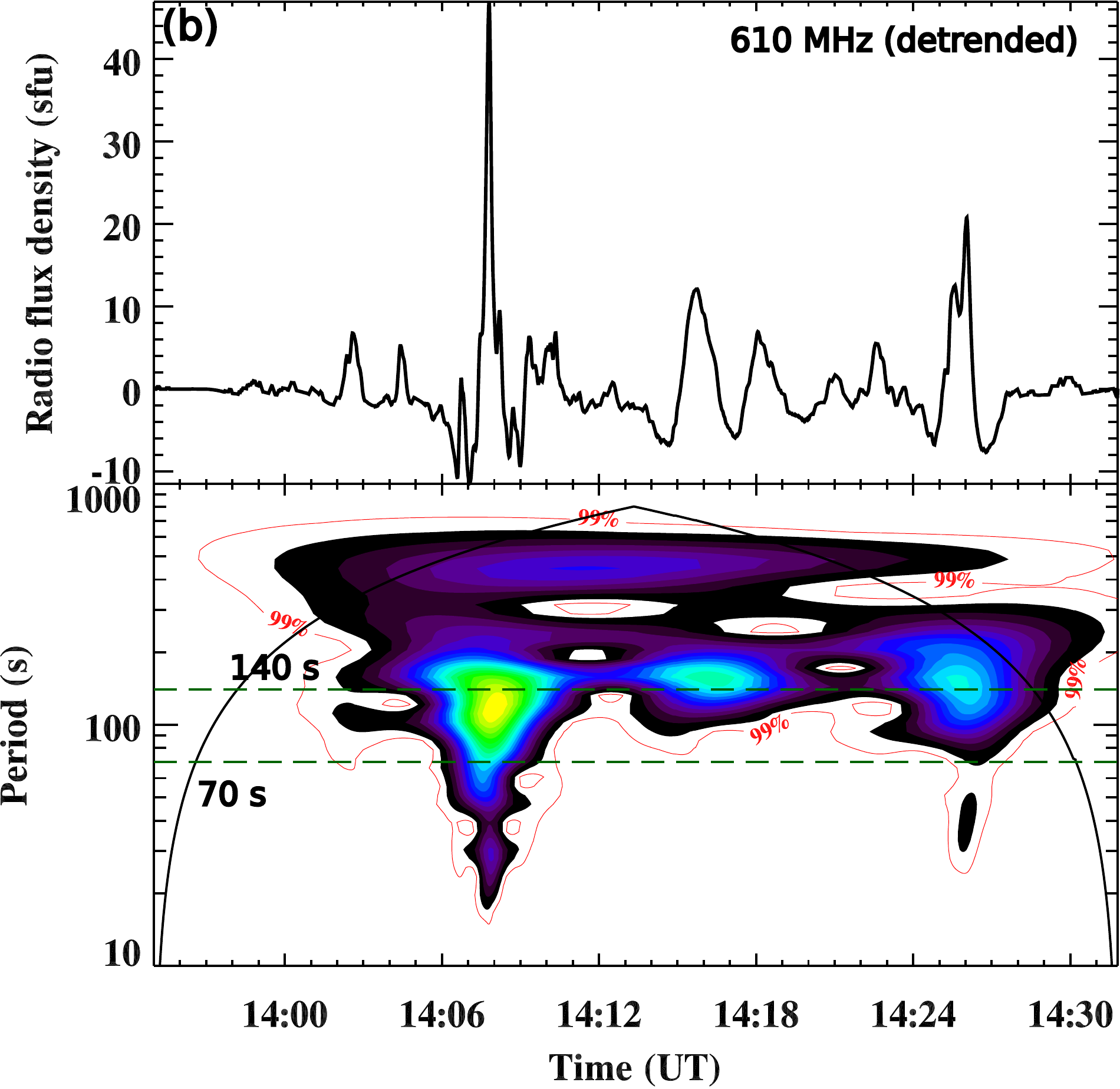}

\includegraphics[width=7.5cm]{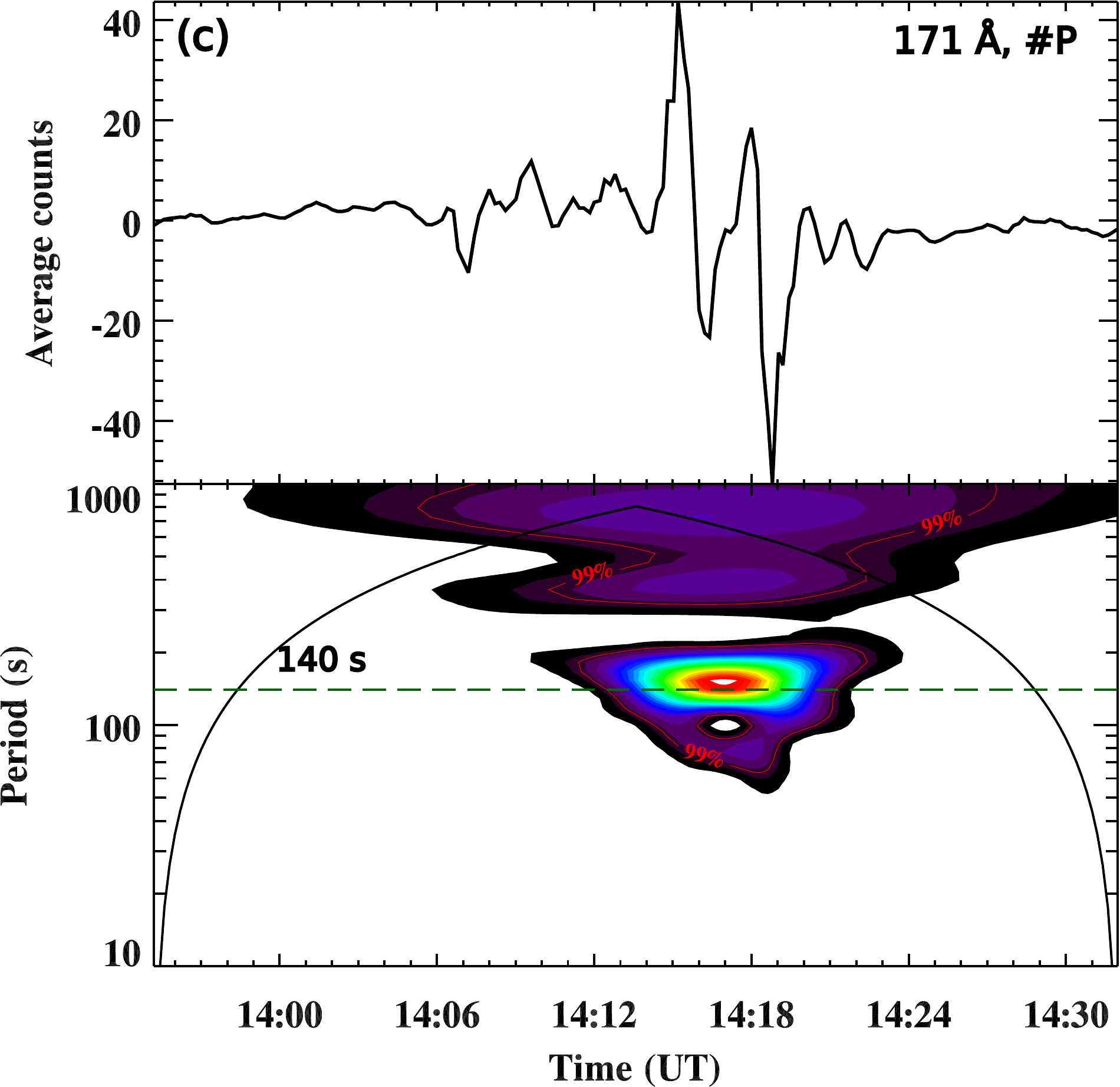}
\includegraphics[width=7.5cm]{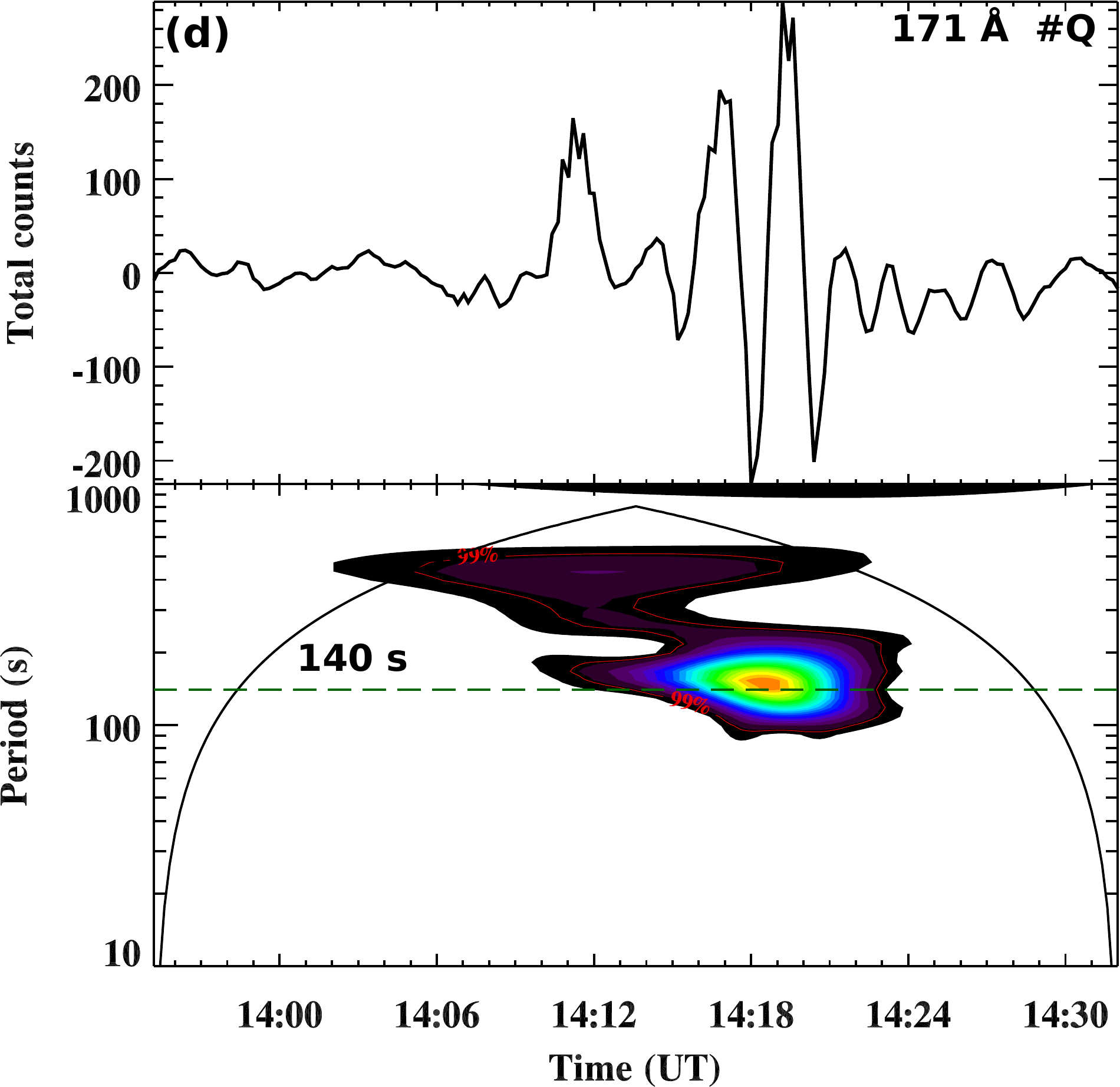}
}
\caption{(a,b) RSTN 245 and 610~MHz detrended flux profiles, and their wavelet power spectra. (c,d) AIA 171~\AA~ running difference (smoothed) intensity profiles along the selected paths in Figure~\ref{st}(b), and their wavelet power spectrum.} 
\label{rstn1}
\end{figure*}

\section{OBSERVATIONS AND RESULTS}
The {\it Atmospheric Image Assembly} (AIA; \citealt{lemen2012}) onboard the {\it Solar Dynamics Observatory} (SDO) records full disk images of the Sun (field-of-view $\sim$1.3~R$_\odot$) with a spatial resolution of 1.5$\arcsec$ (0.6$\arcsec$~pixel$^{-1}$) and a cadence of 12~s. For the present study, we utilized AIA 171~\AA\ (\ion{Fe}{9}, the temperature $T\approx 0.7$~MK), 94~\AA\ (\ion{Fe}{10}, \ion{Fe}{18}, i.e., $T\approx 1$~MK, $T\approx 6.3$~MK), 131~\AA\ (\ion{Fe}{8}, \ion{Fe}{21},  \ion{Fe}{23}, i.e., $T\approx 0.4$, 10, 16~MK, respectively),  1600~\AA\ (\ion{C}{4} + continuum, $T\approx 0.01$~MK), 335~\AA\ (\ion{Fe}{16}, T$\approx$2.5~MK) and AIA 193~\AA\ (\ion{Fe}{12}, \ion{Fe}{24}, i.e., $T\approx  1.2$~MK and $\approx 20$~MK) images. We also used the Helioseismic and Magnetic Imager (HMI; \citealt{schou2012}) magnetogram to infer the magnetic topology of the active region. We utilised the Reuven Ramaty High-Energy Solar Spectroscopic Imager (RHESSI; \citealt{lin2002}) and Nan\c{c}ay radioheliograph (NRH;\citealt{kerdraon1997}) observations to investigate the hard X-ray (HXR) and radio emissions in different energy bands. NRH provides observations at nine frequencies, i.e., 150, 173, 228, 270, 298, 327, 408, 432 and 445~MHz.  We analysed NRH 10-s cadence data. 

On 7 May 2012 the NOAA active region 11471 (the $\beta$ magnetic configuration, S19W50) was located close to the western limb. The wave train, reported here (Section~\ref{wave train}), was associated with an M1.9 flare that started at $\sim$14:03~UT, peaked at $\sim$14:31~UT and ended at $\sim$14:52~UT. The upward propagating wave train was observed in EUV at 14:06--14:26~UT, for about 20 minutes.

Figure~\ref{goes}(a-b) displays the 2-s cadence GOES soft X-ray (SXR) flux profile in 1--8 \AA~ and its time derivative. Unfortunately, RHESSI did not observe the impulsive phase of the flare.  
We used the Lagrangian interpolation technique (three-point numerical differentiation) to estimate the time derivative of the GOES soft X-ray flux. The SXR derivative profile is smoothed by 8-s. We subtracted a 124-s smoothed curve (blue) to detrend the signal (Figure~\ref{goes}(b)), and decomposed it using the Morelet wavelet technique \citep{torrence1998}. The wavelet power spectrum reveals the presence of statistically significant periodicities, with the periods of $\sim$70~s and 140~s being above the 99\% significance level. 

\subsection{Upward propagating wave train and eruption}
\label{wave train}

We analysed the time sequence of AIA 171~\AA~ images taken during the flare to investigate the quasi-periodic propagating EUV wavefronts (i.e., a wave train). The wave train was best observed in the AIA 171~\AA~ channel. Figure~\ref{aia171} displays  selected AIA 171~\AA~ running-difference images showing multiple wavefronts during the impulsive phase of the M1.9 flare. At $\sim$14:01~UT, we notice a radially outward expanding CME frontal loop and an associated EUV brightening near the limb. The first EUV wavefront (behind the CME frontal loop) was observed at $\sim$14:06~UT.   The AIA 171~\AA~ movie clearly shows a train of rapidly-propagating quasi-periodic wavefronts near the limb at $\sim$14:06~UT onward, in the direction indicated by arrow S1 in Figure~\ref{aia171}, lasting till $\sim$14:26~UT. The wavefronts are seen to move radially outward from the flare site. The wavefronts are of a quite large scale, having a semi-circular convex arc shape, of the size of $\sim$300$\arcsec$ while reaching the limb. Another wave train-like feature is seen to propagate in the southward direction along the arcade loop, in the direction indicated by arrow S2 in Figure~\ref{aia171}.
In addition, there are quasi-periodic disturbances propagating downwards from the same site (Section~\ref{downward}).   

To determine the speed of the wavefronts and associated loop eruption, we created time--distance intensity (running difference) plots along slices S1, S2, and S3 during 13:55--14:40~UT (Figure~\ref{st}).  Slice S1 (red) is chosen to estimate the speed of the wave fronts going in the westward direction, whereas S2 (blue) for the laterally moving fronts in the southward direction. Slice S3 (green) is selected to investigate the CME frontal loop expansion during the flare onset. The speeds of wavefronts are estimated using a linear fit to the data points of each visually tracked propagating wavefront that appears as a diagonal feature in the time--distance plot. The speeds of individual wavefronts along slice S1 are 850$\pm$80, 1026$\pm$52, 1482$\pm$176, 1416$\pm$46, 1192$\pm$56, and 1147$\pm$214~\kms, respectively. The speeds of individual wavefronts along slice S2 are 664$\pm$76, 810$\pm$42, 1140$\pm$288, 910$\pm$170, 726$\pm$85, and 790$\pm$110 \kms, respectively. The speed of the southward moving fronts is lower than that of the westward  (radially outward) moving fronts, which may be attributed either to the line-of-sight projection effect or to the difference in the local fast speed or other properties of the corresponding waveguides. The speed of the wavefronts is found to be close to the expected Alfv\'en speed in coronal active regions \citep{mann1999,mann2003,warmuth2005}. Therefore, the rapidly-propagating compressive disturbances are interpreted as quasi-periodic propagating fast-mode waves. The wavefronts are seen to follow each other quasi-periodically, with the period decreasing from about 240~s to about 120~s, which is best seen in panel (b).  According to the wavelet spectrum shown in Figure~\ref{rstn1}(d), the main power of quasi-oscillatory patten (the wavelet tadpole \lq\lq head\rq\rq) is situated near the period of 140~s.

The multiple EUV fronts were mostly observed behind the erupting CME frontal loop. However, one front was observed ahead of the expanding CME frontal loop in the northward direction. This is similar to a ``classical" (single-pulse) EUV wave (e.g., \citealt{kumar2013ww,warmuth2015}) in front of the expanding CME flanks. Panel (c) reveals the lateral expansion of CME flanks in the 171~\AA~ channel, with the speed of 86--120~\kms~ (14:00--14:12~UT).  

Figure~\ref{aia3} shows AIA 131, 335, and 193~\AA~ running difference images of the field of interest. The AIA 131~\AA~ image at $\sim$14:01~UT reveals the appearance of bright loops. The HMI magnetogram contours of positive (red) and negative (blue) polarity regions indicate the magnetic configuration of the active region  (the magnetic polarities of the loops' footpoints). During $\sim$14:05--14:11~UT we see the simultaneous formation of the underlying flare loop (the red curve) and overlying loop systems (the blue curve). This could be a signature of magnetic reconnection associated with the formation of the hot flux rope. The AIA 131~\AA~ movie clearly shows a clockwise rotation of a small filament close to the southern leg of the flux rope. The filament  rotation suggests the existence of a right-handed twist of the flux rope (e.g., see \citealt{kumar2014}). It should be noted that the untwisting filament (marked by an arrow) was also observed in the cool channels (i.e., AIA 304~\AA~ images). However the other part of the flux rope is seen in the hot channel (i.e., AIA 131~\AA) only. 

The AIA 335 \AA~ running-difference images (Figure~\ref{aia3}(d-f)) reveal a large-scale expanding front (F1) moving in the northwest direction. The bright front appeared at $\sim$14:00~UT, and was observed until 14:11~UT (see the AIA 335 \AA~ movie). Apart from this front, we observe a faint wavefront (F2) moving in the north direction (refer to the AIA 335 \AA~ movie).

The quasi-periodic wave train was also detected in the AIA 193~\AA~ channel (Figure~\ref{aia3}(g-i)). The AIA 193~\AA~ movie shows a wave train observed during 14:00--14:30~UT, similar to the train seen in AIA 171~\AA~ images. We also see the  expanding front (F2) in the north direction at 193~\AA. 

To investigate the kinematics of the loop eruption and wavefronts, we selected slices S4, S5, S6, and S7 to create the time--distance intensity plots using AIA 131, 335, and 193~\AA~ running difference images (Figure~\ref{st}(c-f)) during 13:55--14:30~UT.  The AIA 131 \AA~ time--distance plot shows a bright loop (i.e., a flux rope) eruption with a speed of $\sim$200~\kms~at $\sim$14:08~UT (Figure~\ref{st}(d)). We fitted a second order polynomial function ($d=a+bt+ct^{2}$, where $d$ is the distance, $t$ is the time, and $a$, $b$ and $c$ are constant coefficients) to the time--distance profile of the expanding front (F1) observed in the AIA 335~\AA~ channel in the northwest direction (Figure~\ref{st}(e)). Initially, the front shows a slow-rising motion, until $\sim$14:05~UT, and then reveals a fast rise during $\sim$14:06--14:16~UT. The estimated speed of the front F1 is 64--472~\kms.

The AIA 193~\AA~ time--distance plot made along S6 reveals another fast wavefront (F2) moving in the north direction (Figure~\ref{st}(f)). The estimated speed of F2 is $\sim$657$\pm$22~\kms. The time--distance plot made along slice S7 (Figure~\ref{st}(g)) reveals a quasi-periodic rapidly-propagating wave train similar to the wave train observed in the AIA 171 \AA~ channel. We also plotted the GOES SXR flux time derivative to see the association of the wave train with the quasi-periodic flare energy release. The quasi-periodic energy release starts at $\sim$14:06~UT onward at the same time of the appearance of the EUV wave train.
   
The initial speed of front F1 ($\sim$64~\kms) observed in the AIA 335~\AA~ is comparable to the speed of the lateral expansion of the loop (86--102~\kms) observed at AIA 171~\AA~ (Figure~\ref{st}(c)). In addition, the start time of F1 closely matches the lateral expansion of the 171 \AA~ loop. Front F1 initially reveals slow rise until 14:01 UT and an acceleration in the later phase.  This type of slow acceleration is not expected in the case of primary fast-mode waves ahead of a CME piston. The piston driven fast-mode waves generally shows deceleration after decoupling from the piston. Therefore, front F1 is interpreted as a lateral expansion of the CME flank (i.e., a pseudo wave). The AIA 335 \AA~ running difference images do not show any decoupling from F1. Front F2 (14:06 UT) is fast enough to be interpreted as a fast-mode wave. At 171 and 193~\AA~ we also see a co-temporal fast EUV front ($\sim$14:01~UT) that propagates outward along S1/S2/S7 at the speed higher than 1000~\kms. This could be a part of the wave train too.

To determine the relative intensity and amplitude profiles (e.g., for linear/nonlinear behaviour, \citealt{warmuth2011,muhr2014},) of the wave train, we analysed AIA 171 \AA~ base ratio images ((I-I$_0$)/I$_0$, where I$_0$ is an image before the flare at $\sim$13:45 UT). Figure \ref{base_rat} (b) displays the time-distance intensity plot along a slice cut (cyan) marked in Figure \ref{base_rat} (a).  We chose the cut along the radial direction of the outwards propagating wave train, where the fronts are more clear. Figure \ref{base_rat}(c) displays the relative intensity of outward propagating  multiple wavefronts at different distances (arcsecs) along the horizontal dotted line marked in the time-distance intensity plot (panel (b)).
 The measurement of the relative amplitude of the wave trains, i.e. the ratio of the maximum amplitude of the intensity to the unperturbed intensity, is under 5$\%$. Such an amplitude corresponds to a rather linear behaviour. Even for larger amplitude, the wave train should not show a signature of the nonlinear evolution, as the nonlinear steepening is suppressed by the waveguide dispersion. However, this effect would require a dedicated theoretical analysis.

The Solar Terrestrial Relations Observatory (STEREO-A; \citealt{kaiser2008}) Sun Earth Connection Coronal and Heliospheric Investigation (SECCHI; \citealt{howard2008}) observed the same active region near the eastern limb. We used the extreme-ultraviolet imager (EUVI) 195~\AA~ (up to 1.7~R$_\odot$) and white-light coronagraph COR1 (1.5--4~R$_\odot$) running difference images (5-min cadence) to determine the low-coronal eruption and associated CME. Figure~\ref{stereo} displays AIA 193~\AA~ and EUVI 195~\AA~ images to see the eruption from a different viewing angle. The EUVI 195~\AA~ movie clearly shows the lateral expansion of the CME and associated dimming during the eruption. COR1 images reveal the flux rope structure of the CME. To determine the CME speed in the COR1 field of view (1.5--4~R$_\odot$), we visually tracked the leading edge of the CME during 14:15--15:00~UT. Using a linear fit to the height--time dependence of the leading edge, the estimated speed of the CME was found to be $\sim$718~\kms. According to the COR2 catalogue, the CME speed in the COR2 field of view (2.5--15~R$_\odot$) was $\sim$665~\kms. Therefore, the CME speed is almost comparable in both COR1 and COR2.
We carefully compared the CME structure with the low coronal eruptions observed with AIA and EUVI. Front F1 observed with AIA showed a parabolic path in the kinematic plot, which corresponds to the acceleration from 64~\kms~to 472~\kms, suggesting a loop eruption rather than an EUV wave. Therefore, F1 is interpreted as an expanding CME frontal loop, which we observed later with COR1. Behind the CME frontal loop, we observed an erupting flux rope in the AIA 131~\AA~ channel. COR1 images clearly show a flux rope behind the CME frontal loop.
 The COR1 image reveals the separation of two EUV fronts (north and south, red arrows) from the CME flanks. The front in the north direction (F2) was observed in the EUV images (193~\AA) during the lateral expansion of the CME flank.

\subsection{Quasi-periodic radio bursts}
\label{radiosec}

Figure~\ref{rstn}(a) shows the radio flux density profiles at 245, 410, 610, 1415, 2695, and 8800~MHz observed at the Sagamore Hill station of the Radio Solar Telescope Network (RSTN) with a 1-s cadence. The bottom panel shows the NRH 10-s integrated radio flux profiles in 150, 228 (green), amd 298 (blue) MHz frequencies.
Figure~\ref{rstn}(b) displays the dynamic radio spectrum observed by Green Bank Solar Radio Burst Spectrometer  (GBSRBS; \citealt{white2005}) in the 300--1200~MHz frequency band during 14:00--15:00~UT. The high-quality dynamic radio spectra are generated by GBSRBS with the time resolution of 1-s. High frequency type III radio bursts are seen during 14:02--14:05~UT. We see repetitive intense type III and type IV radio bursts (at 300--1200~MHz)  at $\sim$14:06~UT onward, that continue until 14:40~UT. Low-frequency radio bursts ($\sim$300--400~MHz) were detected until $\sim$15:00~UT. The low-frequency (metric, decametric) bursts are shown in the dynamic radio spectra observed at San-Vito (25--180~MHz) and the Nan\c{c}ay Decametric Array \citep[NDA;][]{lecacheux2000} in 10-80~MHz. We see a series of type III radio bursts in the 20--60~MHz frequency band during $\sim$14:02--14:42~UT in Sanvito spectrum. However, the type-III radio bursts were observed until $\sim$15:00 UT in the Nan\c{c}ay Decametric Array dynamic spectrum.

Figure~\ref{ondr} shows the Ondrejov dynamic radio spectrum (800--1200~MHz, \citealt{jiricka1993}) for 14:01--14:28~UT. Note that the Ondrejov dynamic radio spectrum has better resolution than GBSRBS, therefore, it is useful for studying a fine structure of the burst. 

The repeated radio bursts were observed in the decametric, metric, decimetric, and microwave emissions. The microwave emission (2.6 and 8.8~GHz) was observed almost until $\sim$14:27~UT. The emissions at 245 and 610~MHz show quasi-periodic bursts for up to almost one hour (i.e., 14:00--15:00~UT). 

Figure~\ref{ondr} demonstrates the high-resolution dynamic spectrum of the decimetric radio emission during 14:01--14:28~UT. From 14:15~UT, there are quasi-periodic pulsations of the emission intensity, seen as the vertical lanes, with a time scale decreasing from $\sim$170~s to $\sim$30~s.
Note that the AIA 171/193~\AA~ time--distance intensity plot reveals diffuse wavefronts (more clearly seen in 193~\AA, Figure~\ref{st}(g)) during 14:22--14:26~UT. Therefore, these burst are associated with the rapidly propagating EUV waves. 


To determine the periodicity in the metric/decimetric emission, we selected 245 and 610~MHz flux profiles during the impulsive/maximum phase of the flare (13:55--14:32~UT). The original signal was smoothed by 6 s and then a 150~s smoothed curve was subtracted to detrend the light curve. Figure~\ref{rstn1}(a, b) shows the wavelet power spectra of the detrended light curves at 245 and 610~MHz. The 245~MHz power spectrum reveals  two periods, of $\sim$70 and $\sim$140~s. The 610~MHz power spectrum also shows a $\sim$140~s periodicity. These periods are consistent with the periods detected in the GOES soft X-ray flux time derivative profile.


\subsection{Magnetic configuration of the flare site}
\begin{figure*}
\centering{
\includegraphics[width=8.5cm]{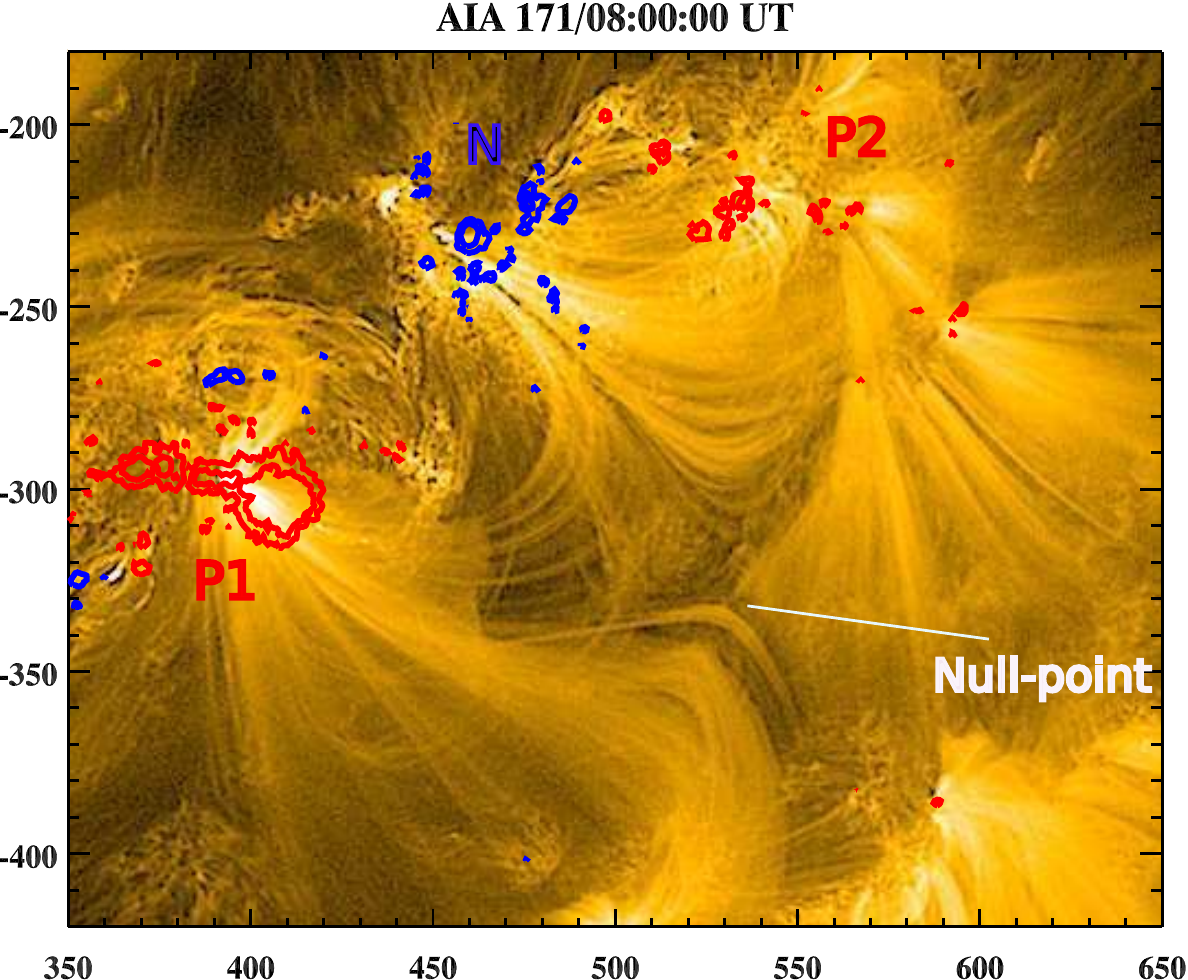}
\includegraphics[width=7.8cm]{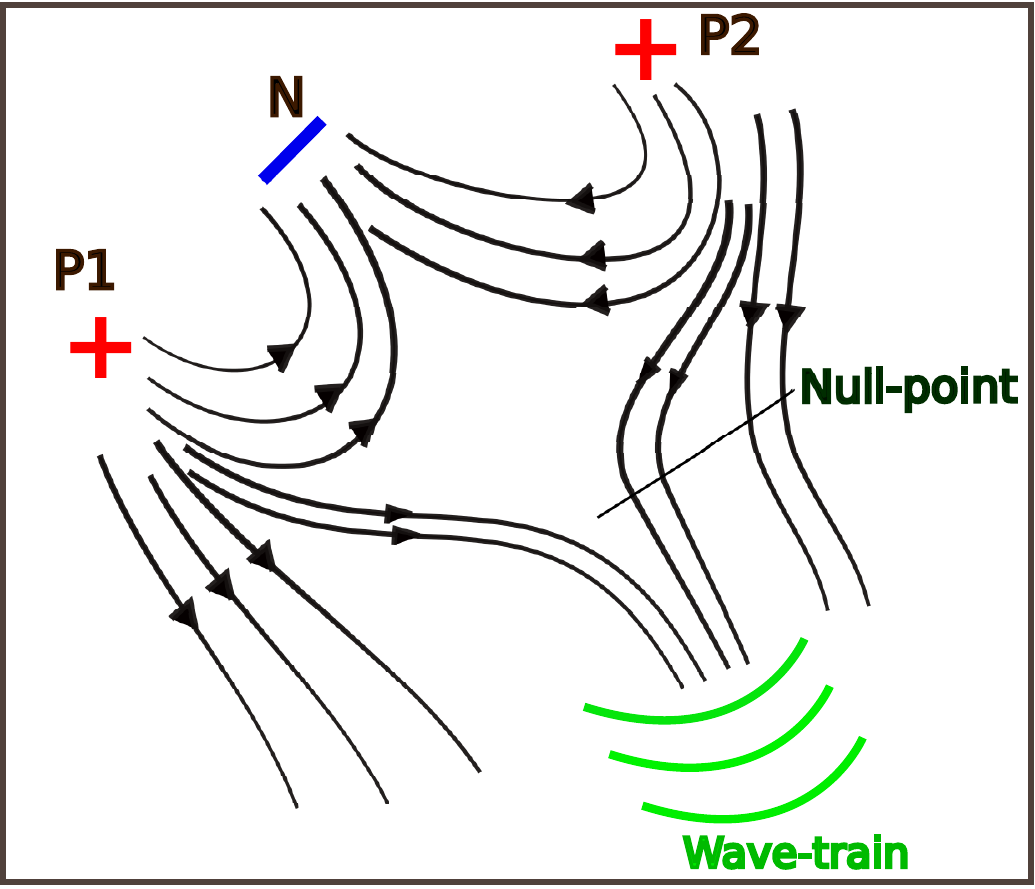}
}
\caption{Left: AIA 171~\AA~ image on 6 May 2012 overlaid by HMI magnetogram contours of positive (red) and negative (blue) polarities. The contour levels are $\pm$500, $\pm$1000, $\pm$2000~G. The X- and Y axes are labeled in arcsecs. Right: Schematic cartoon showing the magnetic field configuration of the flare site. Labels P1, P2, and N indicate positive and negative polarity field regions. The green curves denote the wave-train generated in the vicinity of the magnetic null point.} 
\label{cartoon}
\end{figure*}

To determine the magnetic topology of the flare site, we show the AIA 171~\AA~ intensity image taken one day before the studied event, on 6 May 2012 (Figure~\ref{cartoon} left panel). The image is overlaid by HMI contours of positive (red) and negative (blue) polarities. This could be the best way to avoid the projection effect, allowing for the correct identification of the magnetic polarities, since it was more difficult to obtain this information on 7 May 2012 when the active region was lying closer to the west limb. The AIA 171~\AA~ image shows the connectivity of the active region loops. The active region has a negative polarity region (N) surrounded by positive polarity regions on both sides (P1, P2). We see the EUV loops connecting regions P1/P2 with region N. In addition, we see open plasma structures emanating from P1 and P2. A cusp-shaped structure is also evident. The magnetic field configuration is quite similar to the topology needed for the magnetic breakout reconnection \citep{antiochos1998,antiochos1999,karpen2012} along with a magnetic null point. The right panel of Figure~\ref{cartoon} shows a sketch demonstrating the possible magnetic configuration derived from the AIA 171~\AA~ and HMI magnetogram images. The eruption starts with the expansion of the southward loop system connecting P1 and N. The flux rope eruption originated from the polarity inversion line between P1 and N. Later, most of the post-flare arcade loops were formed between P1 and N. We also observed a remote ribbon (R3, discussed below) at P2 during the flare. 
\begin{figure*}
\centering{
\includegraphics[width=6.4cm]{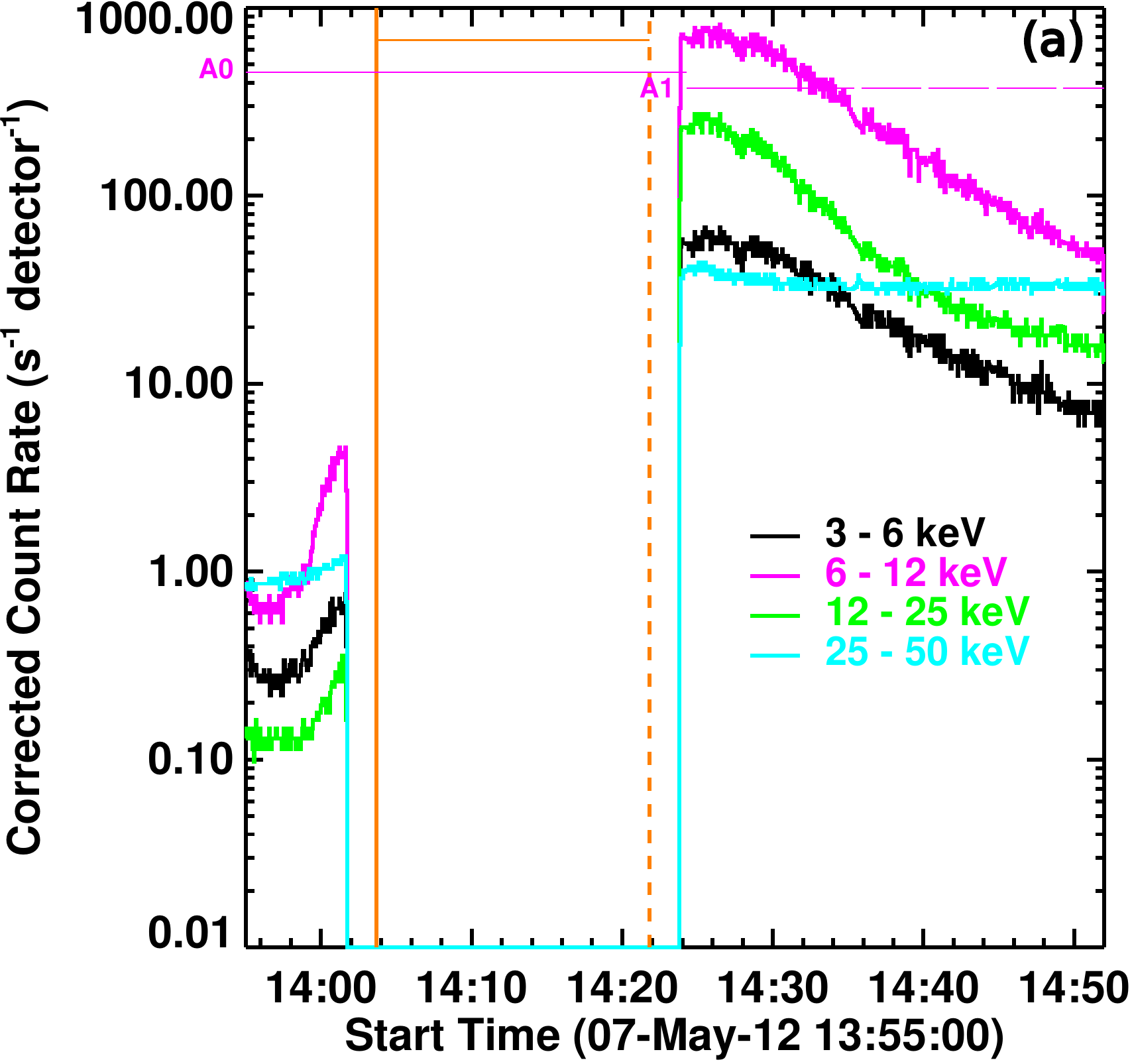}
\includegraphics[width=7.5cm]{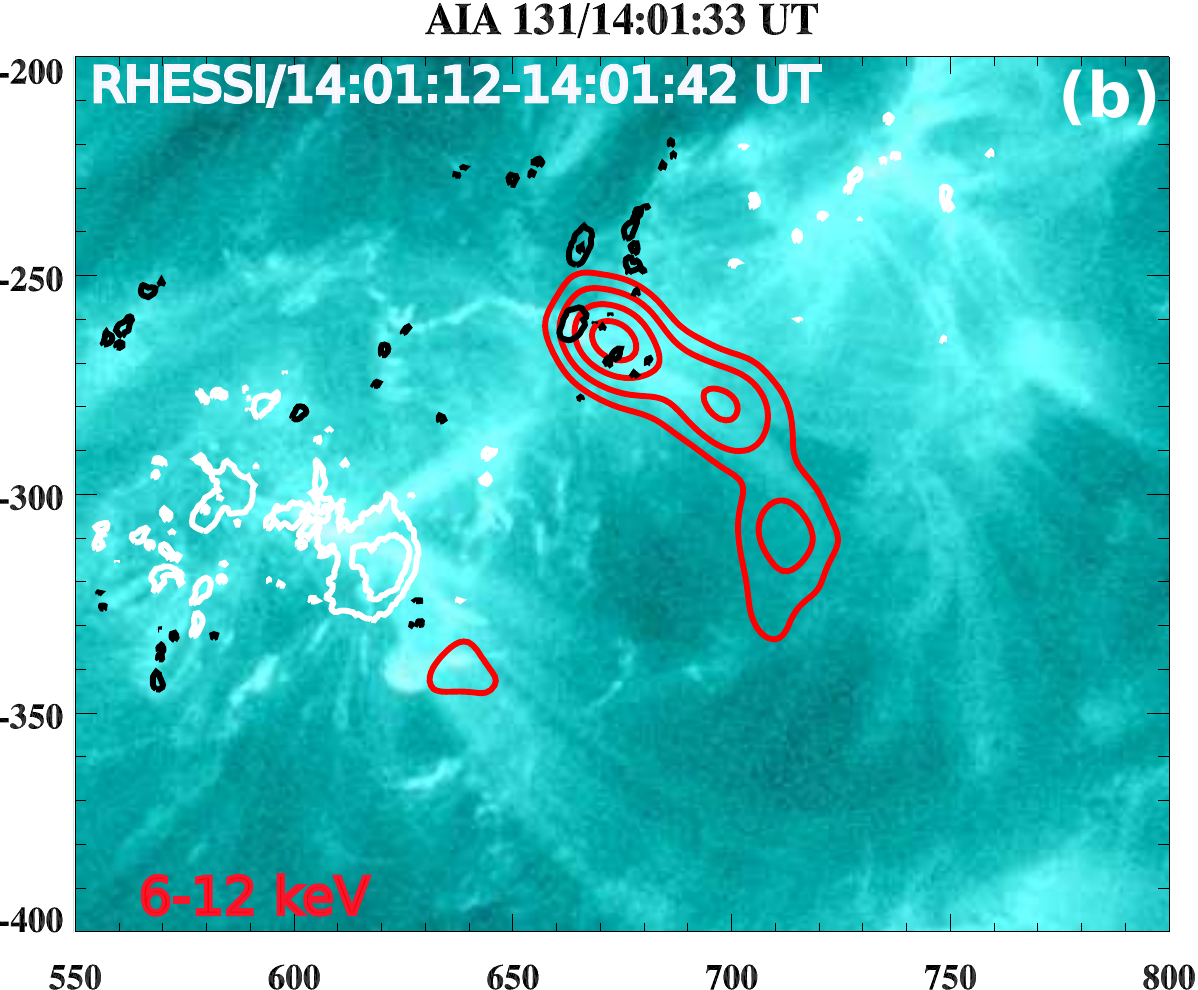}
\includegraphics[width=7.0cm]{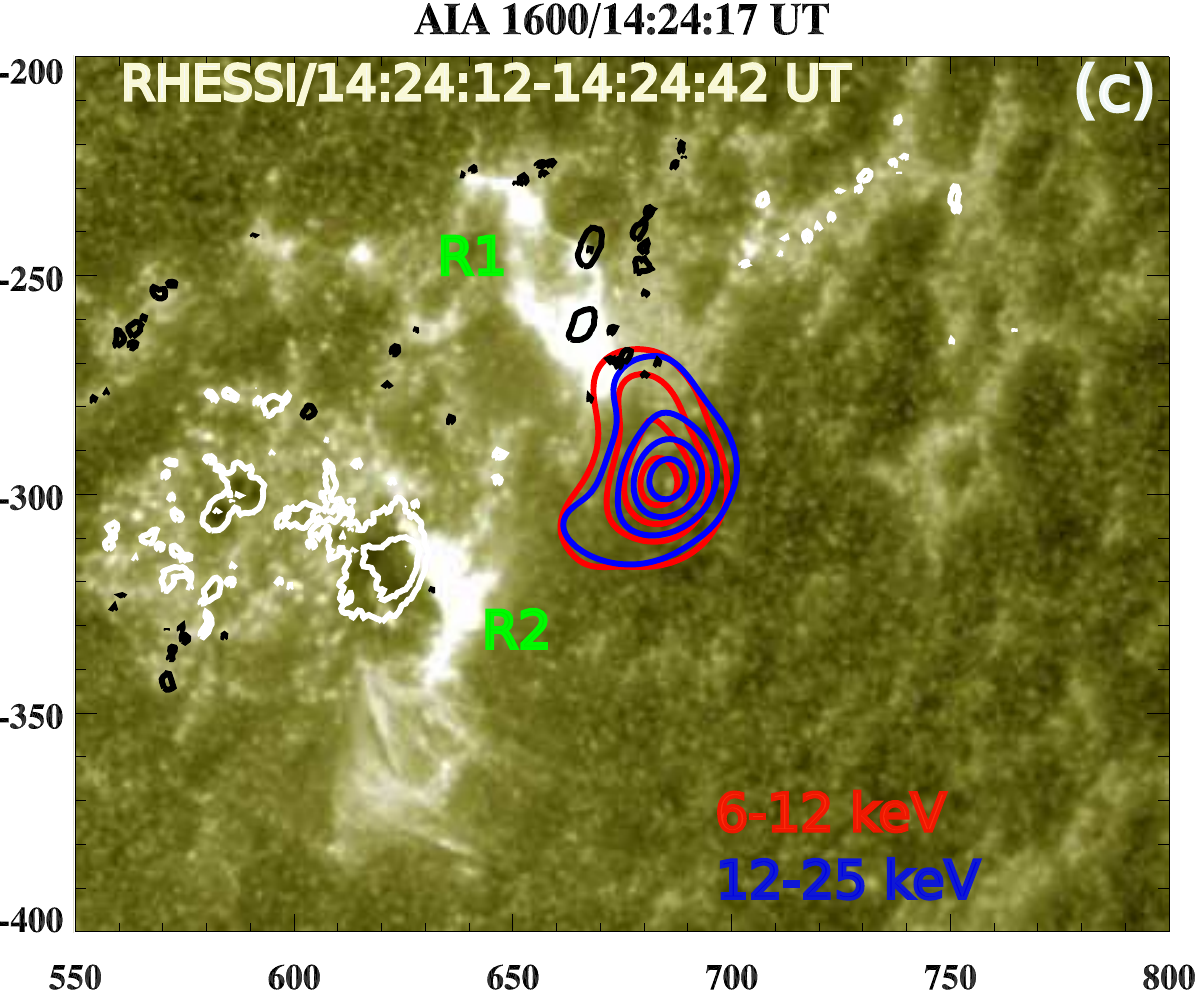}
\includegraphics[width=7.0cm]{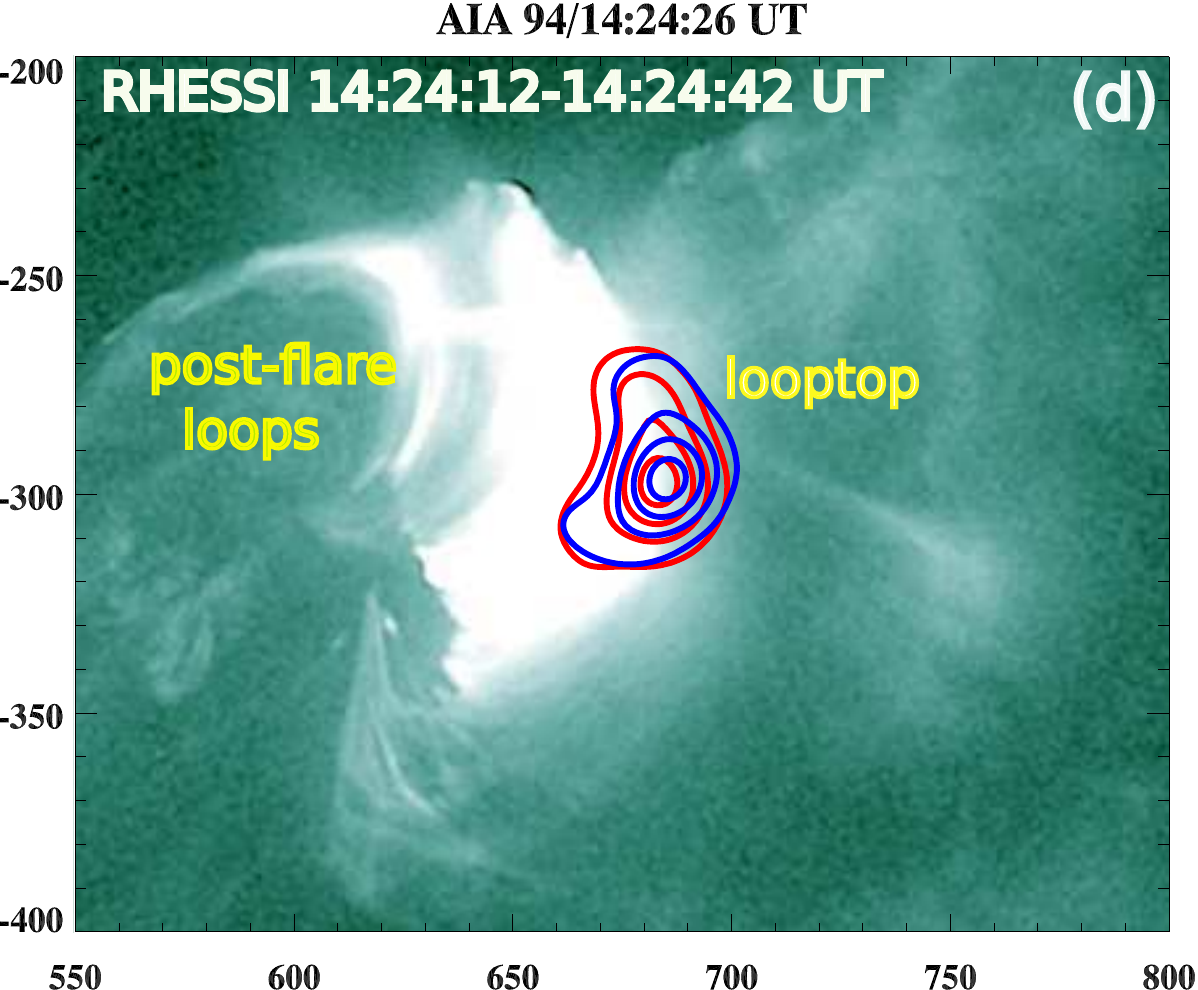}
}
\caption{(a) RHESSI X-ray flux profiles in the 3--6, 6--12, 12--25, 25--50~keV channels. (b,c,d) RHESSI X-ray contours at 6--12~keV (red) and 12-25~keV (blue), overlaid on the AIA 1600, 131, and 94~\AA~ images. The contour levels are 30$\%$, 50$\%$, 70$\%$, 90$\%$ of the peak X-ray intensity. Panels (b,c) are overlaid by HMI magnetogram contours of positive (white) and negative (black) polarities. The contour levels are $\pm$400, $\pm$1000, $\pm$1500~G. Labels R1 and R2 indicate the flare ribbons. The X- and Y axes are labeled in arcsecs.} 
\label{hessi}
\end{figure*}

\begin{figure*}
\centering{
\includegraphics[width=6.5cm]{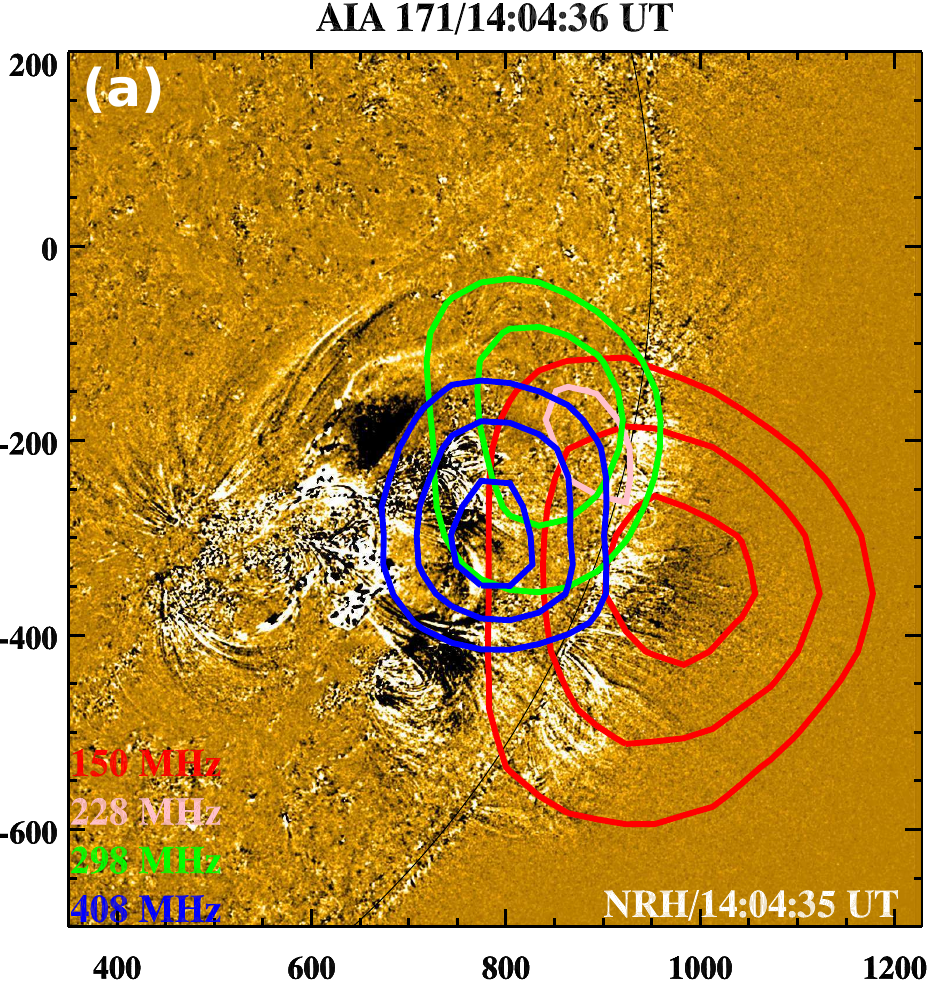}
\includegraphics[width=6.5cm]{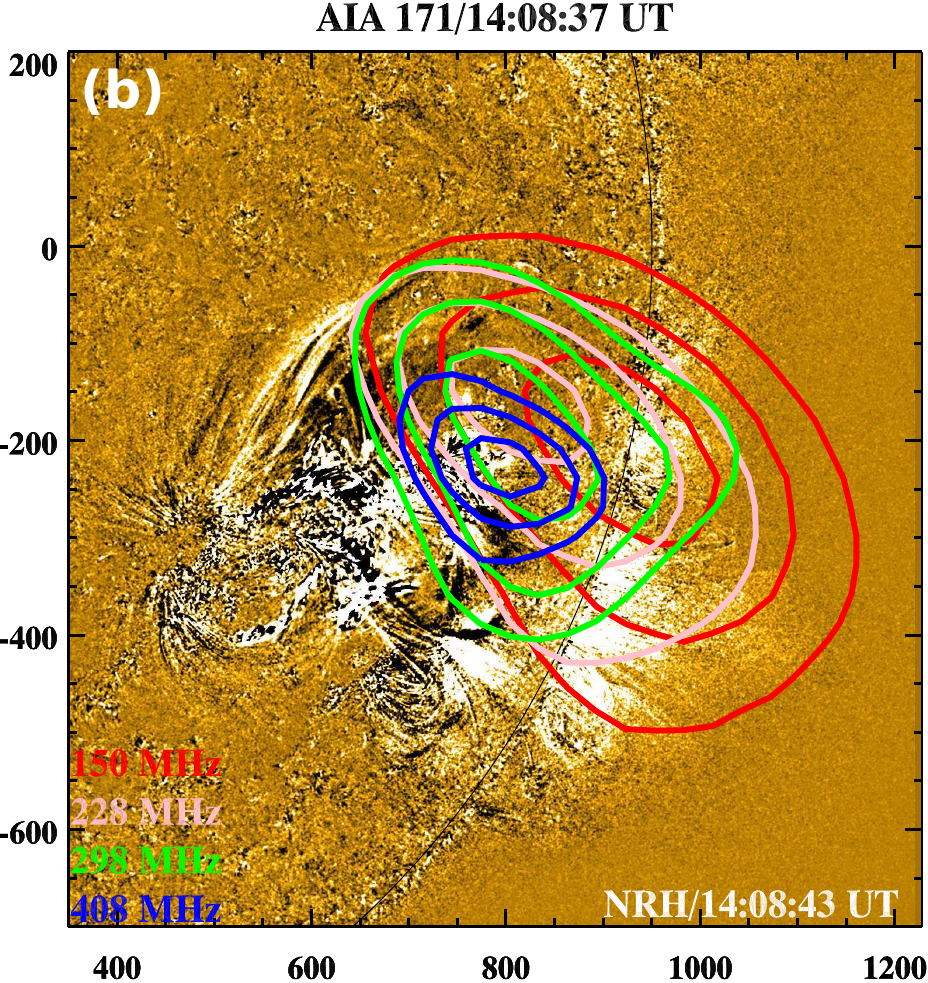}
\includegraphics[width=6.5cm]{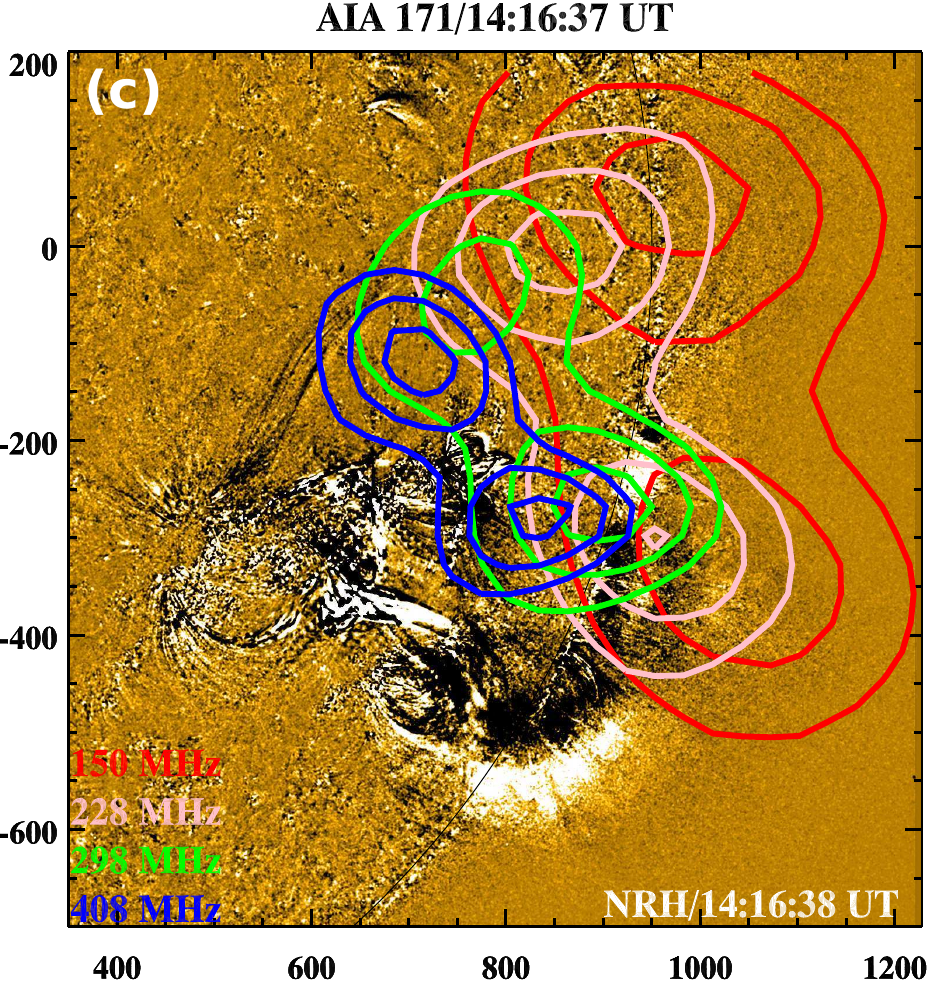}
\includegraphics[width=6.5cm]{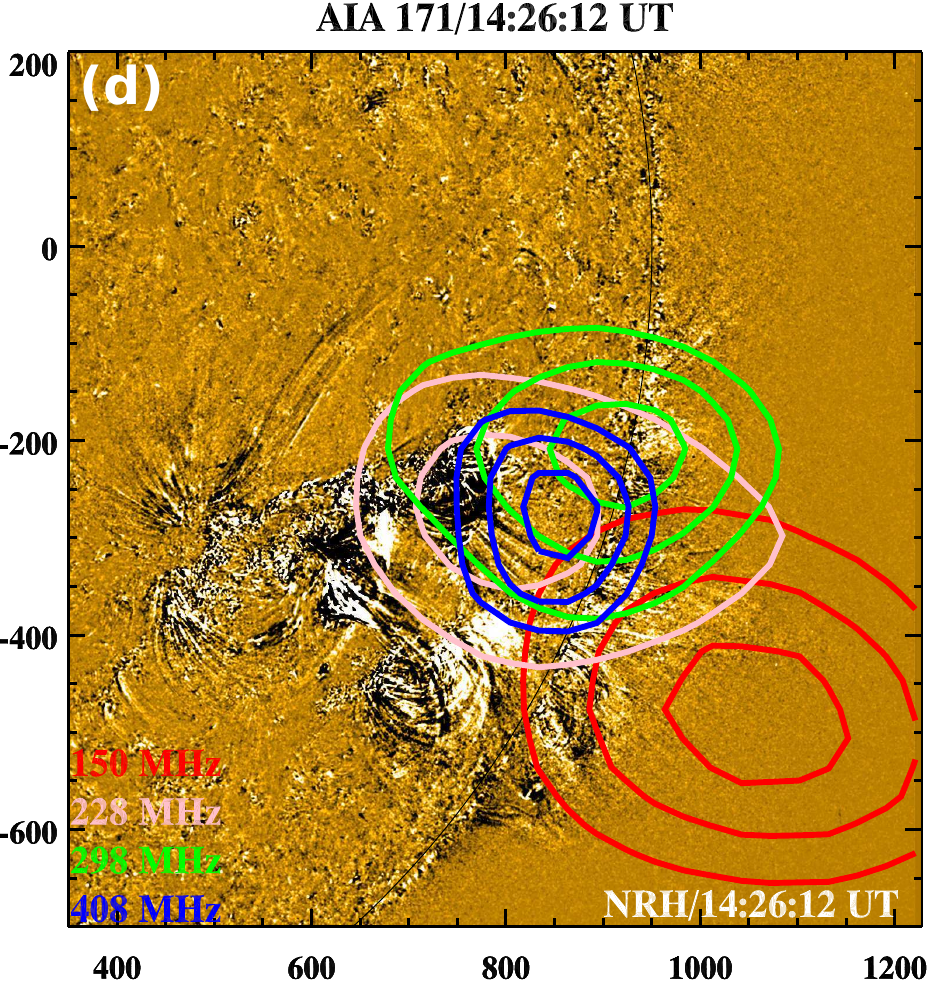}
\includegraphics[width=6.5cm]{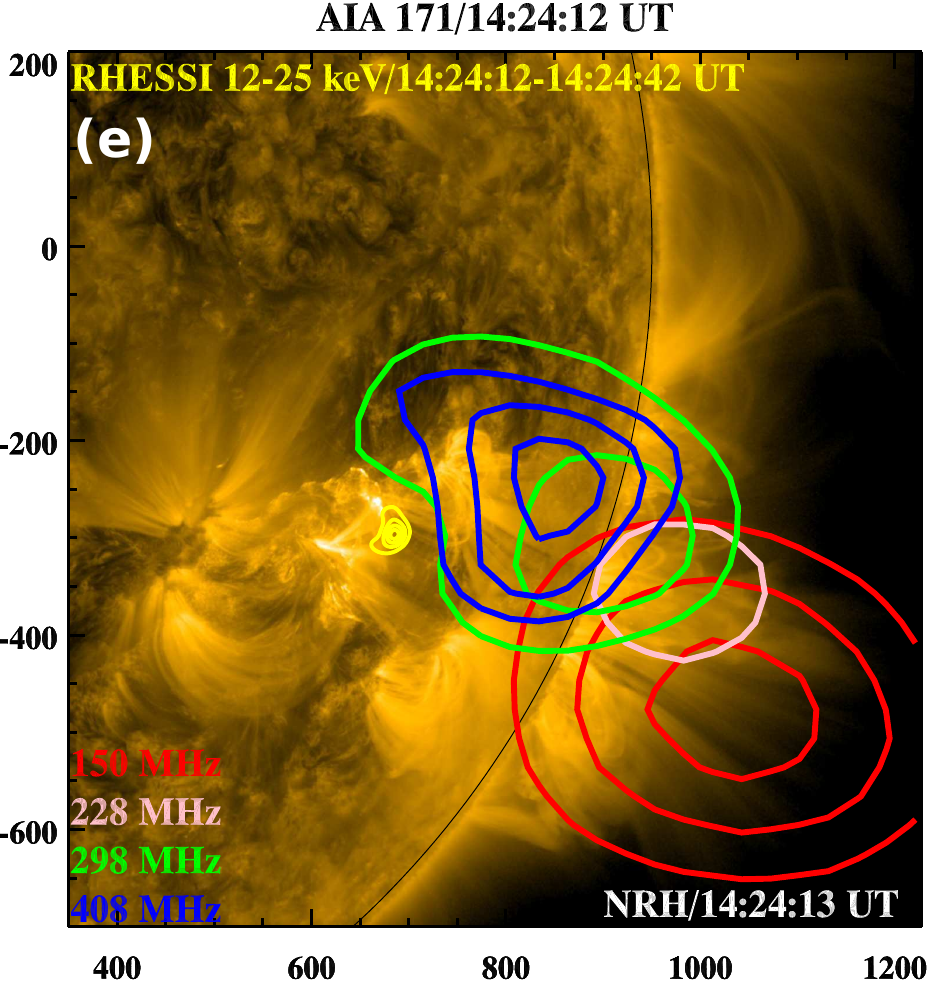}
\includegraphics[width=6.5cm]{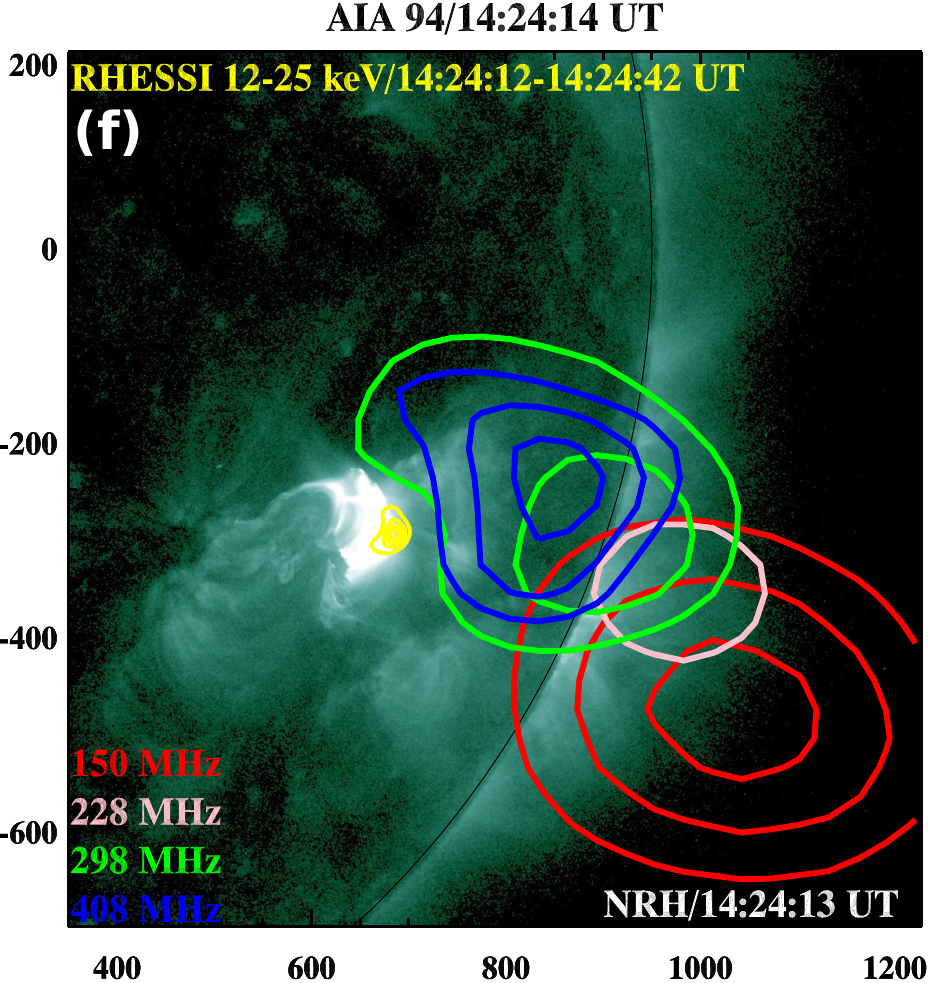}
}
\caption{(a-d) AIA 171 \AA~ running difference ($\Delta$t=1 min) images overlaid by NRH radio contours (contour levels=50$\%$, 70$\%$, 90$\%$ of the peak intensity) at different frequencies (150, 228, 298, and 408~MHz). (e-f) NRH radio and RHESSI X-ray (yellow) contours overlaid on the AIA 171 and 94 \AA~ images. The contour levels for RHESSI are 30$\%$, 50$\%$, 70$\%$, 90$\%$ of the peak intensity.} 
\label{nrh}
\end{figure*}
\begin{figure*}
\centering{
\includegraphics[width=6.0cm]{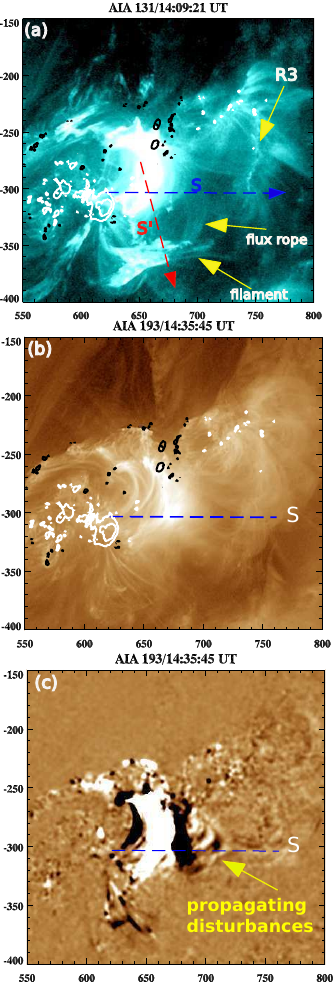}
\includegraphics[width=8.9cm]{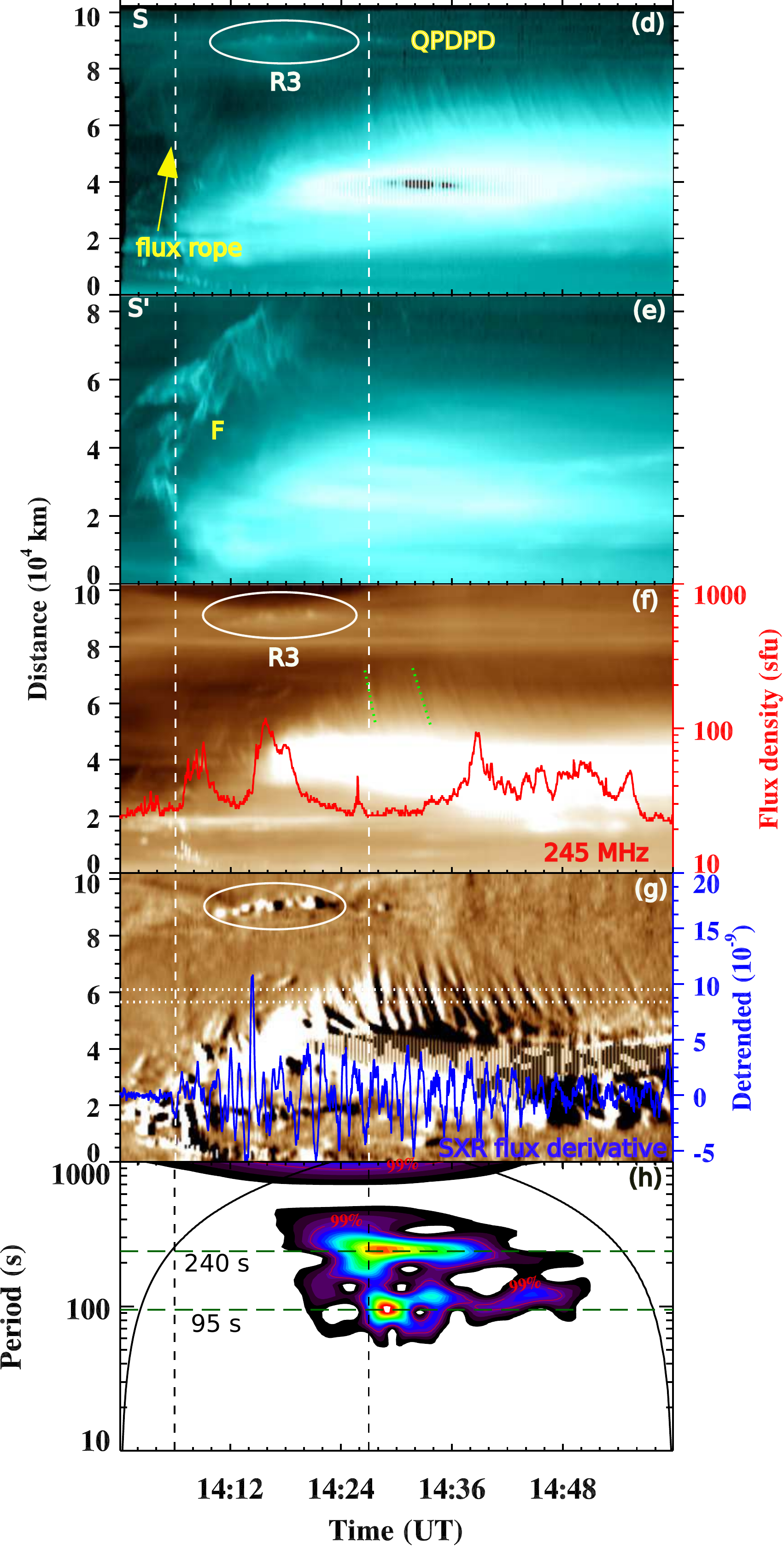}
}
\caption{(a-c) AIA 131 and 193~\AA~ intensity and running-difference images. The blue dashed line indicates the slice (S) used to create the time-distance intensity plots for the study of quasi-periodic downward propagating disturbances above the flare loops. (d-h) Time-distance AIA 131 and 193~\AA~ intensity and running difference ($\Delta$t=1 min)  plots along slices S and S'. The red curve is the 245~MHz radio flux density profile from RSTN. The two white vertical dashed lines indicate the timing of the quasi-periodic disturbances propagating upwards. The dotted green lines represent the paths used for the estimations of the typical speed of the downward propagating disturbances. The blue curve is the GOES soft X-ray flux time derivative. The wavelet power spectrum of the running-difference EUV intensity extracted between two horizontal dotted lines is shown in panel (g).} 
\label{sad}
\end{figure*}
\subsection{X-ray and radio sources}
Figure~\ref{hessi}(a) shows the RHESSI X-ray flux profile in the 3--6, 6--12, 12--25, 25--50~keV energy bands. RHESSI missed the impulsive phase of the M1.9 flare. However, initiation and decay phases of the flare were observed. We utilised the RHESSI images in the 6--12 and 12--25~keV channels during the flare initiation and decay phase. We selected detectors \# 3--9,  and used the `CLEAN' algorithm \citep{hurford2002} for image reconstruction with an integration time of 30-s. 

Figure~\ref{hessi}(b) shows the AIA 131~\AA~ image overlaid by RHESSI 6--12~keV contours during the flare initiation (14:01~UT). We see a loop-like structure in 6--12~keV channel, connecting N and P1. This is the site where the flare was triggered, associated with the eruption of the hot plasma (i.e., the flux rope seen at AIA 131~\AA) and a frontal loop system (335~\AA). During the flare decay phase ($\sim$14:24~UT), we observe a loop-top source at 6--12 (red) and 12--25~keV (blue), flare ribbons (R1 and R2, AIA 1600~\AA), and associated post-flare loops at AIA 94~\AA~ (Figure~\ref{hessi}(c,d)).

The quasi-periodic energy release during the solar flare was associated with the acceleration of nonthermal electrons in the solar corona, producing multiple radio bursts (e.g., type III, metric, decimetric and microwave emission, see Section~\ref{radiosec}).  To identify the particle acceleration sites in the corona, we analysed 10-s cadence radio imaging data obtained with NRH at the 150, 228, 298, 408~MHz frequencies. We  correlated the radio sources in time and frequency with the NRH 10-s integrated flux profiles (e.g., 150 MHz, 208, 298 MHz). The flux profiles are shown in the bottom panel of Figure 7(a). 
 Figures~\ref{nrh}(a-d) display NRH radio contours overlaid on the AIA 171 running difference ($\Delta$t=1~min) images. These contours are selected almost at the peak time of each metric/decimetric burst during $\sim$14:04--14:26~UT. 
 The radio emission is basically the plasma emission depending on the local electron density ($f [\mathrm{Hz}] \approx 9\times 10^3 \sqrt{n_e[\mathrm{cm}^{-3}]}$, where $n_e$ is the electron concentration). The radio sources are seen to appear approximately at/above the cusp. The evolution of the radio sources can be seen in the movie available online. We see a single radio source during the flare initiation (14:04~UT onward). Later on, at $\sim$14:16~UT, a double component/source was observed, which extends northward. Simultaneously, we observed an intense decimetric/microwave emission in the dynamic radio spectrum (14:16~UT) and 1-s cadence RSTN flux density profile ( Section~\ref{radiosec}). It seems to be produced in a broad acceleration region behind the erupting CME frontal loop observed at AIA 335~\AA. A single source was observed at $\sim$14:20~UT onward.

Figures~\ref{nrh}(d-e) show the AIA 171 and 94~\AA~ intensity images along with NRH/RHESSI contours at $\sim$14:24~UT. The 12--25~keV source is located above the hot arcade loops seen at 94~\AA. The AIA 171~\AA~ image shows a system of two loops, as well as open structures in the flaring active region. The metric type III emission is likely to be excited by the electron beam propagating along the open structures. The particle acceleration site should be somewhere between the X-ray loop-top and the NRH sources, which matches approximately the location of the cusp region (the null point) in the figure. Interestingly, multiple radio bursts and wave fronts originate at the same site, suggesting a common origin and common injection site for HXR and metric/decimetric emissions \citep[e.g.][]{vilmer2002}. 

\subsection{Quasi-periodic downward propagating disturbances}
\label{downward}

At 14:14--15:00~UT, in AIA 193 and 131~\AA~ images, we observed another long duration quasi-periodic propagating pattern. After the escape of the flux rope, we see quasi-periodic disturbances that propagate apparently downwards, towards the flare arcade. These quasi-periodic downward propagating disturbances (QPDPD) are rather persistent, occurring during the impulsive, maximum, and decay phases of the flare. An AIA 131~\AA~ image shows the formation of underlying flare loops associated with the eruption of the hot flux rope (marked by the arrow) and rotating filament near the southern footpoint of the flux rope (Figure~\ref{sad}(a)). Panels~(b,c) show snapshots of the 193~\AA~ intensity and its running difference at 14:35:45~UT. The location of the QPDPD coincides approximately with the source of the RHESSI 12--25~keV emission.

The QPDPD fronts have a concave (bent inwards) horseshoe shape, and thus resemble shrinking loops in a cusp-shaped structure above the flare arcade. To estimate the speed and temporal evolution of these QPDPD, we selected a slice S (eastward direction) to create the time--distance intensity plots (Figure~\ref{sad}(d, f)) and a running difference plot (Figure~\ref{sad}(g)) during 14:00--15:00~UT. Another cut, S', was chosen in the southward direction to show the rotating filament during the flux rope eruption. The time--distance plots show the occurrence of  QPDPD in both the AIA 131 and 193~\AA~ channels. Note that the AIA 193~\AA~ bandpass includes the emission of the \ion{Fe}{24} line (T$\sim$20~MK), indicating that the QPDPD are likely to be disturbances of the emission of a hot plasma with a temperature of $\sim$10--20~MK. Figure~\ref{sad}(e) shows the rotating filament along slice S'. Also, we included a 245~MHz radio flux profile (red curve, Figure~\ref{sad}(f), c.f. Figure~\ref{rstn1}) to show the repeated type-III radio bursts during the flare quasi-periodic energy release. Two vertical white dashed lines indicate the timing (14:06--14:27~UT) of the fast wave train propagating upwards, discussed in Section~\ref{wave train}. The typical apparent propagation speeds of QPDPD, tracked along the two dotted lines (Figure~\ref{sad}(f)) were $\sim$172--273$~(\pm$15)~\kms, which is much lower than the speed of the upward propagating wave train ($\sim$664--1615~\kms). A remote ribbon (R3) was also observed during 14:10--14:22~UT. It is evident that the fast wave train propagating upwards was detected during the rise of the flux rope, occurrence of the rotating filament F, the appearance of remote brightening R3, and the beginning of the QPDPD. On the other hand, we see the QPDPD for a much longer time interval, $\sim$14:14--15:00~UT. 

We took the average running difference intensity extracted between two horizontal dotted lines in Figure~\ref{sad}(g) to determine the periodicity of the QPDPD with the wavelet transform (Figure~\ref{sad}(h)). Interestingly, the QPDPD have periods of 95~s and 240~s with the significance level of 99\%. These periods are not exactly the same as observed simultaneously in the SXR flux time derivative (70--140~s), and in the upward rapidly-propagating wave train (decreasing from 240~s  to 120~s). However, the periods are sufficiently close to each other to indicate that they may have the same origin.       

\section{DISCUSSION AND CONCLUSION}

We presented multi-wavelength observations of a long duration M1.9 eruptive flare with quasi-periodic processes. The main findings of this study are as follows:

(1) We detected a quasi-periodic pulsation (QPP) in the GOES SXR flux time derivative with periods of $\sim$70~s and $\sim$140~s during the flare impulsive phase (14:06--14:27~UT), and $\sim$100~s after the flare maximum (14:30--15:00~UT). A similar periodicity ($\sim$140~s) was detected in the metric/decimetric radio bursts (i.e., 245/610~MHz) during 14:06--14:27~UT. 

(2) A quasi-periodic wave train propagating rapidly upward was detected in the AIA 171 and 193~\AA~ channels during the flare impulsive phase (14:06--14:27~UT). The wave front consists of a sequence of multiple wavefronts ($\sim$6--7) propagating outwards from the apparent energy-release site in the corona. The speeds of the fronts range in $\sim$664--1416~\kms, and the instant period decreasing from $\sim$240~s to 120~s. This value of the period is close to the period of the metric/decimetric radio bursts and SXR flux time derivative, suggesting their common origin. 

(3) The flare started at $\sim$14:00~UT with a lateral/radial expansion of the overlying active region loops. The speed of the lateral expansion of the CME (F1) in the low corona was $\sim$64--472~\kms~ (as seen at 171/335~\AA). In addition, a fast EUV wavefront (F2, $\sim$657~\kms) was also detected ahead of the CME frontal loops (F1), which may be interpreted as a piston-driven EUV fast-mode wave. The COR-1 image shows the front F2 ahead of the CME flanks (north and south), suggesting the presence of EUV wave. However, no type-II radio burst was observed during the eruption. The AIA 131~\AA~ images reveal the formation/eruption of a hot flux rope structure with a speed of $\sim$200~\kms, which is linked to the core of the CME observed by STEREO COR1.  

(4) The magnetic field configuration of the flare site was similar to a breakout topology. AIA 171~\AA~ images clearly show a cusp structure (a possible magnetic null point) and underlying closed loops well before the flare. RHESSI images reveal a loop-top source above the flare arcade, whereas NRH images show the metric/decimetric type III sources almost above the cusp. Therefore, the particle acceleration site is likely to be in the cusp region between the NRH source and the RHESSI loop-top source. The type III radio sources (NRH) and the EUV quasi-periodic rapidly-propagating wave train originate nearly at the same region. 

(5) We observed a filament (F) rotating in the clockwise direction near the southern footpoint of the flux rope. The flux rope eruption and filament rotation were seen to be associated with the formation of a remote ribbon R3. It is likely that the erupting flux rope interacts with the cusp region or the null point, causing reconnection. The particle acceleration during the reconnection may produce R3 by a downward electron beam that follows the magnetic field with the footpoint at R3. Likewise, the metric/decimetric type III radio bursts are produced by the corresponding upward electron beam during the flare impulsive phase. Note that quasi-periodic  upward propagating wave train of the EUV emission was detected during the flux rope eruption and formation of the remote ribbon (R3), suggesting their close connection. 

(6) In addition to the upward propagating wave train, we see quasi-periodic downward propagating EUV intensity disturbances, QPDPD. The disturbances consist of multiple propagating fronts too, and last from about 14:14~UT till about 15:00~UT, during both the impulsive and decay phases of the flare. The QPDPD have two periodicities, 95~s and 240~s. The wave fronts propagate at the speed of 172--273~\kms. The wave fronts have the concave horseshoe shape, which is different from the convex shape of the wave fronts in the upward propagating wave train. The disturbances are seen in the 131~\AA\ and 193\AA\ channels of SDO, and hence occur in the hot plasma of the supra-arcade of the flaring site.

(7) We observed tadpole-like structures in the wavelet power spectra of metric/decimetric bursts (245, 610 ~MHz). The periods of the multiple quasi-periodic radio bursts was found to be $\sim$70~s, 140~s.\\

Thus, the time sequence of the main phenomena detected in this flare is as follows: \\

\begin{tabularx}{\linewidth}{lX}

14:00   & The flare begins.\\
14:06--14:27 & Quasi-periodic variations of the SXR emission with the periods of about 70~s and 140~s. \\
14:06--14:27 & Quasi-periodic variations of the radio signals at 245/610~MHz with the period of about 140~s.\\
14:06--14:27 & Quasi-periodic wave train with the period decreasing from 240~s  to 120~s, rapidly propagating upward at the speed 664--1416~\kms, in 171~\AA\ and 193~\AA.\\
14:10--14:22 & Remote ribbon forms.\\
14:14--15:00 & CME is tracked.\\
14:14--15:00 & Quasi-periodic downward propagating disturbances with the periods of 95~s and 240~s at the speed ranging in 172--273~\kms, at 131~\AA\ and 193~\AA.\\
14:16--14:24 & Quasi-periodic pulsations at 800--2000~MHz, with the period decreasing from 170~s to 30~s.\\
14:30--15:00 & Quasi-periodic variations of the SXR emission with a period of about 100~s.\\
\end{tabularx}

The QPP in SXR may also arise from a modulation of the source volume and density, resulting in a modulation of the emission measure.
 The quasi-periodic variation of the thermal, e.g., SXR signal could be produced by the variation of the plasma density (and/or temperature) in a compressive wave, e.g. fast or slow magnetoacoustic wave, propagating through the emitting plasma.
This possibility is unlikely because in this case the quasi-periodic modulation of the nonthermal emission should not be present there. Since, we observed quasi-periodic nonthermal radio emission (metric/decimetric/microwave and coronal type-III bursts) during the QPP in SXR. Therefore, QPP in SXR (derivative) is a result of quasi-periodic energy release. In addition, the observation of QPDPD during the SXR QPP also supports the quasi-periodic energy release.\\

The quasi-periodic wave train of the EUV emission, which propagates upward at the speed exceeding 1000~\kms is most likely another example of a coronal fast magnetoacoustic wave train, similar to those detected by \citet{Liu2011} and in the follow-up works. We see that the periodicity of the quasi-periodic EUV fronts is similar as the quasi-periodic variation of the SXR derivative and radio emissions. These quasi-periodic processes could be produced by quasi-periodic magnetic reconnection. Furthermore, a numerical model by \citet{longcope2007} states that fast-mode waves can be launched by non-stationary current-sheet reconnection.  However, it is not clear whether the quasi-periodicity of the energy release is spontaneous, or the wave train induces the quasi-periodic energy release by the mechanism developed by \citet{nakariakov2006}. In the latter case, the wave train periodicity appears because of the geometrical dispersion of fast magnetoacoustic waves \cite[e.g.][]{roberts1984, nakariakov2004, pascoe2013}. The apparent decrease in the period (observed in EUV and decimetric emission) is consistent with the dispersive evolution. Moreover, the observed difference of the instant periodicities of the quasi-periodic signals detected in different bands is also consistent with this scenario, as the different emissions may be quasi-periodically modulated by the fast wave trains in different phases of its dispersive evolution. Our results confirm that quasi-periodic pulsations of the metric/decimetric radio emission are associated with the quasi-periodic fast wave train, supporting the findings of \citet{mes2009b, mes2009a, mes2013}. To the best of our knowledge, the simultaneous observation of a coronal fast wave train and tadpoles in decimetric radio bursts has not been reported before. 

It is more difficult to explain the concave horseshoe-shaped quasi-periodic EUV intensity disturbances that propagate apparently downward at the speed of 172--273~\kms, QPDPD. These features are seen in the EUV emission of the hot plasma, beginning during/after the launch of the flux rope. On one hand, the apparent downward propagation and the appearance in the hot plasma of the supra-arcade of the flaring site suggest that QPDPD could be a manifestation of the supra-arcade downflows.

Supra-arcade dark lanes, sometimes called supra-arcade downflows are hot (T$\sim$10~MK) features above the flare loops, moving apparently downward, often observed in long-duration eruptive flares. They have been interpreted as shrinking loops/flux tubes, voids behind retracting loops in a vertical current sheet, or the development of the Rayleigh--Taylor instability at the head of reconnection low-density jets \citep[e.g.][and references therein]{mckenzie1999}, \citealt{savage2012}, \citealt{innes2014}. Supra-arcade dark lanes are found to correlate with repeated HXR bursts, and are hence linked with the magnetic reconnection high in the corona \citep{asai2004}. An alternative interpretation based on seismological information provided by transverse oscillations of the dark lanes suggests that the plasma flows in the supra-arcade dark lanes are actually directed upward \citep[e.g.][]{2005A&A...430L..65V}.

Independently of the actual nature of the supra-arcade dark lanes, the concave horseshoe shape of the propagating fronts seen at AIA 193~\AA~ may indicate the sequence of shrinking loops that are shed quasi-periodically from the quasi-periodic reconnection site, and move downward from the cusp region. This interpretation is supported by the time coincidence of the occurrence of QPDPD and the quasi-periodic pulsations in the SXR time derivative. The numerical simulation of the breakout model \citep{karpen2012} revealed similar multiple wave fronts (propagating with Alf\'venic speed) associated with the bursty magnetic reconnection (i.e., along with plasmoids) in the current-sheet behind an erupting flux rope. However, a detail study of multiple, quasi-periodic wavefronts in a breakout topology has not been carried out yet. On the other hand, supra-arcade dark lanes are usually seen as narrow wiggly dark lanes extended in the radial direction, which are gradually apparently moving downwards. This picture does not correspond to the detected in our study. 

Another possibility would be to interpret the QPDPD as the downward propagating counterpart of the quasi-periodic upward rapidly-propagating wave train. In this case the difference in the oscillation period could be attributed to the difference in the dispersive evolution of fast waves in the oppositely directed fast magnetoacoustic waveguides. In this scenario the concave horseshoe shape of the propagating fronts could be associated with the refraction determined by the non-uniformity of the fast speed. Indeed, the wavefront deformation was clearly seen in the numerical simulations of this phenomenon, see, e.g., the inverse-bent fast wave fronts in the internal part of the waveguiding loop in  Fig.~5 of \citet{nistico2014}. However, the relatively low apparent speed of QPDPD, which is rather in the slow magnetoacoustic range, contradicts this interpretation. Another question would be why the upward fast waves are seen during the impulsive phase only, while the downward waves of the same nature are seen for much longer. In principle, the discrepancy between the apparent speed ($\sim$200~\kms) and the expected fast speed (several hundred \kms~at least) could be attributed to the line-of-sight projection effect. 

The QPDPD could also be associated with the leakage of a sausage oscillation from a flaring loop \citep[e.g.][]{2012ApJ...761..134N}. It would naturally explain the concave horseshoe shape of the wave fronts, which resembles the shape of the oscillating loop. Also, the duration of the wave leakage could be much longer than the impulsive phase, and hence last till 15:00~UT. However, this interpretation has the same difficulty as the downward propagating fast magnetoacoustic wave train scenario, the low apparent phase speed. Moreover, the typical periods of sausage oscillations are usually estimated as a few tens of seconds maximum, which is about one order of magnitude shorter than the observed periodicity. 

It is also possible that QPDPD are slow magnetoacoustic waves that travel downward from the reconnection site along the magnetic field of the supra-arcade. In this case the observation is similar to a number of observations of slow magnetoacoustic waves in magnetic fan structures above sunspots \citep[e.g.][]{demoortel2012}, with the only difference that this time these waves are seen to propagate downward. The apparent phase speed is determined by the high temperature of the emitting plasma and by the line-of-sight projection effect. The shape of the wave fronts, similarly to the slow waves in coronal active region fans, is determined by the timing of the excitation of the waves in different magnetic flux tubes, and the local tube speed in each flux tube, and also possibly by the local values of the plasma downflows or upflows. This scenario is also consistent with the time coincidence of the occurrence of QPDPD and the quasi-periodic pulsations in the soft X-ray time derivative. Unfortunately, the available set of observational data, and the lack of relevant modelling do not allow us to discriminate among the above-mentioned interpretations of QPDPD. Our study indicates the need for a dedicated study of this phenomenon. Possibly, a combination of imaging and spectral data would allow one to discriminate between different interpretation of this phenomenon. 

\acknowledgments
We thank the referee for the constructive comments/suggestions that improved the manuscript considerably.
PK thanks Prof. J. Karpen, C.R. DeVore, and S.E. Guidoni for several fruitful discussions during the NASA/GSFC visit. SDO is a mission for NASA Living With a Star (LWS) program. We are grateful to the SDO, SOHO, SWAP, RHESSI, FERMI, and GOES teams for open access to their data. We are grateful to the RSTN, STEREO, RHESSI, and GOES teams for open access to their data. The NRH is funded by the French Ministry of Education and the R\'{e}gion Centre. The authors acknowledge the Nan\c{c}ay Radio Observatory / Unit\'{e} Scientifique de Nan\c{c}ay of the
Observatoire de Paris (USR 704-CNRS, supported by Universit\'{e} d'Orl\'{e}ans, OSUC, and R\'{e}gion Centre in
France) for providing access to NDA observations. National Radio Astronomy Observatory (NRAO) is operated for the NSF by Associated Universities, Inc., under a cooperative agreement. We thank Laurent Lamy and Christian Renie for their help in handeling the NDA data. This research was supported by the Korea Astronomy and Space Science Institute under the R\&D program `Development of a Solar Coronagraph on International Space Station (Project No. 2017-1-851-00)' supervised by the Ministry of Science, ICT and Future Planning. VMN acknowledges the support from the European Research Council under the \textit{SeismoSun} Research Project No. 321141, and the BK21 plus program through the National Research Foundation funded by the Ministry of Education of Korea. Wavelet software was provided by C. Torrence and G. Compo, and is available at \href{http://paos.colorado.edu/research/wavelets/}{http://paos.colorado.edu/research/wavelets/}. This research was supported by an appointment to the NASA Postdoctoral Program at the Goddard Space Flight Center, administered by Universities Space Research Association through a contract with NASA.


\bibliographystyle{apj}
\bibliography{reference}

\begin{thebibliography}{}

\bibitem[\protect\citeauthoryear{{Antiochos}}{{Antiochos}}{1998}]{antiochos1998}
{Antiochos}, S.~K. 1998, \apjl, 502, L181

\bibitem[\protect\citeauthoryear{{Antiochos}, {DeVore}, \&
  {Klimchuk}}{{Antiochos} et~al.}{1999}]{antiochos1999}
{Antiochos}, S.~K., {DeVore}, C.~R.,  \& {Klimchuk}, J.~A. 1999, \apj, 510, 485

\bibitem[\protect\citeauthoryear{{Asai} et~al.}{{Asai} et~al.}{2004}]{asai2004}
{Asai}, A., {Yokoyama}, T., {Shimojo}, M.,  \& {Shibata}, K. 2004, \apjl, 605,
  L77

\bibitem[\protect\citeauthoryear{{Aschwanden}}{{Aschwanden}}{2004}]{asc2004}
{Aschwanden}, M.~J. 2004, {Physics of the Solar Corona. An Introduction}
  (Praxis Publishing Ltd)

\bibitem[\protect\citeauthoryear{{Benz}}{{Benz}}{2017}]{2017LRSP...14....2B}
{Benz}, A.~O. 2017, Living Reviews in Solar Physics, 14, 2

\bibitem[\protect\citeauthoryear{{Chen} et~al.}{{Chen}
  et~al.}{2016}]{2016ApJ...833..114C}
{Chen}, S.-X., {Li}, B., {Xiong}, M., {Yu}, H.,  \& {Guo}, M.-Z. 2016, \apj,
  833, 114

\bibitem[\protect\citeauthoryear{{De Moortel} \& {Nakariakov}}{{De Moortel} \&
  {Nakariakov}}{2012}]{demoortel2012}
{De Moortel}, I.,  \& {Nakariakov}, V.~M. 2012, Royal Society of London
  Philosophical Transactions Series A, 370, 3193

\bibitem[\protect\citeauthoryear{{Goddard} et~al.}{{Goddard}
  et~al.}{2016}]{goddard2016}
{Goddard}, C.~R., {Nistic{\`o}}, G., {Nakariakov}, V.~M., {Zimovets}, I.~V.,
  \& {White}, S.~M. 2016, \aap, 594, A96

\bibitem[\protect\citeauthoryear{{Howard} et~al.}{{Howard}
  et~al.}{2008}]{howard2008}
{Howard}, R.~A., et~al. 2008, \ssr, 136, 67

\bibitem[\protect\citeauthoryear{{Hurford} et~al.}{{Hurford}
  et~al.}{2002}]{hurford2002}
{Hurford}, G.~J., et~al. 2002, \solphys, 210, 61

\bibitem[\protect\citeauthoryear{{Innes} et~al.}{{Innes}
  et~al.}{2014}]{innes2014}
{Innes}, D.~E., {Guo}, L.-J., {Bhattacharjee}, A., {Huang}, Y.-M.,  \&
  {Schmit}, D. 2014, \apj, 796, 27

\bibitem[\protect\citeauthoryear{{Jiricka} et~al.}{{Jiricka}
  et~al.}{1993}]{jiricka1993}
{Jiricka}, K., {Karlicky}, M., {Kepka}, O.,  \& {Tlamicha}, A. 1993, \solphys,
  147, 203

\bibitem[\protect\citeauthoryear{{Kaiser} et~al.}{{Kaiser}
  et~al.}{2008}]{kaiser2008}
{Kaiser}, M.~L., {Kucera}, T.~A., {Davila}, J.~M., {St.~Cyr}, O.~C.,
  {Guhathakurta}, M.,  \& {Christian}, E. 2008, \ssr, 136, 5

\bibitem[\protect\citeauthoryear{{Karpen}, {Antiochos}, \& {DeVore}}{{Karpen}
  et~al.}{2012}]{karpen2012}
{Karpen}, J.~T., {Antiochos}, S.~K.,  \& {DeVore}, C.~R. 2012, \apj, 760, 81

\bibitem[\protect\citeauthoryear{{Kerdraon} \& {Delouis}}{{Kerdraon} \&
  {Delouis}}{1997}]{kerdraon1997}
{Kerdraon}, A.,  \& {Delouis}, J.-M. 1997, in Lecture Notes in Physics, Berlin
  Springer Verlag, Vol. 483, Coronal Physics from Radio and Space Observations,
  ed. G.~{Trottet}, 192

\bibitem[\protect\citeauthoryear{{Kumar} \& {Cho}}{{Kumar} \&
  {Cho}}{2014}]{kumar2014}
{Kumar}, P.,  \& {Cho}, K.-S. 2014, \aap, 572, A83

\bibitem[\protect\citeauthoryear{{Kumar} et~al.}{{Kumar}
  et~al.}{2013}]{kumar2013ww}
{Kumar}, P., {Cho}, K.-S., {Chen}, P.~F., {Bong}, S.-C.,  \& {Park}, S.-H.
  2013, \solphys, 282, 523

\bibitem[\protect\citeauthoryear{{Kumar} \& {Innes}}{{Kumar} \&
  {Innes}}{2015}]{Kumar2015w}
{Kumar}, P.,  \& {Innes}, D.~E. 2015, \apjl, 803, L23

\bibitem[\protect\citeauthoryear{{Kumar} \& {Manoharan}}{{Kumar} \&
  {Manoharan}}{2013}]{Kumar2013blob}
{Kumar}, P.,  \& {Manoharan}, P.~K. 2013, \aap, 553, A109

\bibitem[\protect\citeauthoryear{{Kumar}, {Nakariakov}, \& {Cho}}{{Kumar}
  et~al.}{2016}]{kumar2016}
{Kumar}, P., {Nakariakov}, V.~M.,  \& {Cho}, K.-S. 2016, \apj, 822, 7

\bibitem[\protect\citeauthoryear{{Lecacheux}}{{Lecacheux}}{2000}]{lecacheux2000}
{Lecacheux}, A. 2000, Washington DC American Geophysical Union Geophysical
  Monograph Series, 119, 321

\bibitem[\protect\citeauthoryear{{Lemen} et~al.}{{Lemen}
  et~al.}{2012}]{lemen2012}
{Lemen}, J.~R., et~al. 2012, \solphys, 275, 17

\bibitem[\protect\citeauthoryear{{Lin} et~al.}{{Lin} et~al.}{2002}]{lin2002}
{Lin}, R.~P., et~al. 2002, \solphys, 210, 3

\bibitem[\protect\citeauthoryear{{Liu} et~al.}{{Liu} et~al.}{2012}]{liu2012}
{Liu}, W., {Ofman}, L., {Nitta}, N.~V., {Aschwanden}, M.~J., {Schrijver},
  C.~J., {Title}, A.~M.,  \& {Tarbell}, T.~D. 2012, \apj, 753, 52

\bibitem[\protect\citeauthoryear{{Liu} et~al.}{{Liu} et~al.}{2011}]{Liu2011}
{Liu}, W., {Title}, A.~M., {Zhao}, J., {Ofman}, L., {Schrijver}, C.~J.,
  {Aschwanden}, M.~J., {De Pontieu}, B.,  \& {Tarbell}, T.~D. 2011, \apjl, 736,
  L13

\bibitem[\protect\citeauthoryear{{Longcope} \& {Priest}}{{Longcope} \&
  {Priest}}{2007}]{longcope2007}
{Longcope}, D.~W.,  \& {Priest}, E.~R. 2007, Physics of Plasmas, 14, 122905

\bibitem[\protect\citeauthoryear{{Mann} et~al.}{{Mann} et~al.}{1999}]{mann1999}
{Mann}, G., {Aurass}, H., {Klassen}, A., {Estel}, C.,  \& {Thompson}, B.~J.
  1999, in ESA Special Publication, Vol. 446, 8th SOHO Workshop: Plasma
  Dynamics and Diagnostics in the Solar Transition Region and Corona, ed. J.-C.
  {Vial} \& B.~{Kaldeich-Sch{\"u}}, 477

\bibitem[\protect\citeauthoryear{{Mann} et~al.}{{Mann} et~al.}{2003}]{mann2003}
{Mann}, G., {Klassen}, A., {Aurass}, H.,  \& {Classen}, H.-T. 2003, \aap, 400,
  329

\bibitem[\protect\citeauthoryear{{McKenzie} \& {Hudson}}{{McKenzie} \&
  {Hudson}}{1999}]{mckenzie1999}
{McKenzie}, D.~E.,  \& {Hudson}, H.~S. 1999, \apjl, 519, L93

\bibitem[\protect\citeauthoryear{{M{\'e}sz{\'a}rosov{\'a}}
  et~al.}{{M{\'e}sz{\'a}rosov{\'a}} et~al.}{2013}]{mes2013}
{M{\'e}sz{\'a}rosov{\'a}}, H., {Dud{\'{\i}}k}, J., {Karlick{\'y}}, M.,
  {Madsen}, F.~R.~H.,  \& {Sawant}, H.~S. 2013, \solphys, 283, 473

\bibitem[\protect\citeauthoryear{{M{\'e}sz{\'a}rosov{\'a}}
  et~al.}{{M{\'e}sz{\'a}rosov{\'a}} et~al.}{2009a}]{mes2009b}
{M{\'e}sz{\'a}rosov{\'a}}, H., {Karlick{\'y}}, M., {Ryb{\'a}k}, J.,  \& {Ji{\v
  r}i{\v c}ka}, K. 2009a, \aap, 502, L13

\bibitem[\protect\citeauthoryear{{M{\'e}sz{\'a}rosov{\'a}}
  et~al.}{{M{\'e}sz{\'a}rosov{\'a}} et~al.}{2009b}]{mes2009a}
{M{\'e}sz{\'a}rosov{\'a}}, H., {Karlick{\'y}}, M., {Ryb{\'a}k}, J.,  \& {Ji{\v
  r}i{\v c}ka}, K. 2009b, \apjl, 697, L108

\bibitem[\protect\citeauthoryear{{Muhr} et~al.}{{Muhr} et~al.}{2014}]{muhr2014}
{Muhr}, N., {Veronig}, A.~M., {Kienreich}, I.~W., {Vr{\v s}nak}, B., {Temmer},
  M.,  \& {Bein}, B.~M. 2014, \solphys, 289, 4563

\bibitem[\protect\citeauthoryear{{Nakariakov} et~al.}{{Nakariakov}
  et~al.}{2004}]{nakariakov2004}
{Nakariakov}, V.~M., {Arber}, T.~D., {Ault}, C.~E., {Katsiyannis}, A.~C.,
  {Williams}, D.~R.,  \& {Keenan}, F.~P. 2004, \mnras, 349, 705

\bibitem[\protect\citeauthoryear{{Nakariakov} et~al.}{{Nakariakov}
  et~al.}{2006}]{nakariakov2006}
{Nakariakov}, V.~M., {Foullon}, C., {Verwichte}, E.,  \& {Young}, N.~P. 2006,
  \aap, 452, 343

\bibitem[\protect\citeauthoryear{{Nakariakov}, {Hornsey}, \&
  {Melnikov}}{{Nakariakov} et~al.}{2012}]{2012ApJ...761..134N}
{Nakariakov}, V.~M., {Hornsey}, C.,  \& {Melnikov}, V.~F. 2012, \apj, 761, 134

\bibitem[\protect\citeauthoryear{{Nistic{\`o}}, {Pascoe}, \&
  {Nakariakov}}{{Nistic{\`o}} et~al.}{2014}]{nistico2014}
{Nistic{\`o}}, G., {Pascoe}, D.~J.,  \& {Nakariakov}, V.~M. 2014, \aap, 569,
  A12

\bibitem[\protect\citeauthoryear{{Ofman} et~al.}{{Ofman}
  et~al.}{2011}]{2011ApJ...740L..33O}
{Ofman}, L., {Liu}, W., {Title}, A.,  \& {Aschwanden}, M. 2011, \apjl, 740, L33

\bibitem[\protect\citeauthoryear{{Oliver}, {Ruderman}, \& {Terradas}}{{Oliver}
  et~al.}{2015}]{2015ApJ...806...56O}
{Oliver}, R., {Ruderman}, M.~S.,  \& {Terradas}, J. 2015, \apj, 806, 56

\bibitem[\protect\citeauthoryear{{Pascoe}, {Nakariakov}, \&
  {Kupriyanova}}{{Pascoe} et~al.}{2013a}]{2013A&A...560A..97P}
{Pascoe}, D.~J., {Nakariakov}, V.~M.,  \& {Kupriyanova}, E.~G. 2013a, \aap,
  560, A97

\bibitem[\protect\citeauthoryear{{Pascoe}, {Nakariakov}, \&
  {Kupriyanova}}{{Pascoe} et~al.}{2013b}]{pascoe2013}
{Pascoe}, D.~J., {Nakariakov}, V.~M.,  \& {Kupriyanova}, E.~G. 2013b, \aap,
  560, A97

\bibitem[\protect\citeauthoryear{{Roberts}, {Edwin}, \& {Benz}}{{Roberts}
  et~al.}{1983}]{roberts1983}
{Roberts}, B., {Edwin}, P.~M.,  \& {Benz}, A.~O. 1983, \nat, 305, 688

\bibitem[\protect\citeauthoryear{{Roberts}, {Edwin}, \& {Benz}}{{Roberts}
  et~al.}{1984}]{roberts1984}
{Roberts}, B., {Edwin}, P.~M.,  \& {Benz}, A.~O. 1984, \apj, 279, 857

\bibitem[\protect\citeauthoryear{{Savage}, {McKenzie}, \& {Reeves}}{{Savage}
  et~al.}{2012}]{savage2012}
{Savage}, S.~L., {McKenzie}, D.~E.,  \& {Reeves}, K.~K. 2012, \apjl, 747, L40

\bibitem[\protect\citeauthoryear{{Schou} et~al.}{{Schou}
  et~al.}{2012}]{schou2012}
{Schou}, J., et~al. 2012, \solphys, 275, 229

\bibitem[\protect\citeauthoryear{{Shen} et~al.}{{Shen} et~al.}{2013}]{shen2013}
{Shen}, Y.-D., {Liu}, Y., {Su}, J.-T., {Li}, H., {Zhang}, X.-F., {Tian}, Z.-J.,
  {Zhao}, R.-J.,  \& {Elmhamdi}, A. 2013, \solphys, 288, 585

\bibitem[\protect\citeauthoryear{{Shestov}, {Nakariakov}, \& {Kuzin}}{{Shestov}
  et~al.}{2015}]{2015ApJ...814..135S}
{Shestov}, S., {Nakariakov}, V.~M.,  \& {Kuzin}, S. 2015, \apj, 814, 135

\bibitem[\protect\citeauthoryear{{Torrence} \& {Compo}}{{Torrence} \&
  {Compo}}{1998}]{torrence1998}
{Torrence}, C.,  \& {Compo}, G.~P. 1998, Bulletin of the American
  Meteorological Society, 79, 61

\bibitem[\protect\citeauthoryear{{Verwichte}, {Nakariakov}, \&
  {Cooper}}{{Verwichte} et~al.}{2005}]{2005A&A...430L..65V}
{Verwichte}, E., {Nakariakov}, V.~M.,  \& {Cooper}, F.~C. 2005, \aap, 430, L65

\bibitem[\protect\citeauthoryear{{Vilmer} et~al.}{{Vilmer}
  et~al.}{2002}]{vilmer2002}
{Vilmer}, N., {Krucker}, S., {Lin}, R.~P.,  \& {Rhessi Team}. 2002, \solphys,
  210, 261

\bibitem[\protect\citeauthoryear{{Warmuth}}{{Warmuth}}{2015}]{warmuth2015}
{Warmuth}, A. 2015, Living Reviews in Solar Physics, 12, 3

\bibitem[\protect\citeauthoryear{{Warmuth} \& {Mann}}{{Warmuth} \&
  {Mann}}{2005}]{warmuth2005}
{Warmuth}, A.,  \& {Mann}, G. 2005, \aap, 435, 1123

\bibitem[\protect\citeauthoryear{{Warmuth} \& {Mann}}{{Warmuth} \&
  {Mann}}{2011}]{warmuth2011}
{Warmuth}, A.,  \& {Mann}, G. 2011, \aap, 532, A151

\bibitem[\protect\citeauthoryear{{White} et~al.}{{White}
  et~al.}{2005}]{white2005}
{White}, S.~M., {Bastian}, T.~S., {Bradley}, R., {Parashare}, C.,  \& {Wye}, L.
  2005, in Astronomical Society of the Pacific Conference Series, Vol. 345,
  From Clark Lake to the Long Wavelength Array: Bill Erickson's Radio Science,
  ed. N.~{Kassim}, M.~{Perez}, W.~{Junor}, \& P.~{Henning}, 176

\bibitem[\protect\citeauthoryear{{Williams} et~al.}{{Williams}
  et~al.}{2002}]{williams2002}
{Williams}, D.~R., {Mathioudakis}, M., {Gallagher}, P.~T., {Phillips},
  K.~J.~H., {McAteer}, R.~T.~J., {Keenan}, F.~P., {Rudawy}, P.,  \&
  {Katsiyannis}, A.~C. 2002, \mnras, 336, 747

\bibitem[\protect\citeauthoryear{{Williams} et~al.}{{Williams}
  et~al.}{2001}]{williams2001}
{Williams}, D.~R., et~al. 2001, \mnras, 326, 428

\bibitem[\protect\citeauthoryear{{Yu} et~al.}{{Yu}
  et~al.}{2016}]{2016ApJ...833...51Y}
{Yu}, H., {Li}, B., {Chen}, S.-X., {Xiong}, M.,  \& {Guo}, M.-Z. 2016, \apj,
  833, 51

\bibitem[\protect\citeauthoryear{{Yu} et~al.}{{Yu}
  et~al.}{2017}]{2017ApJ...836....1Y}
{Yu}, H., {Li}, B., {Chen}, S.-X., {Xiong}, M.,  \& {Guo}, M.-Z. 2017, \apj,
  836, 1

\bibitem[\protect\citeauthoryear{{Yuan} et~al.}{{Yuan} et~al.}{2013}]{yuan2013}
{Yuan}, D., {Shen}, Y., {Liu}, Y., {Nakariakov}, V.~M., {Tan}, B.,  \& {Huang},
  J. 2013, \aap, 554, A144

\bibitem[\protect\citeauthoryear{{Zhang} et~al.}{{Zhang}
  et~al.}{2015}]{2015A&A...581A..78Z}
{Zhang}, Y., {Zhang}, J., {Wang}, J.,  \& {Nakariakov}, V.~M. 2015, \aap, 581,
  A78

\end{thebibliography}
\clearpage

\end{document}